\documentclass[12pt,preprint]{aastex}
\usepackage{amssymb}
\usepackage{amsmath}
\usepackage{graphicx}
\usepackage{apjfonts}

\newcommand\HII{H\,{\sc ii}~}

\newcommand\kms{km~s$^{-1}$}

\newcommand\cc{cm$^{-3}$}

\newcommand\mjb{mJy~beam$^{-1}$}
\newcommand\jb{Jy~beam$^{-1}$}
\newcommand\jyb{Jy~beam$^{-1}$}

\newcommand\pp{$^{\prime\prime}$}
\newcommand\um{$\mu$m}

\newcommand\q{$\sim$}

\newcommand\h{H$_{2}$}
\newcommand\msun{M$_{\odot}$}  

\newcommand\lsun{L$_{\odot}$}

\newcommand{\methanol}{CH$_3$OH}
\newcommand{\ammonia}{NH$_3$}

\newcommand{\elow}{$E_{l}$}
\newcommand{\hco}{HCO$^{+}$}
\newcommand{\hisoco}{H$^{13}$CO$^{+}$}
\newcommand\water{H$_{2}$O}

\newcommand{\meth}{CH$_3$OH}

\newcommand{\formald}{H$_{2}$CO}
\newcommand{\vlsr}{$v_{LSR}$}

\newcommand{\noprint}[1]{}

\begin{document}
\shortauthors{Cyganowski et al.}

\title{A Class I and Class II \methanol\/ Maser Survey of Extended Green Objects (EGOs) from the GLIMPSE Survey}
\author{C.J. Cyganowski\altaffilmark{1}, C.L. Brogan\altaffilmark{2},
  T.R. Hunter\altaffilmark{2}, E. Churchwell\altaffilmark{1}}

\email{ccyganow@astro.wisc.edu}

\altaffiltext{1}{University of Wisconsin, Madison, WI 53706}
\altaffiltext{2}{NRAO, 520 Edgemont Rd, Charlottesville, VA 22903}

\begin{abstract}
We present the results of a high angular resolution Very Large Array
(VLA) Class I 44 GHz and Class II 6.7 GHz \methanol\/ maser survey of
a sample of \q 20 massive young stellar object (MYSO) outflow
candidates selected on the basis of extended 4.5 \um\/ emission in
\emph{Spitzer} Galactic Legacy Infrared Mid-Plane Survey
Extraordinaire (GLIMPSE) images.  These 4.5 \um-selected candidates
are referred to as extended green objects (EGOs), for the common
coding of this band as green in three-color IRAC images.  The
detection rate of 6.7 GHz Class II \methanol\/ masers, which are
associated exclusively with massive YSOs, towards EGOs is
$\gtrsim$64\%--nearly double the detection rate of surveys using other
MYSO selection criteria.  The detection rate of Class I 44 GHz
\methanol\/ masers, which trace molecular outflows, is \q89\% towards
EGOs associated with 6.7 GHz \methanol\/ masers.  The two types of
\methanol\/ masers exhibit different spatial distributions: 6.7 GHz
masers are centrally concentrated and usually coincide with 24 \um\/
emission, while 44 GHz masers are widely distributed and generally
trace diffuse 4.5 \um\/ features.  We also present results of a
complementary James Clerk Maxwell Telescope (JCMT) single-pointing
molecular line survey of EGOs in the outflow tracers \hco(3-2) and
SiO(5-4).  The \hco\/ line profiles and high SiO detection rate (90\%)
are indicative of the presence of active outflows.  No 44 GHz
continuum emission is detected at the 5 \mjb\/ (5$\sigma$) level
towards 95\% of EGOs surveyed, excluding bright ultracompact \HII\/
regions as powering sources for the 4.5 \um\/ outflows.  The results
of our surveys constitute strong evidence that EGOs are young, massive
YSOs, with active outflows, presumably powered by ongoing accretion.
\end{abstract}


\section{Introduction}\label{introduction}

The role and physics of accretion in dense (proto)cluster environments are
central to understanding massive star formation, yet remain poorly understood
\citep[c.f.][]{ZY07}.  This is in part attributable to the rapid evolution of
massive young stellar objects (MYSOs): sources in a phase of active accretion
are expected to be rare \citep[e.g.][]{ZY07}, and are observationally
difficult to identify in distant, crowded, and heavily extincted massive star
forming regions (MSFRs).

The 4.5 \um\/ band of the \emph{Spitzer Space Telescope's} InfraRed
Array Camera \citep[IRAC,][]{Fazio04} offers a promising new approach
for identifying MYSOs with outflows that are, presumably, still
actively accreting.  The 4.5 \um\/ IRAC band (``band 2'') is notably
lacking in PAH features \citep[c.f. Fig. 1 of][]{Reach06} and contains
molecular lines that may be shock-excited in protostellar outflows,
including both the CO (v=1-0) bandhead and \h\/ lines, most notably
the v=0-0, S(9,10,11) transitions.  Weaker \h\/ lines can also, in
aggregate, contribute significantly to the 4.5 \um\/ broadband flux under some
shock conditions, or if photodissociation region (PDR) emission is present
\citep[e.g.][]{SmithRosen05,Smith06,Ybarra09}.  There are also \h\/
lines in the other IRAC bands, and indeed, in the more extreme shock
environment of supernova remnants (SNR), \citet{NeufeldYuan} have
found that all IRAC bands may be dominated by \h\/ emission.  In
MSFRs, however, PAH emission generally dominates the 5.8 and 8.0 \um\/
bands.  
Thus extended 4.5 \um\/ emission with morphology distinct from that in
the other IRAC bands is a common and conspicuous feature of MSFRs.
Such features are known as ``Extended Green Objects (EGOs)''
\citep{egocat} or ``green fuzzies'' \citep{Chambers09}, for the common
coding of the [4.5] band as green in three-color IRAC images.  In the
massive DR21 outflow, comparison to \emph{ISO} spectra and narrowband
near-infrared (NIR) \h\/ images support the interpretation of extended
4.5 \um\/ emission as tracing shocked \h\/ in the outflow \citep[][and
  references therein]{Smith06,Davis07}.

Reported size scales for outflows in MSFRs span more than an order of
magnitude, from $<$0.1 pc to $>$1 pc \citep[e.g.][]{Hunter95,Beuther02}, and
the apparent extent of an outflow may depend on the angular resolution and
tracer \citep[e.g.][]{Hunter08,Klaassen08}.  The resolution of IRAC
\citep[$<$2\pp\/ in all bands,][]{Fazio04} is sufficient to resolve MYSO
outflows of length $\gtrsim$0.1 pc as extended emission, distinguishable from
point sources, in MSFRs nearer than the Galactic Center (2\pp\/\q16000
AU\q0.08 pc at 8 kpc).  \citet{egocat} catalogued over 300 EGOs in the
Galactic Legacy Infrared Mid-Plane Survey Extraordinaire survey area
\citep[GLIMPSE,][]{Benjamin03}.  Catalogued EGOs were divided into ``likely''
and ``possible'' MYSO outflow candidates based primarily on the angular extent
and morphology of the 4.5 \um\/ emission \citep{egocat}.  EGOs for which the
MIR morphology may be attributable to multiple nearby point sources or to
image artifacts around bright sources, instead of to truly extended 4.5 \um\/
emission, were categorized as ``possible'' outflow candidates \citep[see
also][]{egocat}.  The GLIMPSE survey is also sensitive to another class of
objects identified with the early stages of massive star formation: infrared
dark clouds (IRDCs), seen in silhouette against MIR Galactic background
emission, particularly at 8 \um\/ \citep{Churchwell09}.  First identified and
catalogued in \emph{Infrared Space Observatory (ISO)} and \emph{Midcourse
Space Experiment (MSX)} surveys of the Galactic Plane, IRDCs are cold (T$<$20
K), dense (n(\h)$>$10$^{5}$ \cc) clouds of molecular gas and dust
\citep[e.g.][]{Carey98,Egan98,Hennebelle01,Simon06a,Simon06b}.  Recent studies
have shown that IRDCs host both protostellar cores and intermediate and
massive YSOs, pinpointing IRDCs as sites of the earliest stages of massive
star and cluster formation
\citep[e.g.][]{Rathborne05,Pillai06,Rathborne06,Rathborne07,Wang08,Ragan09}.
Based on the midinfrared (MIR) colors of EGOs and their correlation with IRDCs
and published 6.7 GHz \methanol\/ maser data, \citet{egocat} argued that EGOs
constitute a population of actively accreting MYSO outflow candidates.

There are two elements to this hypothesis: (1) EGOs are \emph{massive} YSOs
and (2) extended 4.5 \um\/ emission traces active molecular outflows from
these MYSOs (which may then be presumed to be actively accreting).  Recent
advances in the theoretical and observational understanding of \methanol\/
masers, particularly the differing conditions under which Class I and Class II
transitions are excited, suggest that \methanol\/ masers are well
suited for testing this hypothesis and establishing the nature of EGOs.

Collisionly excited Class I \methanol\/ masers are observationally
well-correlated with molecular outflows in MSFRs
\citep{PlambeckMenten90,Johnston92,Cragg92,Kurtz04}.  In the largest
interferometric survey to date, \citet{Kurtz04} found that 44 GHz Class I
\methanol\/ masers were common in MSFRs and well correlated with shocked gas
in outflows as traced by SiO and \h.  Based on the relation of 95 GHz Class I
\methanol\/ maser emission to molecular cores traced by CS and outflow shocks
traced by \h\/ in DR21 and on the close-to-systemic velocities of the masers
(v$_{maser}$\q v$_{LSR,gas}$), \citet{PlambeckMenten90} suggested that Class I
\methanol\/ masers may be excited specifically in interface regions where
outflows encounter surrounding molecular material.  Class I \methanol\/ masers
in several transitions are associated with shocked \h\/ knots in the outflow
from IRAS 16547-4247 \citep{Brooks03,Voronkov06}, a well-studied MYSO
that is also an EGO catalogued by \citet{egocat}.  \citet{Chen09} find that
67\% of the EGOs catalogued by \citet{egocat} are within 1\arcmin\/ of a 44 or
95 GHz Class I \methanol\/ maser, but that the Class I \methanol\/ maser data
available in the literature--mostly from single dish surveys with poor angular
resolution ($\ge$52\pp)--is insufficient to establish physical association
between the maser and MIR emission.  The single dish surveys used by
\citet{Chen09} are also shallow (detection limits $\gtrsim$5 Jy), and
comparison with the \citet{Kurtz04} results suggests that shallow surveys miss
a substantial fraction of the maser population, at least for 44 GHz Class I
\methanol\/ masers.  Only \q29\% of the masers reported by \citet{Kurtz04}
have peak flux densities $\gtrsim$5 Jy, although \q33\% of Molinari sources
and \q59\% of all sources with 44 GHz detections are associated with at least
one $\gtrsim$5 Jy maser.  There is evidence that relatively weak ($\lesssim$11
Jy) 44 GHz \methanol\/ masers can occur in bipolar outflows from low-mass
protostars \citep{Kalenskii06}, but the authors caution that the maser nature
of the observed emission requires confirmation with interferometric
observations.

Class II 6.7 GHz \methanol\/ masers, radiatively pumped by IR emission
from warm dust \citep[][and references therein]{Cragg92,Cragg05}, are
associated exclusively with massive YSOs \citep[e.g.][]{Minier03}.
Sensitive searches towards low-mass YSOs
\citep[L$\lesssim$10$^{3}$ \lsun, M$\lesssim$3 \msun;][3$\sigma\lesssim$0.2
  Jy]{Minier03,Bourke05,Xu08}, including hot corinos
\citep[][3$\sigma$\q0.003 Jy]{Pandian08}, have uniformly yielded null
results.  In addition to being an observationally robust result, the
lack of 6.7 GHz \methanol\/ masers towards low-mass YSOs may be
understood theoretically in the context of the excitation models: the
energetics of low-mass YSOs do not produce regions with the necessary
combinations of dust temperature, density, and \methanol\/ abundance
for 6.7 GHz maser emission \citep{Minier03,Pandian08}.

Unaffected by extinction and accessible at high angular resolution with the
Expanded Very Large Array (EVLA), 6.7 GHz Class II and 44 GHz Class I
\methanol\/ masers provide an observationally efficient avenue for testing
whether EGOs are massive YSOs with outflows.  This paper presents the results
of a survey for 6.7 GHz Class II and 44 GHz Class I \methanol\/ masers towards
a sample of \q 20 EGOs selected from the catalog of \citet{egocat}, chosen to
cover a range of MIR properties and be visible from the northern hemisphere.
We also report results from simultaneous EVLA 44 GHz psuedo-continuum
observations and from a James Clerk Maxwell Telescope (JCMT) molecular line
survey undertaken to complement the maser observations by providing
information about the gas kinematics and LSR velocities.  In \S\ref{obs} we
describe the observations, in \S\ref{results} we present the
results of the surveys, in \S\ref{discussion} we discuss the implications of
our results for understanding the nature of EGOs, and in \S5 we summarize our main conclusions.


\section{Observations}\label{obs}

\subsection{Very Large Array (VLA)}\label{vlaobs}

The 6.7 GHz observations reported here were made possible by the expanded
tuning capability of the NRAO\footnote{The National Radio Astronomy
Observatory is a facility of the National Science Foundation operated under
agreement by the Associated Universities, Inc.} Expanded Very Large Array
(EVLA) at C band.  A sample of 28 EGOs was initially observed in the 6.668519
GHz \methanol\/ (5$_{1}$-6$_{0}$)~A$^{+}$ (\elow=48.7 K) maser transition on
2007 October 31 in the B configuration using Doppler tracking: 11 EVLA
antennas were available for these observations.  Unfortunately, we discovered
that Doppler tracking with the EVLA antennas produced large phase jumps
between the targets and calibrators for all but one source (G11.92$-$0.61) for
which the sky frequency of the target and calibrator were sufficiently close
to permit calibration (this problem has since been fixed). For the other
sources the phase stability was sufficiently good during the short
observations that maser emission could be detected in uncalibrated
vector-averaged u-v plots (it is notable that weak masers or masers far from
the phase center would not have been apparent in these plots).  The 18
uncalibratable sources with maser features apparent in the original {\em u-v}
data were reobserved at 6.7 GHz in February 2008 in fixed-frequency mode; the
larger overheads associated with fixed-frequency observing did not permit
reobserving all 28 sources in the allotted time.  The fixed-frequency 6.7 GHz
EVLA observations were obtained on 2008 February 1 in the B configuration and
2008 February 9, 17, and 19 in the CnB configuration: 13 EVLA antennas were
available for these observations.  The sources were split into two groups
based on their declinations, and each group of sources was observed twice.
The bandwidth was 1.56 MHz (70 \kms); aliasing on EVLA-EVLA baselines using
the interim system (EVLA antennas and VLA correlator) meant that the usable
bandwidth was \q 30 \kms.  The aliasing is worse for narrow bandwidths and in
the presence of broadband continuum emission.  The spectral resolution was
3.05 kHz (0.14 \kms), prior to offline Hanning smoothing.  The data were taken
in single polarization mode.  For each source, the phase calibrator, bandpass
calibrator, flux calibrator, observation date, velocity range searched for
maser emission, and synthesized beam size are listed in
Table~\ref{67obstable}.

Our 44 GHz VLA observations, of the 18 EGOs observed with fixed frequency at
6.7 GHz plus G11.92$-$0.61, were obtained on 2008 February 16 and 18 in the
CnB configuration, in fixed frequency mode: 26 antennas were available for
these observations.  One IF was tuned to the 44.069410 GHz \methanol\/
(7$_{0}$-6$_{1}$)~ A$^{+}$ (\elow=62.9 K) maser transition; the bandwidth was
3.125 MHz (21 \kms, \q 19 \kms\/ usable due to the aliasing described above)
and the spectral resolution 24.4 kHz (0.17 \kms), prior to offline Hanning
smoothing.  Pseudo-continuum observations were obtained simultaneously
($\nu$=44.1 GHz, 25 MHz bandwidth divided into 16 channels).  The data were
taken in single polarization mode.  The bandpass calibrator was J2253+161 and
the flux calibrator was 3C286.  Fast switching was used with a cycle time of 2
minutes, and the pointing model was updated once per hour during the
observations.  For each source, the phase calibrator, observation date,
velocity range searched for maser emission, synthesized beam size, and
continuum rms from the wideband data are listed in Table~\ref{44obstable}.

In the calibration of both the 6.7 GHz and 44 GHz datasets, a model
was used for 3C286.  All maser data were self-calibrated using the
channel with the strongest maser emission (except for the
G37.48$-$0.10 44 GHz data, which did not have sufficiently strong
masers).  We estimate that the absolute flux calibration is accurate
to \q10\%, and that the absolute positional uncertainty is \q0\farcs1.
Our VLA observations were obtained prior to the single dish molecular
line observations of our EGO sample described in \S\ref{jcmtobs}.
Source velocities for tuning the VLA correlator were obtained from a
broad range of data from archives and the literature.  As subsequent
followup shows, this was not always ideal (see \S\ref{results}).  The
primary beam (FWHP) of the VLA is 6\farcm7 at 6.7 GHz and 1\farcm0
at 44 GHz.

The search for maser emission in the 6.7 and 44 GHz image cubes was automated
using the AIPS task SAD.  These data are dynamic range limited such that the
rms noise in channels with bright maser emission is significantly increased
compared to non-maser channels (this is due primarily to relatively poor {\em
u-v} coverage). To account for this limitation, the rms noise cutoff for SAD
was adjusted from a conservative value of $\sim 6\sigma_{channel}$ for the
majority of the channels in each cube to as high as $\sim$10$\sigma_{channel}$
for those channels containing strong emission. This was a greater issue for
the 6.7 GHz data than the 44 GHz data because the former had significantly
fewer operational antennas.  Tables~\ref{67obstable} and \ref{44obstable} list the
minimum, maximum, and median threshold values used for each source.  Only
spectral features above these thresholds in at least two channels are
reported.  The median 6$\sigma$ detection limit for the whole sample of 19
EGOs is \q 0.16 Jy (brightness temperature T$_{B}$\q1247 K) for the 6.7 GHz
data and \q 0.15 Jy (T$_{B}$\q330 K) for the 44 GHz
data.  Reported maser flux densities and intensities have been corrected for primary
beam attenuation.

\subsection{James Clerk Maxwell Telescope (JCMT)}\label{jcmtobs}

Observations of \hco(\emph{J}=3-2) and \hisoco(\emph{J}=3-2) were obtained at
the James Clerk Maxwell Telescope (JCMT)\footnote{The James Clerk Maxwell
Telescope is operated by The Joint Astronomy Centre on behalf of the Science
and Technology Facilities Council of the United Kingdom, the Netherlands
Organisation for Scientific Research, and the National Research Council of
Canada.} on 2008 May 28-30 towards 18 of the EGOs surveyed for \meth\/ masers
with the VLA (G10.34$-$0.14 excepted).  Two 1 GHz IFs were observed
simultaneously: one IF was centered on \hco(\emph{J}=3-2) at 267.558 GHz
(\elow=12.8 K) in the USB, while the other IF covered \hisoco(\emph{J}=3-2) at
260.255 GHz (\elow=12.5 K) in the LSB.  Also included in the \hisoco\/ tuning
was the \methanol(5$_{2,3}$-4$_{1,3}$) transition at 266.838 GHz (\elow=44.3
K) in the USB. Position-switching mode was used, with each EGO observed for
3.5 minutes on-source.  In addition, a subset of 10 EGOs was observed in
position-switching mode with a single, 1 GHz IF centered on SiO (5-4) at
217.105 GHz (\elow=20.8 K) in the LSB: 8 sources were observed during the 2008
May 28-30 observing run, and one source each on 2008 June 12 and 13.  The EGOs
in the SiO subsample were chosen to have associated 44 GHz Class I \methanol\/
masers and strong, broad \hco\/ line wings.  Integration times for the SiO
observations were adjusted based on weather and source elevation and ranged
from 10 to 15 minutes on-source.  The ``off'' positions were chosen to avoid
IRDCs and MSFRs, using 8 and 24 \um\/ \emph{Spitzer} images; the angular
separation between the ``off'' and target position ranged from \q
3\arcmin\/-7\arcmin.  We did not perform dedicated observations to check the
off positions for emission; however, no evidence for emission in the off beam
is seen in the \hisoco\/ spectra.  The spectral resolution was \q 488 kHz for
both observing setups, corresponding to a velocity resolution of \q 0.55
\kms\/ for the \hco\/ observations and \q 0.67 \kms\/ for the SiO
observations.  The SiO spectra were subsequently smoothed to a resolution of
1.5 \kms\/ to increase their signal-to-noise.  The data were obtained using
the A3 receiver and the ACSIS digital autocorrelation spectrometer, and
reduced using the STARLINK software package.  Line-free regions near each line
of interest were selected for baseline-fitting, and a linear baseline was
removed from each spectrum.

The observation dates, rms, and fitted line properties for each source are
listed in Table~\ref{jcmtfitstable}.  During the 2008 May run, T$_{sys}$ was
typically \q 250 K for the \hco\/ observations, and $\tau_{225}$ \q
0.06, 0.07, and 0.09 on 2008 May 28,29, and 30, respectively.
On 2008 June 12 and 13, T$_{sys}$ was \q 370 K and $\tau_{225}$ \q 0.15.  The half-power beamwidth of the
JCMT is \q 19\pp\/ at the frequency of \hco(3-2) and \q 23\pp\/ at
the frequency of SiO(5-4).

Archival data were downloaded from the JCMT archive for G10.34$-$0.14.
Maps (\q40\pp$\times$40\pp\/ in extent) of \hco(3-2)
and \hisoco(3-2), centered at $18^{\rm h}08^{\rm m}59^{\rm
  s}.9$, $-20^{\circ}$03'42\pp\/ (J2000.0) and including the EGO
position, were obtained on 24 April 2002 and 1 May 2002
respectively as part of project M02AU26.  T$_{sys}$ was \q470 K on 24
April 2002 and 345 K on 1 May 2002 at the time of the observations;
$\tau_{225}$ was \q0.14 (24 April 2002) and \q0.09 (1 May 2002).  The
data were taken with the A3 receiver and reduced as described above.
The spectrum of the pixel containing the EGO position
(Table~\ref{samplelit}) was extracted for each line, and the data
smoothed from its native resolution of \q 156 kHz (\q 0.175 \kms) to a
resolution of \q 0.55 \kms.  Fitted line properties for \hisoco from
the archival G10.34$-$0.14 data are included in
Table~\ref{jcmtfitstable}.

\section{Results}\label{results}

The sample of 19 EGOs successfully observed with the VLA consists
entirely of ``likely'' MYSO outflow candidates (see \S\ref{introduction}).  Our
sample otherwise spans a range of MIR properties and known
associations with other tracers of MYSOs, detailed in
Table~\ref{samplelit}.  About half the EGOs in the sample (10/19,
\q52\%) are associated with IRDCs \citep{egocat}, and a majority are
associated with (sub)mm cores (12/19, \q63\%, Table 4).  Since EGOs as
a class have not been systematically studied at (sub)mm wavelengths,
most available data is from maps targeting other sources that happen
to cover the EGO position or from studies of previously identified
\methanol\/ masers (see \S\ref{litcomparison}).  The early results of
the ATLASGAL 870 \um\/ survey illustrate the promise of upcoming blind
surveys of the Galactic Plane at (sub)mm wavelengths: a partial
catalog of compact sources in a 1$\times$1.5 deg$^{2}$ field near
l=19$^{\circ}$ includes the first reported submm counterparts for two
EGOs in our sample (G18.89$-$0.47 and G19.01$-$0.03).

Three-color IRAC images for the 19 EGOs in our sample showing 8.0 \um\/ (red),
4.5 \um\/ (green), and 3.6 \um\/ (blue), with MIPSGAL 24 \um\/ contours
overlaid, are presented in Figure~\ref{3color}.  The levels for the 24 \um\/
contours in Figure~\ref{3color} are given in Table~\ref{mipstable}, as are
estimated J2000 positions and fluxes for 24 \um\/ EGO counterparts.  Most
targeted EGOs have MIPS 24 \um\/ counterparts (see Fig.~\ref{3color} and
Table~\ref{mipstable}), as expected for MYSOs
\citep[e.g.][]{Robitaille06,Carey09}.  Three (G10.29$-$0.13, G28.28$-$0.36,
G49.42+0.33), however, are located near the edges of 8 and 24 \um\/-bright
nebulae, and have no detected discrete 24 \um\/ counterparts (see
Fig.~\ref{3color}a,m,s). In addition, three EGOs included in the
\citet{egocat} catalog that were not targeted in our maser survey
serendipitously fell within the VLA primary beam at 6.7 GHz (6\farcm7 FWHP),
though outside the much smaller 44 GHz primary beam (1\farcm0 FWHP).  These
sources are listed at the bottom of Tables~\ref{samplelit}, \ref{mipstable},
and \ref{summarytable} and include one ``likely'' and two ``possible''
candidates as categorized by \citet{egocat}.  Table~\ref{summarytable}
summarizes the observational properties of EGOs from our surveys.

Tables~\ref{maserfitparams_67} and \ref{maserfitparams_44}, available online
in their entirety, present fitted parameters for every maser detected in our
survey.  Table~\ref{maserfitparams_67} presents fits for the 6.7 GHz
\methanol\/ masers, and Table~\ref{maserfitparams_44} presents fits for the 44
GHz \methanol\/ masers.  The maser positions from
Tables~\ref{maserfitparams_67} and ~\ref{maserfitparams_44} are overplotted on
the three-color GLIMPSE images of targeted EGOs in Fig.~\ref{3color} (diamonds
represent 6.7 GHz \methanol\/ masers, crosses 44 GHz \methanol\/ masers).  In
these images, the symbol sizes are larger than the positional uncertainties.
No thermal emission from either the 6.7 or the 44 GHz \methanol\/ line was
detected in our VLA survey.  Continuum emission at 44 GHz was detected towards
only one EGO in the sample (G35.03+0.35); this continuum detection is
discussed in \S\ref{g3503}.

\subsection{Maser Detection Rates}\label{detectionrates}

The detection rate of 6.7 GHz Class II \meth\/ masers towards EGOs in our
survey is at least 64\%: of the 28 sources in our original sample, 18 have
maser emission coincident with the EGO.  This detection rate is a lower limit
because vector-averaged {\em u-v} plots select for strong masers near the
phase center (see \S\ref{vlaobs}): the 9 sources that were not reobserved
could have weak 6.7 GHz masers.  Including the three EGOs not targeted but
observed serendipitiously (see \S\ref{results}), the detection rate is
essentially the same (20/31, $\ge$ 64.5\%).  The masers range in strength from
\q 0.5 to 469.4 \jb\/ (T$_{B}$\q7390 to 1743038 K, see
Table~\ref{maserfitparams_67}).  This detection rate for 6.7 GHz masers
towards EGOs is considerably higher than those of single-dish surveys with
comparable sensitivity (rms $\lesssim$ 0.3 Jy) towards \emph{IRAS} or water
maser selected samples, which are \q 37\% and 35\%, respectively
\citep{Walsh97,Xu08}.  The detection rate toward EGOs is also much higher than
for a single-dish survey of GLIMPSE point sources selected on the basis of 8
\um\/ intensity and [3.6]-[4.5] colors, which is less than 20\%
\citep{Ellingsen07}.

The detection rate of 44 GHz Class I \meth\/ masers towards EGOs in our survey
is \q 90\% (17/19).  The masers span a broad range in intensity, from \q 0.2
\jb\/ for the weakest masers to 89.5 \jb\/ for the strongest (T$_{B}$\q403 to
131994 K, see Table~\ref{maserfitparams_44}).  Very few sensitive searches for
44 GHz \meth\/ masers in star-forming regions have been done to date.  Large
single-dish surveys \citep[e.g.][]{Bach90,Hasch90,Slysh94} have been
comparatively shallow (detection limits of $\gtrsim$ 5 Jy), yielding detection
rates of \q 13-32\%.  An exception is the study of \citet{Kurtz04}, who
searched 44 MSFRs for 44 GHz \meth\/ maser emission with the VLA to a median
6$\sigma$ detection limit of 0.24 Jy (the median 6$\sigma$ detection limit of
the current survey is 0.15 Jy).  The sample of \citet{Kurtz04} included
compact and ultracompact (UC) \HII regions \emph{with single dish 44 GHz
\meth\/ maser detections} (VLA detection rate 28/32, 88\%) and young massive
protostellar objects from the Molinari sample (VLA detection rate 9/12, 75\%).
Like the sources from the Molinari sample targeted by \citet{Kurtz04}, EGOs
are thought to be MYSOs in an early stage of formation, prior to the
development of UC \HII regions.  Our detection rate for 44 GHz \methanol\/
masers toward EGOs is greater than that of \citet{Kurtz04} toward Molinari
sources.  This higher detection rate toward EGOs is not attributable to the
greater sensitivity of our survey; for each EGO with associated 44 GHz
\methanol\/ masers in our survey, at least one maser is above the
\citet{Kurtz04} detection limit.

\subsection{Previous Observations of Masers toward the EGO Sample}\label{litcomparison}

\water\/ and 6.7 GHz \methanol\/ masers, both established tracers of massive
star formation, have been detected previously in the vicinity of some of the
EGOs in our sample, either through blind surveys or targeted searches of
nearby MSFRs (see Table~\ref{samplelit}).  However, most of the previous data
are from single dish telescopes, and consequently have relatively poor angular
resolution (0\farcm7-7\arcmin\/ for \methanol\/ and 0\farcm7-1\farcm9 for
\water). This resolution makes it difficult to establish a physical
association between an EGO and a maser reported in the literature.
Table~\ref{samplelit} presents a list of all the possible associations that we
have found, though there are a number of caveats.  For example, the spectrum
of the 6.7 GHz maser G24.93+0.08 \citep[][a blind survey with 5\farcm5
resolution and 30\pp\/ positional accuracy]{Sz02} is a blend of two masers
resolved by our VLA observations (G24.943+0.074, coincident with the EGO
G24.94+0.07, and G24.920+0.088), and the published position is \q1 \arcmin\/
from the EGO.  The position for G35.03+0.35 from \citet{Pandian07}, the
single-dish survey with the highest angular resolution (0\farcm7; a blind
survey with 7\pp\/ pointing rms), is offset from 4.5 and 24 \um\/ emission and
is 13\pp\/ from our interferometric position.  The \water\/ maser surveys
generally have smaller beams, but target maser and \emph{IRAS} positions that
are themselves uncertain.  Only three targeted EGOs (G11.92$-$0.61,
G23.01$-$0.41, G35.03+0.35) are definitely associated with known \water\/
masers (maser positions known to $<$1\pp\/ from interferometric observations,
see Table~\ref{samplelit}).


In addition to the targeted EGOs with possible 6.7 GHz \methanol\/ maser
associations from single dish observations (Table 4), we reobserved 8 EGOs
with previous interferometric detections of 6.7 GHz \methanol\/ masers to
obtain improved sensitivity and positional accuracy.  Our sample includes 5
EGOs with 6.7 GHz \methanol\/ maser detections in the \citet{Walsh98}
Australia Telescope Compact Array (ATCA) survey, and one EGO (G11.92$-$0.62)
that fell within the field of view of that survey but was not detected.  The
absolute uncertainty in the ATCA maser positions is $\gtrsim$ 1\pp (compared
to 0\farcs1 for our VLA observations).  The nominal detection limit of the
\citet{Walsh98} survey (0.5 Jy) is about three times poorer than that of our
VLA observations (0.16 Jy), but these authors targeted \emph{IRAS} sources
(0\farcm4 to 2\farcm2 from our EGO targets), reducing the sensitivity at the
EGO positions (ATCA primary beam \q8\farcm5 FWHP at 6.7 GHz).  In all cases,
we detect maser emission over a wider velocity range than reported by
\citet{Walsh98}, and our positions differ by up to 1\farcs8 (for the weak
maser associated G28.28$-$0.36 at -03$^{\circ}$ declination).  We also detect
two weak ($\lesssim$1.1 \jyb) 6.7 GHz maser spots towards G11.92$-$0.61
(\S\ref{g1192}), which were undetected by \citet{Walsh98}.  Since the
kinematics of 6.7 GHz \methanol\/ maser emission are of interest for
understanding the nature of EGOs, we also reobserved three sources for which
only positions and peak velocities were reported by \citet{Sz07} (and
references therein).

\subsection{Spatial Relationship of Masers and MIR emission}\label{spatial}

While detection rates for both 6.7 GHz Class II and 44 GHz Class I \methanol\/
masers are very high towards the EGOs in our survey, the spatial distributions
of the two maser types are quite different.  Figure~\ref{3color} illustrates
several general, salient points: (a) 6.7 GHz Class II \methanol\/ masers are
spatially concentrated, typically within one synthesized beam (\q 2\pp), (b)
6.7 GHz \methanol\/ masers are usually, but not invariably, coincident with
discrete 24 \um\/ sources, (c) each EGO is usually associated with only one
(at most two) locus of 6.7 GHz \methanol\/ maser emission, (d) 44 GHz Class I
\methanol\/ masers are spatially distributed, often across many tens of
arcseconds (10\pp\q0.2pc at a typical distance of 4 kpc), (e) most EGOs are
associated with two to many tens of locii of 44 GHz \methanol\/ maser
emission, and (f) many 44 GHz \methanol\/ masers are coincident with extended
4.5 \um\/ emission (seen as green in Figure~\ref{3color}) or appear to trace
edges where extended 4.5 \um\/ emission meets the surrounding environment
(e.g. an IRDC).  A corollary is that the positions of 6.7 and 44 GHz masers
are, in general, anticorrelated: 6.7 GHz masers are centrally concentrated,
towards 24 \um\/ sources, while 44 GHz masers are distributed across (and in
some cases, beyond) the area of extended 4.5 \um\/ emission.  These general
statements are discussed in more detail below and in
\S\ref{individualsources}.

\subsubsection{6.7 GHz Masers}\label{spatial67}

Each diamond plotted in Figure~\ref{3color} represents the fitted position of
a 6.7 GHz maser in a single velocity channel.  Although in general there are
many diamonds plotted for each source in Figure~\ref{3color}, they tend to
overlap at the sizescale of these images into one or a few maser groups.  To
account for this we have two approaches: (1) each cluster of masers (typically
at different velocities) that lie within a synthesized beam (\q 2\pp, see
Table~\ref{67obstable}) is defined as a ``maser group'', with its position
defined by the intensity weighted mean (Table~\ref{67iweighted_egos}); and (2)
the fitted position of each maser spot is reported in
Table~\ref{maserfitparams_67}.  The sub-synthesized-beam kinematic analysis
facilitated by these fits will be discussed in \S\ref{maserkin} and
\S\ref{individualsources}.  For each EGO in our survey within the FWHP of the
VLA primary beam at 6.7 GHz (6\farcm7; 19 targeted EGOs plus 3
serendipitiously observed), Table~\ref{67iweighted_egos} lists the properties
of associated 6.7 GHz \methanol\/ maser group(s) and the angular separation of
the 6.7 GHz maser group(s) and 24 \um\/ EGO counterpart.  Only two EGOs are
potentially associated with multiple 6.7 GHz maser groups: G11.92$-$0.61 and
G49.42+0.33.  The case of G49.42+0.33 is ambiguous: two spatially and
kinematically distinct maser spots are detected, separated by \q9\pp\/ (\q0.5
pc at 12.3 kpc, see also \S\ref{g4942}).  The western maser spot
(G49.416+0.326) is associated with the EGO, but lacks a discrete 24 \um\/
counterpart; however, G49.42+0.33 is much more distant than the other EGOs in
our sample, so the sensitivity of the MIPSGAL survey corresponds to
an intrinsically brighter 24 \um\/ upper limit.  The eastern maser spot
(G49.417+0.324) is associated with a 24 \um\/ source, but may not be
associated with the EGO.  

Table~\ref{67iweighted_egos} quantifies the correlation of 6.7 GHz
\methanol\/ masers and 24 \um\/ emission that is visually apparent in
Figure~\ref{3color}.  Excluding the ambiguous case of G49.42+0.33, of
the 21 remaining EGOs in Table~\ref{67iweighted_egos} (18 targeted
plus 3 serendipitously observed), 2 do not have 6.7 GHz maser
detections (G49.27$-$0.34 and G49.27$-$0.32) and 2 do not have
discrete 24 \um\/ counterparts (G10.29$-$0.13 and G28.28$-$0.36).  Of
the remainder, the angular separation between a 6.7 GHz \methanol\/
maser and the nearest 24 \um\/ counterpart is $\lesssim$1\pp\/ for
13/17 EGOs (76\%), and $\lesssim$3\pp\/ for all sources with 24 \um\/
detections.  \citet{Carey09} characterize the absolute positional
accuracy of the MIPSGAL survey by cross-correlating sources in 8 \um\/
GLIMPSE and 24 \um\/ MIPSGAL images and examining the distribution of
positional offsets.  The tail of the distribution extends up to 3\pp,
with the median offset being 0\farcs85.  The larger maser-24 \um\/
angular separations (\q2-3\pp) that we find in regions where the 24
\um\/ emission is saturated and/or likely consists of blended
contributions from several sources (G11.92$-$0.61, G19.36$-$0.03,
G35.03+0.35) are consistent with the greater uncertainties in
attempting to measure positions in such regions.  With the exception
of the two sources identified without discrete 24 \um\/ EGO counterparts
(G10.29$-$0.13, G28.28$-$0.36) and the ambiguous case of G49.42+0.33,
the positions of 24 \um\/ EGO counterparts and 6.7 GHz \methanol\/
masers from our survey are coincident within the combined
uncertainties of the MIPSGAL absolute astrometry \citep{Carey09} and
our measurements of the 24 \um\/ source positions (see note (a) to
Table~\ref{mipstable}).

\subsubsection{44 GHz Masers}\label{spatial44}
As shown in Figure~\ref{3color}, there is diversity in the
relationship of 44 GHz \methanol\/ masers to extended 4.5 \um\/
emission.  As for the 6.7 GHz \methanol\/ masers, each individual
magenta cross plotted represents a fitted position of a 44 GHz maser
in a single velocity channel.  Details of the 44 GHz maser kinematics
are discussed in \S\ref{maserkin} and \S\ref{individualsources}.  In
stark contrast to the Class II 6.7 GHz masers, the Class I 44 GHz
masers are, in general, widely distributed over tens of arcseconds.
Of the 17 EGOs with 44 GHz maser emission, only one is associated with
a single locus of maser emission.  For the majority of sources, the
distribution of maser emission is spatially and kinematically complex,
and not well characterized as an ensemble of mean ``spots'' (in
contrast to the 6.7 GHz masers discussed above).

A striking feature of the images in Figure~\ref{3color} is that many
sources ($\gtrsim$1/3) exhibit filamentary arcs of 44 GHz \methanol\/
maser emission (e.g. G11.92$-$0.61, G19.01$-$0.03, G19.36$-$0.03,
G22.04+0.22, G24.94+0.07, G35.03+0.35).  In these
sources, the maser arcs appear to trace out the edges of 4.5 \um\/
arcs or lobes.  Similarly, in G28.83$-$0.25, 44 GHz masers are
concentrated near the edges of the extended 4.5 \um\/ emission, though
the masers themselves do not trace out an arc.  Some 44 GHz
\methanol\/ masers are apparently dissociated from any obvious MIR
emission (e.g. the masers in IRDCs, offset from EGOs, in the
G18.89$-$0.47 and G28.83$-$0.25 fields).  Individual sources will be
discussed in more detail in \S\ref{individualsources}.

\subsection{Serendipitous Maser Detections}\label{serendipitous}

The three EGOs that were not targets of our survey but that
serendipitously fell within the VLA primary beam at 6.7 GHz are
discussed in \S\ref{detectionrates} and \S\ref{spatial67} and included
in Tables~\ref{samplelit}, \ref{mipstable}, \ref{maserfitparams_67} and
\ref{67iweighted_egos}.  In addition, we serendipitously detect seven
6.7 GHz masers unrelated to EGOs.  For each of these serendipitous
detections, Table~\ref{67iweighted_serendip} includes the maser
properties, the name of the nearest \emph{IRAS} point source, and
previous reports of 6.7 GHz maser emission in the literature.  Most of
the serendipitously detected masers in
Table~\ref{67iweighted_serendip} are associated with MIR-bright
regions of massive star formation and/or are identifiable with
\emph{IRAS} point sources; about half were reported in the study of
Walsh et al. (1998; see Table~\ref{67iweighted_serendip}).  The present
survey is about three times more sensitive than that of
\citet{Walsh98}, and so detects some masers over a wider velocity
range.  In addition to the intensity-weighted mean positions in
Table~\ref{67iweighted_serendip}, we include fitted parameters for
these sources in Table~\ref{maserfitparams_67}.

All 44 GHz \methanol\/ masers detected in our survey are reported in
Table~\ref{maserfitparams_44}, including those outside the half-power
point of the VLA primary beam at 44 GHz (FWHP 1\farcm0), because these
are, to our knowledge, the first reports of these masers in the
literature and hence the best data available.  The serendipitously
detected 6.7 GHz masers are widely separated from the target EGOs
(\q0\farcm6-4\arcmin\/ from the pointing center of the observations),
readily identifiable as distinct sources.  In many fields the case is
less obvious for 44 GHz masers which, as discussed above
(\S\ref{spatial44}), are widely distributed along and around diffuse
MIR emission.  In the absence of high-resolution observations in a
direct tracer of molecular outflows (e.g. CO or SiO), it is impossible
to know in a given case whether 44 GHz masers offset from the EGO are
tracing a larger-scale flow or identifying distinct sources.  Except
in the G18.67+0.03 and G28.28$-$0.36 fields (discussed in
\S\ref{individualsources}), all 44 GHz masers in an EGO field are
treated as potentially associated with that EGO.

\subsection{44 GHz Continuum Emission}\label{44cont}

The vast majority (18/19,\q95\%) of EGOs in our sample are not
detected in our 44 GHz psuedo-continuum data (\S\ref{vlaobs}).
Continuum emission detected towards the exception, G35.03+0.35, is
discussed in \S\ref{g3503}.  The 44 GHz wideband data are shallow,
with a typical 5$\sigma$ detection limit of \q5 \mjb\/ (see
Table~\ref{44obstable}).  The data are also most sensitive to compact
sources: the typical angular resolution is \q0\farcs5 (\q0.01 pc\q2000
AU at a typical distance of 4 kpc for the EGOs in our sample) and the
interferometer is not sensitive to smooth structures larger than
\q20\pp.  The implications of our 44 GHz continuum limits for the
powering sources of EGOs will be discussed further in
\S\ref{discussion_evolution}.

\subsection{Thermal Molecular Line Emission}\label{mollines}

Our JCMT molecular line observations provide context for our maser surveys by
establishing the thermal gas \vlsr\/ for each EGO and providing independent
outflow indicators.  Figure~\ref{jcmtlineplots} presents the JCMT \hco(3-2)\/
and \hisoco(3-2)\/ spectra towards each EGO in our sample.  The integrated 6.7
GHz maser spectrum for each EGO-associated maser group (see \S\ref{spatial67}
and Table~\ref{67iweighted_egos}) and the velocity range of 44 GHz maser
emission are overplotted on the JCMT spectra to illustrate the relationship of
the thermal gas, Class I, and Class II \methanol\/ maser velocities, discussed
in \S\ref{maserkin} and \ref{individualsources}.  For each source, the
velocity range shown is that searched for 6.7 GHz \methanol\/
maser emission (Table~\ref{67obstable}); the velocity range searched for 44
GHz \methanol\/ maser emission (Table~\ref{44obstable}) is delimited by dotted
vertical lines.  The exception is G49.42+0.33, which was observed at 44 GHz
using two different tunings centered on the two kinematically distinct 6.7 GHz
maser components; for this source, the velocity range searched for 6.7 GHz
maser emission is delimited by vertical dashed lines.

Every EGO in our targeted sample was detected in \hco(3-2)\/ (\elow=12.8 K):
as shown in Figure~\ref{jcmtlineplots}, the \hco\/ profiles are, in general,
non-Gaussian, and are characterized by self-absorption dips and broad line
wings.  Due to the self-absorption, we have not fit the \hco\/ profiles.  The
majority of sources in our sample have full widths to zero intensity
($\Delta$v$_{FWZI}$) in excess of 20 \kms\/ in \hco.  Of the 17 EGOs in our
survey with self-absorbed or asymmetric \hco(3-2)\/ profiles, the blue peak is
stronger in about half (7), while in the other half the red peak is stronger
(9), with 1 indeterminate (Table~\ref{summarytable}).  Contracting cloud
models predict ``blue-skewed'' (asymmetric, with the blue peak brighter)
profiles for optically thick lines such as low-energy \hco\/ transitions
\citep{Myers96,DeVries05}.  The approximately equal distribution of red and
blue-skewed profiles in our sample is consistent with the results of \hco(1-0)
surveys of 6.7 GHz \methanol\/ maser sources by \citet{Purcell06} and
\citet{Sz07}, although \citet{Klaassen07} suggest that higher J transitions
may be better suited to tracing infall.  A more detailed analysis of the JCMT
spectra will be presented in a future work.

\hisoco(3-2)\/ (\elow=12.5 K) emission is also detected at $\gtrsim$3 $\sigma$
in all observed EGOs.  For each EGO, Table~\ref{jcmtfitstable} lists the
amplitude, line centroid velocity, $\Delta$v$_{FWHM}$, and integrated line
intensity obtained by a single Gaussian fit to the \hisoco(3-2)\/ profile, and
the kinematic distance based on the prescription of \citet{Reid09} and the
\hisoco\/ line center velocity.  The distance listed is the near kinematic
distance, unless otherwise noted; the angular extent of EGOs (see
\S\ref{introduction}) and the association of EGOs with IRDCs \citep[see
also][]{egocat} support the adoption of the near kinematic distance.
Parameters from single Gaussian fits to the \methanol(5$_{2,3}$-4$_{1,3}$,
\elow=44.3 K) and SiO (5-4, \elow=20.8 K) lines are also listed in
Table~\ref{jcmtfitstable}, and the spectra are presented in Figures
~\ref{thermalmethspectra} (\methanol(5$_{2,3}$-4$_{1,3}$)) and
~\ref{siospectra} (SiO (5-4)).  If a transition was observed but not detected
at the 3$\sigma$ level, the 3$\sigma$ upper limit is listed in
Table~\ref{jcmtfitstable}.  We detect thermal \methanol(5$_{2,3}$-4$_{1,3}$)
emission toward 83\% (15/18) of surveyed EGOs.  The median line FWHM for our
sample is 3.3 \kms\/ for \hisoco(3-2) and 4.8 \kms\/ for
\methanol(5$_{2,3}$-4$_{1,3}$).


We detect SiO (5-4) emission towards 9/10 (90\%) of EGOs surveyed for SiO
emission, with a median FWHM of 10.5 \kms.  The weather was relatively poor
during the SiO observations of the source (G39.10+0.49) that was not detected,
and the 3$\sigma$ upper limit is comparable to or greater than the fitted
amplitudes of SiO lines towards several EGOs observed in better weather.  This
detection rate is much higher than that in a JCMT SiO(5-4) survey of MYSOs
with known CO outflows \citep[5/12\q42\%,][]{Gibb07}.  The sensitivity of the
\citet{Gibb07} survey is variable, however, and ranges from 1.3 to 5 times
our median rms; four of our detections would not have met a threshold of
3$\times$ their lowest rms.  Comparisons of our SiO(5-4) detection rate to
detection rates of surveys in other SiO transitions must be treated with
caution, because the relative line strengths of the low vs. high J transitions
depend on density and shock velocity \citep[e.g. Fig. 7 of][]{Schilke97}.  For
completeness, we note that \citet{Harju98} detected SiO(2-1) towards \q37\%
(137/369) of a sample of sources with \water\/ and OH masers and UC \HII
regions and SiO(3-2) towards \q52\% (95/183) of a subsample observed in both
transitions.  \citet{Klaassen07} detected SiO(8-7) towards \q61\% (14/23) of a
sample of UC \HII regions with evidence of molecular outflows (in CO or CS).
Our SiO(5-4) detection rate towards EGOs is comparable to that of \citet{debuizer09} in
SiO(6-5) towards a sample of MYSOs with linear distributions of 6.7 GHz
\methanol\/ masers and \h\/ emission in narrowband images (9/10; 90\%).

Eleven EGOs targeted in our survey have previously been searched for molecular
line wings indicative of outflow as part of single-dish surveys (resolution
10-60\pp) targeting \water\/ or \methanol\/ masers (in \hco, CO, CS, or SiO,
see Table~\ref{samplelit}).  Of these, \q73\% (8/11) were reported to have at
least tentative evidence of outflow activity.

\subsection{Maser and Thermal Gas Kinematics}\label{maserkin}

Figure~\ref{jcmtlineplots} presents a direct visual comparison of the
velocity ranges of thermal gas emission (\hco\/ and \hisoco), 6.7 GHz
\methanol\/ maser emission, and 44 GHz \methanol\/ maser emission for
each EGO.  The sheer kinematic diversity of our
sample is illustrated by
Figure~\ref{jcmtlineplots}.  The velocity extents of both 44 GHz Class I and 6.7 GHz Class
II \methanol\/ maser emission vary widely from source-to-source ($<$1
to \q12 \kms\/ and \q1.5 to 18 \kms\/ respectively), as does whether
maser emission is continuous across a velocity range or comprised of
multiple, kinematically separated components.  In general, 6.7 GHz
maser emission spans a wider velocity range than 44 GHz maser
emission, but this is not universally the case (exceptions include
G11.92$-$0.61, G19.01$-$0.03, G19.36$-$0.03, and G24.94+0.07).  The
usable bandwidth of the 6.7 GHz maser observations is also \q1.5
$\times$ greater (in \kms) than that of the 44 GHz maser observations
(\S\ref{vlaobs},Tables~\ref{67obstable}-\ref{44obstable}), and there
are sources in which 44 GHz maser emission extends to, and most likely
beyond, the edge of the observed velocity range (e.g. G10.29$-$0.13,
Figure~\ref{jcmtlineplots}a, G18.89$-$0.47,
Figure~\ref{jcmtlineplots}e).

One generalization that may be made based on Figure~\ref{jcmtlineplots} is
that 44 GHz Class I \methanol\/ maser emission is invariably (but not
exclusively) present at the gas \vlsr\/ (as indicated by the \hisoco\/
emission).  The velocity of Class II 6.7 GHz \methanol\/ maser emission, in
contrast, exhibits every possible permutation with respect to the velocity of
the thermal gas: coincident, anticorrelated, redshifted, blueshifted, and
every combination of these characteristics.  The relative
velocities of the thermal gas and 6.7 GHz \methanol\/ maser emission for each
source are listed in Table~\ref{summarytable}.


To investigate the spatial structure of the kinematic complexity of
the maser spectra, Figure~\ref{kinplots} presents plots of maser
positions color-coded by velocity.  For each source, the left-hand
panel of Fig.~\ref{kinplots} shows 44 GHz \methanol\/ maser positions
overplotted on the 4.5 \um\/ image from the GLIMPSE survey.  On the
scale of these plots the uncertainty in the 44 GHz maser positions is
smaller than the symbol size, and every fitted position from
Table~\ref{maserfitparams_44} is plotted individually.  For each
source, the black rectangle(s) overplotted on the 4.5 \um\/ image is
the field of view of the right-hand panel(s), which show the positions
of 6.7 GHz masers color-coded by velocity.  In the 6.7 GHz maser
plots, the sizes of the crosses correspond to the relative positional
uncertainties for each maser in Table~\ref{maserfitparams_67} (from
the SAD fitting).  While individual 6.7 GHz maser spots are unresolved
by the VLA synthesized beam (\q 2\pp), the positional uncertainty of
an unresolved source in a well-calibrated image (such as our
self-calibrated 6.7 GHz data), relative to other such sources in the
same image, goes as $\Delta\theta\sim\theta_{syn.beam}/(2\times SNR)$,
where SNR is the signal-to-noise ratio of the source
\citep{Fomalont99}.  For our 6.7 GHz data, a 10$\sigma$ detection
corresponds to a relative positional uncertainty of \q0\farcs15 for
the low-declination sources ($\delta\le-08^{\circ}$) and \q0\farcs10
for the high-declination sources ($\delta>-08^{\circ}$) in our sample
(see Table~\ref{67obstable}), while the strongest masers have relative
positional uncertainties at the milliarcsecond level
(Table~\ref{maserfitparams_67}).  To simplify the plots and
concentrate on the best-determined features, 6.7 GHz maser fits with
positional uncertainties corresponding to a SNR$\lesssim$10 are not
shown in Fig.~\ref{kinplots} and 6.7 GHz maser fits with positional
uncertainties corresponding to a SNR$\lesssim$30 are plotted as
narrower lines (0\farcs05$\le\Delta\theta<$0\farcs15 for the
low-declination sources and 0\farcs033$\le\Delta\theta<$0\farcs10 for
the high-declination sources).

For each source in Fig.~\ref{kinplots}, the absolute limits in \kms\/
of the velocity bins for that source are given in a legend (far
right): the bin color-coded as green is approximately centered on the
thermal gas \vlsr\/ as determined from the \hisoco\/ observations
(Table~\ref{jcmtfitstable}).  The EGO G49.27$-$0.34 is not shown in
Figure~\ref{kinplots} because no 6.7 GHz masers were detected and all
detected 44 GHz masers fall within a single (66.5$<$v$<$67.5 \kms)
velocity bin.  

In most cases, the targeted EGO is the only or the dominant MIR source
within the JCMT beam.  Mm-wavelength interferometric observations have
revealed protoclusters, on $\lesssim$10\pp\/ scales, in young MSFRs
without obvious MIR multiplicity in \emph{Spitzer} images
\citep[e.g. NGC6334I(N) and S255N,][Brogan et al. in
  prep.]{Hunter06,s255n,Klein09}.  Additional sources either
undetected or unresolved by MIPS at 24 \um\/ may likewise be present
within our JCMT beam.  In our EGO spectra (Fig.~\ref{jcmtlineplots}),
however, the profile of the (optically thin) \hisoco\/ is generally
single-peaked and at the velocity of the dip in the profile of the
(optically thick) \hco\/, suggesting that the \hco\/ profile shape
is likely due to dynamics and not superposition.  While 6.7 GHz
\methanol\/ maser emission is believed to originate near the
MYSO, there is little consensus in the literature on the
kinematics of the maser environment--e.g. disk, outflow, or other (see
\S\ref{discussion_masersinmsfr}).  For EGOs in which both thermal
\methanol\/ and \hisoco\/ emission are detected in our JCMT spectra,
the median offset between the fitted centroid velocities of the two
lines is $\lesssim$0.3 \kms.  We thus take the \vlsr\/ of the
\hisoco\/ emission as the best indicator of the EGO systemic velocity
that is available for all sources in our sample.  The preponderance of
44 GHz masers at or near the gas \vlsr, noted above with reference to
Figure~\ref{jcmtlineplots}, is reflected in Figure~\ref{kinplots}.

\subsection{Notes on Individual Sources}\label{individualsources}

\subsubsection{G10.29$-$0.13}\label{g1029}

This EGO is located in an IRDC, west of an extremely MIR-bright complex of
sources in the W31 giant molecular cloud (Fig.~\ref{3color}a).  The extended
4.5 \um\/ emission is roughly linear and is extended along an E-W axis.  Class
I 44 GHz \methanol\/ masers are detected at the ends of the linearly extended
4.5 \um\/ emission: the 44 GHz masers at the western tip are blueshifted and
those at the eastern tip are redshifted relative to the thermal gas \vlsr\/
(Fig.~\ref{kinplots}a).  The 6.7 GHz \methanol\/ maser group is located about
midway between the red and blue-shifted 44 GHz \methanol\/ masers and near the
southern edge of the extended 4.5 \um\/ emission.  \citet{Walsh98} report a
N-S linear distribution of 6.7 GHz masers, but report masers over a much
smaller velocity range (v=2.2-7.8 \kms) than detected in our survey
(v=1.7-19.9 \kms, Table~\ref{67iweighted_egos}).  In aggregate, the 6.7 GHz
\methanol\/ masers detected in our survey are oriented along a NE-SW axis,
while the brightest red and blueshifted masers are oriented along an E-W axis
(Fig.~\ref{kinplots}a), similar to the orientation of the extended 4.5 \um\/
emission.

The 6.7 GHz \methanol\/ masers span a much broader velocity range than the 44
GHz masers (Fig.~\ref{jcmtlineplots}a,~\ref{kinplots}a): to the blue, where
our velocity coverage for the two frequencies is comparable, the 6.7 GHz maser
emission extends \q9 \kms\/ further from the thermal gas \vlsr\/ (13.6 \kms,
Table~\ref{jcmtfitstable}) than the 44 GHz maser emission.  The strongest 6.7
GHz \methanol\/ maser emission is blueward of the \hco\/ emission, while the
stronger line wing in the \hco\/ profile is to the red
(Fig.~\ref{jcmtlineplots}a).
  

This source is unique in our sample in that the EGO lacks a discrete MIPS 24
\um\/ counterpart, but both 6.7 GHz Class II and 44 GHz Class I \methanol\/
masers are detected.  A submm clump is coincident with the EGO \citep[][see
Table~\ref{samplelit}]{Walsh03,DiFran08}, and a weak 24 \um\/ EGO counterpart
could be masked by the wings of the PSF from the adjacent saturated complex.
The submm data, combined with the lack of a clear MIPS 24 \um\/ counterpart,
suggest that the ``central'' source powering the 4.5 \um\/ outflow may be very
young, with its envelope contributing significantly to its SED at long
wavelengths \citep[c.f.][]{Robitaille06}.  Higher resolution observations are,
however, required to determine the number, luminosities and masses of compact
dust core(s) in this region and their relationship to the extended 4.5 \um\/
and \methanol\/ maser emission.

This field is notable for widely distributed 44 GHz \methanol\/ masers
(Figs.~\ref{3color}a,\ref{kinplots}a).  The eastern and western ensembles of
masers are $\gtrsim$30\arcsec\/ ($\gtrsim$ 0.3 pc at 2.2 kpc) from the EGO and
lie along a ridge of 850 \um\/ dust emission.  Without large-scale maps in a
molecular outflow tracer, it is impossible to know whether these 44 GHz masers
are excited by a large-scale outflow emanating from the EGO or by distinct
sources.

\subsubsection{G10.34$-$0.14}\label{g1034}

This EGO is located north of the MIR-bright complex in the W31 molecular cloud
described above (\S\ref{g1029}), and is located in an IRDC.
The extended 4.5 \um\/ emission of this EGO is predominantly linear, along a
NW-SE axis, with an additional patch of more diffuse 4.5 \um\/ emission
\q15\pp\/ to the SE (Figs.~\ref{3color}b,\ref{kinplots}b).  The spatial
relationship of the Class I \methanol\/ masers and the extended 4.5 \um\/
emission is complex; this is also true of the velocity distribution of the 44
GHz masers.  The most coherent features are (from E to W in the left panel
Figure~\ref{kinplots}b): (1) a concentration of slightly blueshifted masers
coincident with the diffuse 4.5 \um\/ emission to the SE; (2) two lines of
blueshifted masers to the south of the EGO, roughly parallel to each other and
\q5\pp\/ apart; and (3) a concentration of redshifted masers near the NW edge
of the linear 4.5 \um\/ emission.

The velocity distribution of the 6.7 GHz \methanol\/ maser emission is also
complex.  To the north is a dense concentration of masers spanning \q 9 \kms\/
(v\q7-16 \kms, Table~\ref{maserfitparams_67}).  To the south are two linear
features, one comprised predominantly of redshifted masers (oriented N-S), and
the other of blueshifted masers (oriented NE-SW).  This 6.7 GHz \methanol\/
maser was previously reported by \citet{Walsh98}, but over a much narrower
velocity range (v=14.3-16.7 \kms, compared to v=3.6-18 \kms\/ in our survey,
Table~\ref{67iweighted_egos}).  The 6.7 GHz maser group is located near
redshifted 44 GHz \methanol\/ maser emission.

\subsubsection{G11.92$-$0.61}\label{g1192}

This EGO is unique in our sample in having two spatially distinct 6.7 GHz
\methanol\/ maser groups, separated by \q5\pp\/
(Figs.~\ref{3color}c,\ref{kinplots}c).  The 4.5 \um\/ emission is bipolar in
appearance, consisting of a NE and a SW lobe.  Both 6.7 GHz \methanol\/ maser
groups, and MIPS 24 \um\/ emission, are coincident with the NE lobe.  The
morphology of the MIPS 24 \um\/ emission is elongated north-south, suggesting
that it may be a blend of emission from multiple sources (Fig.~\ref{3color}c).

In contrast to many of the other EGOs in our sample, the velocity
extent of emission from each 6.7 GHz \methanol\/ maser group is
narrow, and the total velocity extent of 6.7 GHz maser emission falls
well within the \hco\/ profile (Fig.~\ref{jcmtlineplots}c: maser spot
1, northern; maser spot 2, southern).  The emission from both 6.7 GHz
\methanol\/ maser groups is weak, and redshifted by a few \kms\/ from
the thermal gas \vlsr.  A strong (I$_{peak}$\q248 \jb) \water\/ maser
\citep{HC96}, marked by a triangle in Fig.~\ref{kinplots}c (left
panel), is offset by \q0\farcs5 from the southern 6.7 GHz
\methanol\/ maser.  The \water\/ maser emission peaks \q 3.3 \kms\/
redward of the \methanol\/ maser emission, but the \water\/ maser
emission spans a wide velocity range (16.4 \kms\/), and a weaker
secondary peak is present in the spectrum at the velocity of the
\methanol\/ maser \citep[c.f. Fig. 1a of][]{HC96}.

The 44 GHz Class I \methanol\/ maser emission in the vicinity of the
NE 4.5 \um\/ lobe is dominated by an arc of masers, near and
around the edge of the extended 4.5 \um\/ emission, from the northwest
to the southeast, and at or near the thermal gas \vlsr\/
(Fig.~\ref{kinplots}c).
This EGO is in an IRDC (Fig.~\ref{3color}c), and this morphology and velocity
structure are consistent with the masers tracing the interface between
outflow(s) (traced by diffuse 4.5 \um\/ emission) and the surrounding
environment (IRDC).  In contrast, the 44 GHz \methanol\/ masers associated
with the SW 4.5 \um\/ lobe are coincident with the 4.5 \um\/ emission (not
along its edges), are arranged linearly (not in an arc), and are predominantly
blueshifted (although a redshifted maser is also present).  The 44 GHz masers
towards the SW lobe are also the strongest in this region.  The JCMT beam was
centered on the NE lobe, so the reported thermal gas \vlsr\/ is that towards the
6.7 GHz masers and MIPS source.  Broad SiO(5-4) emission ($\Delta v_{FWHM}$=
9.6 \kms) was detected in our JCMT survey (Table~\ref{jcmtfitstable}).
Taken together, the evidence suggests that the ``central'' source(s)
responsible for driving the outflow(s) (traced by extended 4.5 \um\/ emission
and 44 GHz \methanol\/ masers) are located in the NE lobe (and hence
contribute to its MIR emission), while the SW 4.5 \um\/ lobe may be dominated
by emission from shocked outflow gas.

\subsubsection{G18.67+0.03}\label{g1867}

The relation of the 44 GHz \methanol\/ masers to the 4.5 \um\/
emission towards this EGO is varied; some masers are coincident with
bright 4.5 \um\/ emission, while others lie near its edge
(Figs.~\ref{3color}d,\ref{kinplots}d).  The velocity range of the 44 GHz maser
emission is extremely narrow ($\lesssim$ 2 \kms, centered on the thermal gas
\vlsr\/; Fig.~\ref{jcmtlineplots}d).  The 6.7 GHz \methanol\/ maser
emission is triple-peaked and spans a wider velocity range (\q5 \kms)
but still falls within the \hco\/ profile.  On 0\farcs1 scales, the
6.7 GHz \methanol\/ masers appear segregated by velocity, with a
concentration of blueshifted masers to the NW and a concentration of
predominantly systemic to slightly redshifted masers to the SE.  Very
little is known about this source beyond the results of the present
survey.  The EGO is not in a clear IRDC, and is \q 40\pp\/ east of a
submm source (G18.66+0.04) reported by \citet{atlasgal}.

This field is notable for detections of 44 GHz \methanol\/ masers that
are clearly associated with other MIR sources in the field and not
with the targeted EGO.  Figure~\ref{3color}d shows three compact MIPS
24 \um\/ sources, each with associated masers: from E to W (left to
right), these are: (1) the targeted EGO, G18.67+0.03, with associated
6.7 and 44 GHz masers; (2) a MIR source with associated 44 GHz masers
(l\q18.665, b\q0.030); and (3) a MIR source with an associated 6.7 GHz
\methanol\/ maser (G18.662+0.035, see
Table~\ref{67iweighted_serendip}) and 44 GHz \methanol\/ masers.  This
westernmost source is encompassed by the dimensions of the submm
source G18.66+0.04 \citep{atlasgal}.  As these sources are not the
focus of our study and information on the gas kinematics comparable to
that provided by our JCMT survey is not available, we do not discuss
the kinematics of these serendipitously detected masers.

\subsubsection{G18.89$-$0.47}\label{g1889}

This EGO is in an IRDC, SW of a multiband IRAC and 24 \um\/ source
(Fig.~\ref{3color}e).  Both MIR sources, and much of the IRDC, are encompassed
by the dimensions of a submm source reported by \citet{atlasgal}
(G18.89$-$0.47, M=2040 \msun).  The extended 4.5 \um\/ emission of this EGO is
predominantly linear, and is extended along a N-S axis.
The 6.7 GHz \methanol\/ maser emission towards this EGO is kinematically
complex ($\Delta$v\q4.3 \kms\/) and lies on the far blue wing of the \hco\/
profile (Fig.~\ref{jcmtlineplots}e).  The 6.7 GHz maser emission is offset to
the blue by \q 9-13 \kms\/ from the thermal gas \vlsr\/ (66.6 \kms,
Table~\ref{jcmtfitstable}), with the maser emission peak offset by \q 10 \kms.
The velocity coverage of our 6.7 GHz maser observations does not extend a
comparable amount to the red of the gas velocity, so whether the maser
emission is symmetric with respect to the thermal gas emission is not known.
Our 44 GHz maser observations (centered on the 6.7 GHz maser velocity) do not
extend redward of the gas \vlsr.  The velocities of the 44 GHz \methanol\/
masers \q20\pp\/ and 25\pp\/ to the SW, in the IRDC, are comparable to those
of the 44 GHz masers near the EGO (Fig.~\ref{kinplots}e).  The masers in the
IRDC may be associated with an extended outflow from the EGO or may be excited
by other sources.



\subsubsection{G19.01$-$0.03}\label{g1901}

This source is in many ways the MIR prototype of the EGO class: it has a clear
bipolar morphology at 4.5 \um, an identifiable central point source at IRAC
wavelengths and at 24 \um\/ (Fig.~\ref{3color}f), and had not been studied
prior to the EGO catalog of \citet{egocat}.  This EGO is located in an IRDC,
and \citet{atlasgal} have subsequently reported a submm clump of \q 1000
\msun\/ (dimensions 40\pp\/$\times$34\pp\/) coincident with the EGO.  Our
detection of 6.7 GHz \methanol\/ maser emission towards the central source
(Figs.~\ref{3color}f,\ref{kinplots}f) and the presence of a massive dust core
both support the identification of the EGO as a \emph{massive} YSO.  The 6.7
GHz \methanol\/ maser emission is kinematically complex
(Figs.~\ref{jcmtlineplots}f,\ref{kinplots}f).  The most blueshifted maser
emission is concentrated to the NW and has an arclike morphology, while
redshifted maser emission is concentrated to the east (Fig.~\ref{kinplots}f).

Many 44 GHz Class I \methanol\/ masers are detected towards this EGO; the masers are
mostly coincident with the extended 4.5 \um\/ emission and, to the north of
the central source, trace out a striking arc or loop.  This morphology is
similar to
that sketched by \citet{Kalenskii06} for Class I \methanol\/ masers excited in
an outflow with its axis perpendicular to the observer (e.g. in the plane of
the sky).  The arc-like morphology of the 44 GHz \methanol\/ masers observed
in G19.01$-$0.03 is also similar to that of 44 GHz \methanol\/ masers
attributed to bow shocks in a bipolar outflow in DR21(OH) \citep{Araya09}.
 
The velocities of the 44 GHz \methanol\/ masers in G19.01$-$0.03 span \q9
\kms\/, and the range is asymmetric with respect to the thermal gas \vlsr\/ (59.9 \kms,
Table~\ref{jcmtfitstable}): the maser emission extends farther to the blue
than to the red (Fig.~\ref{jcmtlineplots}f).  Figure~\ref{kinplots}f shows
that there is a pattern to the 44 GHz maser velocities.  North of the central
source, the masers generally have velocities systemic and blueward (by up to
\q 6.5 \kms).  In contrast, south of the central source there is a dearth of
blueshifted masers; the masers have velocities systemic and redward, though by
a maximum of \q 2.2 \kms\/.  This morphology and velocity structure suggest
that the 44 GHz Class I \methanol\/ masers and the extended 4.5 \um\/ emission
trace a bipolar outflow, with the blue lobe to the north and the red lobe to
the south of the central source.  Broad SiO(5-4) emission ($\Delta v_{FWHM}$=
9.8 \kms) was detected toward this EGO in our JCMT survey
(Table~\ref{jcmtfitstable}).  Further observations at high angular resolution
are required to establish the extent and kinematics of the outflow traced by
4.5 \um\/ emission and Class I 44 GHz \methanol\/ masers and to determine
whether it is excited by a single source or a (proto)cluster.



\subsubsection{G19.36$-$0.03}\label{g1936}

The extended 4.5 \um\/ emission for G19.36$-$0.03 extends to the south
and to the northwest of several multiband IRAC sources, while the MIPS
24 \um\/ emission is elongated along the same axis, suggesting blended
24 \um\/ emission from multiple sources (Fig.~\ref{3color}g).  The
6.7 GHz \methanol\/ maser group is coincident with 24 \um\/ emission,
\q 3\pp\/ south of the nominal MIPS peak position
(Table~\ref{67iweighted_egos}).  The most striking feature of the 44
GHz Class I \methanol\/ maser emission is an arc of masers to the
southeast that is coincident with 4.5 \um\/ emission.  The masers in
this arc span the entire observed velocity range, $\Delta v$\q 7
\kms\/ (Fig.~\ref{kinplots}g).  This EGO is in an IRDC and the arc of
44 GHz \methanol\/ masers may be tracing the interaction of a 4.5
\um\/ outflow with the surrounding environment, similar to
G11.92$-$0.61.  North of the arc, other 44 GHz \methanol\/ masers are
distributed along the edges of 4.5 \um\/ emission, coincident with the
northwestern 4.5 \um\/ extension, and coincident with the 24 \um\/
source.  Several 44~GHz masers are also coincident with filamentary 8
\um\/ emission just west of the EGO. A 44 GHz maser \q 20\pp\/ east of
the EGO is associated with an additional weak source of diffuse 4.5
\um\/ emission.  Very broad SiO(5-4) emission ($\Delta v_{FWHM}$=
20.6 \kms) was detected toward this EGO in our JCMT survey
(Table~\ref{jcmtfitstable}). 

Only one locus of 6.7 GHz \methanol\/ maser emission is detected, implying
that despite the complicated MIR morphology, only one YSO of sufficient mass
and early enough evolutionary state to excite 6.7 GHz \methanol\/ maser
emission is present.  There is structure in the velocity distribution of the
6.7 GHz \methanol\/ masers: redshifted masers are concentrated to the north
and east of blueshifted masers (Fig.\ref{kinplots}g).  There is also an E-W
velocity gradient in the redshifted masers, with the most redshifted masers to
the east.  \citet{Walsh98} report 6.7 GHz \methanol\/ maser emission towards
this source; compared to the present survey, their positions are offset by
\q0\farcs5 to the north.  The structure seen in our survey in the 6.7 GHz
\methanol\/ maser velocities on small spatial scales is not reflected in the
velocity distribution of the 44 GHz \methanol\/ masers on much larger spatial
scales.

\subsubsection{G22.04+0.22}\label{g2204}

G22.04+0.22 is located in an IRDC, south and east of two bright sources of MIR
emission (Fig.\ref{3color}h).  The 6.7 GHz \methanol\/ masers detected toward
this EGO are coincident with bright 4.5 \um\/ emission, and with 24 \um\/
emission (angular separation from 24 \um\/ peak is \q 2\pp\/;
Fig.~\ref{3color}h and Table~\ref{67iweighted_egos}).  In contrast, the 44 GHz
\methanol\/ maser emission tends towards the periphery of the 4.5 \um\/
emission, and is characterized by three main regions of maser emission--to the
north, northeast, and south--and a clump of masers to the southwest.  The
northeastern masers coincide with and trace the extended 4.5 \um\/ emission,
while the other regions of maser emission lie at the edges of or beyond the extended 4.5 \um\/
emission.  The two northern maser regions are kinematically well-mixed
(e.g. include masers across a range of velocities).  The southern region, in
contrast, consists predominantly of a N-S line of masers redshifted relative
to the thermal gas \vlsr; south of this line lies an E-W distribution of masers at the
systemic velocity and blueward (Fig.~\ref{kinplots}h). The 6.7 GHz \methanol\/
masers have a broadly N-S distribution, with the most blueshifted masers to
the south, redshifted masers in the middle, and masers with slightly
blueshifted and systemic velocities to the north.  The strongest, blueshifted,
6.7 GHz masers appear to be distributed in an arc-like pattern.  Very broad
SiO(5-4) emission ($\Delta v_{FWHM}$= 18.3 \kms) was detected toward this EGO
in our JCMT survey (Table~\ref{jcmtfitstable}).

The copious 44 GHz \methanol\/ maser emission, broad line wings of the
\hco\/ profile (Fig.~\ref{jcmtlineplots}h), and detection of SiO all
support the presence of molecular outflow(s) from this previously
unstudied MYSO, but their number, orientation and kinematics remain
unclear from the available data.  High resolution observations of a
direct molecular outflow tracer are required to understand the
kinematics of the outflow(s) in this region and their relation to the
unusual 44 GHz \methanol\/ maser morphology and kinematics.

\subsubsection{G23.01$-$0.41}\label{g2301}

This EGO is host to one of only seven known \formald\/ 6 cm masers, a
maser species which is thought to be associated with very young MYSOs
\citep{Araya08}, and is unusual among our sample in having been the
subject of detailed multiwavelength studies \citep[see for
  example][for a review]{Araya08}.  Zeeman splitting analysis of the
6.7 GHz \methanol\/ maser emission from this source by
\citet{Vlemmings08} indicates that the brightest maser peak
(v$_{peak}$=74.81 \kms, I$_{peak}$=469.36 \jyb\/ in our data) has a
strong negative magnetic field (B$_{\|}$= -30.3 mG), while the
redshifted maser components (v$>$79 \kms) have a weaker, positive
magnetic field (B$_{\|}$\q 14 mG).
Blueshifted (negative magnetic field) and redshifted (positive
magnetic field) 6.7 GHz \methanol\/ maser emission are spatially
distinct (Fig.~\ref{kinplots}i).  Blueshifted 6.7 GHz maser emission is
concentrated to the northeast and consists of two linear features with
broadly NE-SW orientations that meet to form a V-like morphology, with
the opening of the V oriented \q 45$^{\circ}$\/ east of north.
Redshifted 6.7 GHz maser emission is concentrated to the south/southwest, in a
more diffuse structure. The projected offset between the main
concentrations of blueshifted and redshifted 6.7 GHz masers is
$\sim$0\farcs2 ($\gtrsim$920 AU at 4.6 kpc).

The 6.7 GHz \methanol\/ maser emission is coincident with 3 mm
continuum emission observed at \q2\pp\/-5\pp\/ angular resolution by
\citet{Furuya08}. The intensity-weighted 6.7 GHz \methanol\/ maser
position from our observations is within 0\farcs4 of the average 3~mm
continuum peak position. Scaled to the \citet{Brunthaler09} parallax
distance of 4.6 kpc, the dust mass of the 3~mm core reported by
\citet{Furuya08} is \q 70 \msun.

The morphology of the high velocity molecular gas in the region \citep[mapped
at \q5\pp\/ angular resolution by][]{Furuya08} is complicated, and depends on
the tracer, but is generally oriented NE to SW. Both red and blueshifted
$^{12}$CO and $^{13}$CO emission are present to the NE and SW of the 3 mm
continuum source, while HNCO shows only blueshifted emission to the SW
\citep[Figure 6 in][]{Furuya08}.  The 44 GHz Class I \methanol\/ masers lie on
or along the edges of the outflow, but the mix of maser velocities--both red
and blueshifted masers to the SW--is more consistent with the complex velocity
structure seen in the CO observations.  However, red and blueshifted CO
emission extends to much higher relative velocities than the 44~GHz masers.
Broad SiO(5-4) emission ($\Delta v_{FWHM}$= 14.9 \kms) was detected toward
this EGO in our JCMT survey (Table~\ref{jcmtfitstable}), consistent with the
results of \citet{Harju98} (full width 34.4 \kms\/ for SiO(2-1) and 24.7
\kms\/ for SiO(3-2)).

The positions of the \formald\/ maser from \citet{Araya08} (black
diamond) and of \water\/ masers from \citet{FC99} (triangles) are
overplotted on the 4.5 \um\/ image in Figure~\ref{kinplots}i.  The
\formald\/ maser is located \q0\farcs1 to the east of the blueshifted
cluster of 6.7~GHz masers (Fig.\ref{kinplots}i) at a comparable
blueshifted velocity (73.6 \kms). The velocity range of the \water\/
masers (68.41-81.59 \kms) is comparable to that of the 6.7 GHz
\methanol\/ maser emission, but the \water\/ maser positions are
offset to the west by \q1\farcs5.

\subsubsection{G23.96$-$0.11}\label{g2396}

The relationship of the \methanol\/ masers to the MIR emission towards this
EGO suggest that the EGO may in fact be comprised of at least two distinct
sources: (1) a northern source, associated with a 6.7 GHz \methanol\/ maser
group, a cluster of 44 GHz \methanol\/ masers, and a weaker 24 \um\/ source,
and (2) a southern multiband IRAC source with a brighter 24 \um\/ counterpart
and extended 4.5 \um\/ emission to its south.  Most of the 44 GHz \methanol\/
masers are coincident with or at the edges of extended 4.5 \um\/ emission
(Fig.~\ref{3color}j,\ref{kinplots}j).  The brighest 6.7~GHz masers are
blueshifted with respect to the thermal gas \vlsr.  A position and peak
velocity for this 6.7 GHz maser were previously reported by \citet{Sz07} (and
references therein).  The JCMT beam encompassed both the northern and southern
24 \um\/ emission, and the \hisoco\/ profile is asymmetric, with a weaker,
redder secondary bump.  High resolution molecular line observations are
required to clarify the number, nature, and kinematics of sources in this
region.

\subsubsection{G24.94+0.07}\label{g2494}

This EGO is extended roughly north-south and lies east of
two multiband MIR point sources (Fig.~\ref{3color}k). The 6.7~GHz masers are coincident
with 24~\um\/ emission, and have an usually narrow velocity extent
($\Delta$v \q 2 \kms\/), with the peak being \q 5 \kms\/ redward of
the thermal gas \vlsr\/ and \q 2 \kms\/ redward of the reddest 44 GHz
\methanol\/ maser (Fig.~\ref{jcmtlineplots}k).  The 6.7 GHz maser group is spatially closest to
the reddest 44 GHz \methanol\/ masers (Fig.~\ref{kinplots}k).  Most of
the other 44 GHz \methanol\/ masers are at or near the thermal gas \vlsr\/ and
are distributed on or along the edges of extended 4.5 \um\/ emission.
The arc of masers to the northeast, in particular, traces a 4.5 \um\/
lobe.  SiO(5-4) emission ($\Delta v_{FWHM}$= 7.2 \kms) was detected toward
this EGO in our JCMT survey (Table~\ref{jcmtfitstable}).

\subsubsection{G25.27$-$0.43}\label{g2527}

This EGO, located in an IRDC (Fig.\ref{3color}l), is unique in our sample in
having only one locus of 44 GHz \methanol\/ maser emission.  The extended 4.5
\um\/ emission is roughly linear and is extended along a roughly NW-SE axis; the
44 GHz maser is near the edge of the extended 4.5 \um\/ emission.  The
velocity of the 44 GHz maser emission corresponds well with that of the \hco\/
and \hisoco\/ emission (Fig.~\ref{jcmtlineplots}l) and spans a range of \q 3
\kms.  The 6.7 GHz \methanol\/ maser emission, in contrast, avoids the
velocity range of the \hco\/ emission and is offset to the red and blue by up
to \q 8.5 \kms\/ from the thermal gas \vlsr.  The strongest blueshifted and redshifted
6.7 GHz maser components are spatially segregated, with the blueshifted masers
concentrated \q 0\farcs1 ($\gtrsim$400 AU at 3.9 kpc) north of the redshifted
masers (Fig.~\ref{kinplots}l).  In aggregate, the brightest red and
blueshifted masers are oriented along a NW-SE axis, similar to the orientation
of the extended 4.5 \um\/ emission.

\subsubsection{G28.28$-$0.36}\label{g2828}

This EGO, located just west of an 8 and 24 \um\/-bright MSFR, does not
have a discrete MIPS 24 \um\/ counterpart; as for G10.29$-$0.13, the
ability to identify such a counterpart is compromised by confusion
with the PSF of the adjacent saturated complex of sources.  The 6.7
GHz \methanol\/ maser group is coincident with bright 4.5 \um\/
emission near the eastern end of the EGO (towards the MIR-bright
nebula), and is blueshifted by \q7.5-9 \kms\/ from the thermal gas \vlsr\/
(Fig.~\ref{jcmtlineplots}m,\ref{kinplots}m).  No 44~GHz \methanol\/
masers were detected in the vicinity of the EGO.

\citet{Longmore07} detect two \ammonia\/ cores in the vicinity of the
6.7 GHz \methanol\/ maser group \citep[previously reported by][see
  Table~\ref{samplelit}]{Walsh98}.  As noted by \citet{Longmore07},
neither \ammonia\/ core is coincident with the 6.7 GHz maser emission:
Core 1 is offset \q8\farcs6 east-southeast and Core 2 is offset
\q10\farcs3 west-northwest of our intensity-weighted 6.7 GHz maser
position.  The two \ammonia\/ cores have quite different spectral
characteristics, with Core 2 exhibiting extreme \ammonia(1,1)
hyperfine asymmetries.  \citet{Longmore07} associate the \methanol\/
maser with Core 1, but note that the association is uncertain.  The
6.7 GHz \methanol\/ maser appears to lie on a ``bridge'' of
\ammonia(1,1) emission between the two cores \citep[c.f. Figure 11
  of][]{Longmore07}.  The \ammonia\/ velocities are redshifted by \q
1.1 \kms\/ (Core 1) to \q1.5 \kms\/ (Core 2) from that of the \hisoco\/
observed with the JCMT.  The JCMT beam also encompasses the 8 \um\/
source just northeast of the EGO, but the spectrum does not show
multiple velocity components.

No 44 GHz continuum emission is detected toward the targeted EGO, but
we do serendipitously detect the G28.29-0.36 UC \HII\/ region (at the
center of the MIR-bright MSFR noted above), reported previously at
lower frequencies \citep{Walsh98,Giveon05,Longmore07}.  Fitting a
single Gaussian component, we find an integrated 44 GHz flux density
for G28.29-0.36 of 174$\pm$23 mJy (corrected for the primary beam) at
$18^{\rm h}44^{\rm m}15^{\rm s}.09$, $-04^{\circ}17'54\farcs9$ and a
deconvolved source size of 2$\farcs$3$\times$1$\farcs0$
(PA=103$^{\circ}$).  Our measured 44 GHz integrated flux density is
significantly lower than that predicted by extrapolating the 24 GHz
continuum flux density of \citet{Longmore07} assuming optically thin
free-free emission.  The most likely explanation is that the 24 GHz
observations (\q10\pp\/ beam) are sensitive to extended emission that
is resolved out by our higher resolution (0\farcs5 beam) 44 GHz
observations.  The 44 GHz \methanol\/ masers shown in
Figure~\ref{3color}m appear to be associated with the MIR-bright MSFR
centered on the G28.29-0.36 UC \HII\/ region: fitted parameters for
these masers are reported in Table~\ref{maserfitparams_44}.

\subsubsection{G28.83$-$0.25}\label{g2883}

This EGO is located in an IRDC on the rim of a MIR ``wind-blown bubble''
\citep{bubblescat}, near two other young sources, and may represent an
example of triggered star formation \citep{christerbubbles}.  From north
to south (Fig.~\ref{3color}n), the three compact 24 \um\/ sources in
this field are: a compact \HII\/ region, the EGO counterpart, and a
multiband IRAC source that has been modeled as a massive YSO based on
its MIR SED \citep{christerbubbles}.

The 4.5 \um\/ morphology of the EGO has a bipolar structure with a 24 \um\/
counterpart and a 6.7~GHz maser group near the center of the east-west
lobes. A more diffuse ``tail'' of extended 4.5 \um\/ emission also extends
south and east of the two main 4.5 \um\/ lobes
(Figs.\ref{3color}n,\ref{kinplots}n).  Class I 44 GHz \methanol\/ masers are
spatially concentrated at the ends of the 4.5 \um\/ lobes and at the end of
the diffuse eastern 4.5 \um\/ ``tail''.  These 44 GHz masers show some
evidence of a velocity gradient associated with the bipolar lobes: the masers
at the end of the western lobe have velocities systemic and blueward, while
the masers at the end of the eastern lobe have velocities systemic and redward
(Fig.~\ref{kinplots}n).  The eastern 4.5 \um\/ tail, however, is associated
with blueshifted masers, muddying the picture.  The velocity spread of the 44
GHz masers is fairly narrow and symmetric about the gas systemic velocity
($\pm$\q1.8 \kms, Fig.~\ref{jcmtlineplots}n).  SiO(5-4) emission was detected
toward this EGO in our JCMT survey over a velocity range comparable to that of
the 44 GHz \methanol\/ maser emission (SiO $\Delta v_{FWHM}$= 5.9 \kms,
Table~\ref{jcmtfitstable}).

Class I 44 GHz \methanol\/ masers are
also detected in the IRDC: these masers could be associated with an
extended outflow from the EGO, with outflow(s) from the MIR source
south of the EGO, or with sources not visible in the MIR.

The 6.7 GHz \methanol\/ maser emission largely avoids the thermal gas \vlsr\/,
consisting of spectrally complex components that coincide with the red and
blue wings of the \hco\/ emission (Fig.~\ref{jcmtlineplots}n).  \citet{Walsh98} report 6.7 GHz \methanol\/
maser emission over a similar velocity range, but report an order of magnitude
fewer components, and in their data the positional uncertainty is greater than
the separation of the maser spots (see their Fig. 2, 18421-0348 panel).
Spatially, the 6.7 GHz maser emission we detect is generally linearly
distributed along a northwest-southeast axis (P.A. \q 45$\arcdeg$), with
blueshifted maser emission to the northwest, redshifted maser emission to the
southeast, and a \q 0\farcs3 gap in between (Fig.~\ref{kinplots}n).  Notably,
the most highly doppler-shifted maser emission is nearest this central gap,
with less doppler-shifted emission at the extremes of the linear distribution.
This configuration is suggestive of Keplerian rotation, though the linear
extent of the maser distribution ($\gtrsim$3000 AU at 5.0 kpc) is rather large
for the masers to be tracing an accretion disk.  A detailed analysis is beyond
the scope of this work, and will be presented along with high resolution
observations of dust continuum and molecular line emission in a later paper.

\subsubsection{G35.03+0.35}\label{g3503}

The 4.5 \um\/ emission of this EGO has a bipolar morphology, with one
lobe to the NE and the other to the SW. The 6.7 GHz \methanol\/ maser
group 
lies on the ``waist'' between these two lobes, coincident with 24 \um\/
emission (Figs.\ref{3color}o,\ref{kinplots}o).  In contrast to many of the
other EGOs in our sample, the 44 GHz \methanol\/ masers do not lie primarily
on or at the edges of these lobes, but rather trace out an arc to the
east/southeast of the 6.7 GHz maser position.  The 44 GHz maser arc lies
between the main EGO and an eastern patch of 4.5 \um\/ emission; it is unclear
from the MIR emission whether this eastern 4.5 \um\/ emission is part of the
EGO or an unrelated source.  The velocity spread of the 44 GHz \methanol\/
masers is narrow (\q2.5 \kms\/) and all of the masers are either near the
thermal gas
\vlsr\/ or slightly blueshifted (Figs.~\ref{jcmtlineplots}o,\ref{kinplots}o).  The 6.7 GHz
Class II \methanol\/ maser emission is also blueshifted relative to the
thermal gas
\vlsr\/, but by a much greater amount (\q 6-12 \kms).  In contrast, the
\water\/ masers reported by \citet{FC99} (see Fig. ~\ref{kinplots}o)--which
are positionally coincident, within their reported position uncertainty of
0\farcs5, with the 6.7 GHz \methanol\/ maser group--are redshifted relative to
the gas \vlsr\/ by \q13-17 \kms.  The velocity coverage of our observations
does not extend to the \water\/ maser velocities (\q66-70 \kms), but neither
\citet{Sz00} nor \citet{Pandian07} report 6.7 GHz \methanol\/ maser emission
outside the range of \q 40-47 \kms\/ in their much wider bandwidth single-dish
spectra.

\citet{KCW94} found two unresolved cm-wavelength continuum sources in
this region: a western source with a 15 GHz (2 cm) integrated flux
density of 12.5 mJy at $18^{\rm h}54^{\rm m}00^{\rm s}.49$,
$+02^{\circ}01'18\farcs0$ (J2000, positional uncertainty 0\farcs1), and
a second, marginally detected source \q2\pp\/ east for which no
position or fluxes were reported.  We detect both sources at 44 GHz in
our pseudocontinuum data, which have a similar resolution.  The western
source has a 44 GHz integrated flux density of 12.7$\pm$2.4 mJy, based
on fitting the source with a single 2D Gaussian component: the
deconvolved source size is 0$\farcs$88$\times$0$\farcs$26
(PA=5$^{\circ}$) and the fitted position is $18^{\rm h}54^{\rm
  m}00^{\rm s}.49$, $+02^{\circ}01'18\farcs3$ (J2000).
Based on comparison with the 15 GHz integrated flux density from
\citet{KCW94}, our measured 44 GHz flux density is consistent within
errors with an optically thin spectral index of $-$0.1.  The 44 GHz
flux density of the western source is consistent with a single
ionizing star of spectral type B1.5 \citep{Smith02}.
The eastern source, which is not sufficiently resolved in our data to
fit a reliable source size, has a fitted peak intensity of 3.6$\pm$0.8
\mjb\/ (\q4.5$\sigma$) at $18^{\rm h}54^{\rm m}00^{\rm s}.65$,
$+02^{\circ}01'19\farcs5$ (J2000).  The positions of both 44 GHz
continuum sources are overplotted as black asterisks on the 4.5 \um\/
image in Figure~\ref{kinplots}o.  The angular separation between the
two continuum sources is 2\farcs6 ($\gtrsim$ 9000 AU at 3.4 kpc).  The
eastern 44 GHz continuum source is between the two 4.5 \um\/ lobes,
and offset from the intensity-weighted 6.7 GHz maser position by
$\lesssim$0\farcs25 (\q860 AU at 3.4 kpc).  This geometry suggests that the
eastern 44 GHz continuum source may be the powering source of the putative
outflow traced by the extended, bipolar 4.5 \um\/ emission.  In this scenario,
the western 44 GHz continuum source would be another member of a protocluster,
akin to the proto-Trapezia (multiple massive protostars within
$\lesssim$10000AU) observed in NGC6334I and I(N) \citep{Hunter06}.  The
\hisoco\/ line profile provides some additional support for the suggestion
that multiple sources are present, as it appears asymmetric, with a red
shoulder, and the JCMT beam did not encompass MIR sources other than the EGO.
Broad SiO(5-4) emission ($\Delta v_{FWHM}$= 11.1 \kms) was also detected toward
this EGO in our JCMT survey (Table~\ref{jcmtfitstable}).
Additional data are, however, required to constrain the nature of the eastern
44 GHz continuum source (e.g. hypercompact \HII\/ region or dust core) and to
establish whether the bipolar 4.5 \um\/ lobes in fact trace a bipolar
molecular outflow.

\subsubsection{G37.48$-$0.10}\label{g3748}

Only two loci of (weak) 44 GHz \methanol\/ maser emission are detected towards
this EGO, one at each end of linearly extended 4.5 \um\/ emission
(Figs.~\ref{3color}p.\ref{kinplots}p).  The extended 4.5 \um\/ emission is
oriented along a roughly east-west axis; a weaker, more diffuse 4.5 \um\/ tail
extends westward from the main 4.5 \um\/ emission feature.  The velocity
extent of the 44 GHz maser emission is narrow, $<$2 \kms\/ near the thermal
gas \vlsr\/ (Fig.\ref{jcmtlineplots}p).  The 6.7 GHz \methanol\/ maser
emission is stronger and spans a much broader velocity range, extending into
and beyond the wings of the \hco\/ profile.  (The peak velocity and position
of this 6.7 GHz maser were previously reported by \citet{Sz07}.)  The spatial
distribution of the 6.7 GHz maser emission is characterized by arcs, including
a northern arc of predominantly redshifted maser emission and a southern arc of blueshifted
maser emission.  Both arcs resemble parabolas with their opening axes oriented
toward the southeast: maser emission near the systemic velocity is
concentrated between the redshifted and blueshifted arcs.

\subsubsection{G39.10+0.49}\label{g3910}

The morphology of the extended 4.5 \um\/ emission in this EGO is predominantly
linear, and the 44 GHz Class I \methanol\/ masers are (with one exception)
distributed linearly along the 4.5 \um\/ emission
(Figs.~\ref{3color}q,\ref{kinplots}q).  While the velocity distribution of the
44 GHz masers is not monotonic, there is a predominance of redshifted masers
to the southeast of the 6.7 GHz maser group position while the most
blueshifted maser is located northwest of the 6.7 GHz maser (Fig.
~\ref{kinplots}q).  The 6.7 GHz \methanol\/ maser emission spans a broader
velocity range (\q 16 \kms) than the 44 GHz maser emission (\q 6 \kms).  The
6.7 GHz \methanol\/ maser emission largely avoids the velocity range of the
\hisoco\/ emission: the strongest 6.7 GHz maser emission is blueshifted by \q
7 \kms\/ relative to the thermal gas \vlsr\/ (23 \kms,
Table~\ref{jcmtfitstable}) and falls on the blue wing of the \hco\/ profile,
while weaker redshifted maser components are present at v\q25-26 and 28.5-29
\kms (Fig.~\ref{jcmtlineplots}q).  The velocity coverage of our observations
does not include the full extent of the red \hco\/ line wing, but \citet{Sz00}
do not report 6.7 GHz maser emission outside the velocity range of 12-28
\kms\/ in their single-dish spectrum.  (An interferometric position and peak
velocity for this maser were reported in \citet{Sz07} and references therein.)
Blueshifted 6.7 GHz masers are concentrated to the northwest, while a line of
redshifted 6.7 GHz masers extends to the south and east (Fig.~\ref{kinplots}q).
The sense of the red-blue velocity gradient of the 6.7 GHz
masers matches that suggested by the 44 GHz \methanol\/ masers
on a much larger spatial scale.

\subsubsection{G49.27$-$0.34}\label{g4927}

This EGO is in an IRDC, and has an unusual cometary 4.5 \um\/ morphology
(Fig.~\ref{3color}r).  The very strong ($>$700 Jy) 6.7 GHz maser in W51
\citep[G49.49$-$0.39 in][and references therein]{Pandian07} falls within the
sidelobes of our VLA pointing.  This was responsible for the maser feature in
the vector-averaged u-v plot (\S\ref{vlaobs}).  No 6.7 GHz maser is detected
towards the EGO.  Class I 44 GHz \methanol\/ maser emission is detected at two
loci south of the EGO, coincident with and at the edge of extended 4.5 \um\/
emission (Fig.~\ref{3color}r).  The velocity range of the 44 GHz maser
emission is extremely narrow ($<$1 \kms, Fig.~\ref{jcmtlineplots}r).  SiO(5-4)
emission ($\Delta v_{FWHM}$= 6.3 \kms) was detected toward this EGO in our
JCMT survey (Table~\ref{jcmtfitstable}).

\citet{Mehringer94} reports a 1.4 GHz (20 cm) continuum source, G49.27$-$0.34,
with a flux density of 73 mJy and an ionizing flux equivalent to a zero age
main sequence (ZAMS) star of spectral type B0 at $19^{\rm h}23^{\rm m}06^{\rm
s}.84$, $+14^{\circ}20'17\farcs9$ (precessed to J2000, positional
uncertainty of 3\pp).
The size, emission measure, and electron density of the \HII\/ region are not
well constrained by the \citet{Mehringer94} observations, as the source is
unresolved at the \q14\pp\/ angular resolution.  Extrapolating the 1.4 GHz flux
density assuming optically thin free-free emission ($\alpha=-0.1$) predicts a 44 GHz
integrated flux density of \q50 mJy.  This is a lower limit to the
expected flux density, as a compact \HII\/ region is unlikely to be optically
thin over a frequency range of 1.4 to 44 GHz.  This source is, however, undetected in our 44
GHz pseudocontinuum observations.  The most likely explanation is that the 1.4
GHz emission arises from a relatively smooth extended structure that is resolved
out by our much higher resolution observations (0\farcs5 resolution at
44 GHz compared to \q14\pp\/ at 1.4 GHz), but observations at intermediate
frequencies are required to better characterize the nature of the cm continuum
emission.

\subsubsection{G49.42+0.33}\label{g4942}

This field is notable for having two spatially and kinematically distinct 6.7
GHz \methanol\/ maser groups, though it is not clear whether both are
associated with the target EGO.  The western 6.7 GHz maser group
(G49.416+0.326, Table~\ref{67iweighted_egos}, maser group 1 in
Fig.~\ref{jcmtlineplots}s) is coincident with the 4.5 \um\/ emission of the
targeted EGO, on a slight waist between two bulges of 4.5 \um\/ emission
(Figs.~\ref{3color}s,\ref{kinplots}r).  The EGO and the associated western
maser group do not have a clear 24 \um\/ counterpart; the 24 \um\/ source to
the east is \q8\pp\/ distant (\q0.48 pc at 12.29 kpc) and is coincident with
another 6.7 GHz \methanol\/ maser group (G49.417+0.324, Table
\ref{67iweighted_egos}, maser group 2 in Fig.~\ref{jcmtlineplots}s).  An
east-west extension of 4.5 \um\/ emission appears to connect the eastern 24
\um\/ source and \methanol\/ maser with the extended 4.5 \um\/ emission of the
EGO, but it is unclear from the MIR data which source excites this emission.
No 44 GHz Class I \methanol\/ masers are detected in this field.  In light of
the non-detection of 44 GHz \methanol\/ masers, the 4.5 \um\/ morphology of
the EGO--linear, with a north-south orientation, but lumpy (``beads on a
string'') as opposed to smooth (as in G39.10+0.49, for example)--raises the
possibility that the 4.5 \um\/ morphology is due to a superposition of nearby
point sources rather than to diffuse emission.  This source is, however, much
more distant than most of our sample, so the sensitivity limit of our survey
corresponds to intrinsically stronger masers than in less distant regions.

\section{Discussion}\label{discussion}

\subsection{The Nature of EGOs}\label{discussion_natureofegos}

The high detection rate of 6.7 GHz \methanol\/ masers spatially coincident
with extended 4.5 \um\/ emission in our survey is strong evidence for the
identification of EGOs as MYSOs.  The coincidence of detected 6.7 GHz masers
with MIPS 24 \um\/ EGO counterparts for the vast majority of surveyed EGOs is
consistent with the expectation that Class II \methanol\/ maser emission is
excited in the presence of warm dust near an MYSO.  Interestingly,
\citet{debuizer06} found, based on subarcsecond images, that the MIR emission
from the MYSO G35.2$-$0.74 \citep[which is an EGO catalogued by][]{egocat} is
dominated by thermal continuum emission from heated dust on the walls of an
outflow cavity, and suggested that this environment is suitable for pumping
Class II \methanol\/ masers.  The resolution of MIPS is too poor to address
whether the 24 \um\/ emission detected toward EGOs traces the central
protostar or simply warm dust in the inner outflow cavity.  In either case,
the three EGOs with detected 6.7 GHz \methanol\/ maser emission but no
detected discrete 24 \um\/ counterparts (G10.29$-$0.13, G28.28$-$0.36,
G49.42+0.33) are intriguing.  Two sources (G10.29$-$0.13, G28.28$-$0.36) are
in regions of high background adjacent to saturated 24 \um\/ complexes, and
are coincident with (sub)mm clumps (see Table~\ref{samplelit}).  If these EGOs
are the youngest sources in our sample, they may lack discrete 24 \um\/
counterparts because they are cooler and/or more heavily embedded (see also
\S\ref{discussion_evolution}).  Further observations are necessary to clarify
the nature of these sources and of G49.42+0.33, about which little is known
beyond the results of the present study.

The high detection rates of \hco\/ line wings, SiO emission, and 44
GHz Class I \methanol\/ masers in our EGO surveys are all strong
evidence for the presence of molecular outflows.  While \hco\/ may
trace relic outflows in MSFRs \citep[c.f.][]{Klaassen07,Hunter08}, the
mechanisms and timescales governing SiO emission make it well-suited
to tracing active outflows.  Sufficiently fast shocks enhance gas
phase Si abundance via sputtering of grain mantles and cores, and Si
is oxidized to SiO via gas-phase reactions
\citep{Schilke97,Caselli97}; after \q10$^{4}$ years, the gas phase
abundance of SiO drops due to reaccretion onto grains and conversion
to SiO$_{2}$ \citep{pdf97}.  In well-known MSFRs, Class I \methanol\/
masers trace shocked gas, as seen in \h\/ and SiO, and interfaces between
outflows and surrounding molecular gas (\S\ref{introduction}).  The 44 GHz Class I
\methanol\/ masers detected in our survey are predominantly on or along the
edges of extended 4.5 \um\/ emission, which is thought to trace shocked
\h\/ (\S\ref{introduction}).  Massive stars form in clusters, however, surrounded both
by other massive objects and by intermediate and low mass stars, raising the
question of whether different sources might be responsible for the 6.7 and 44
GHz maser emission.  As discussed in \S\ref{spatial44} and
\S\ref{individualsources}, there are examples of far-flung 44 GHz masers that
may be excited by sources other than the EGO.  In general, however,
the 4.5 \um\/ morphology shown in Fig.~\ref{3color} argues that the same source--a
MYSO--is reponsible for exciting both the 6.7 GHz Class II \methanol\/ maser
emission (by heating its surrounding environment) and at least some of the 44
GHz Class I \methanol\/ maser emission (by driving an outflow).  G19.01$-$0.03
is a particularly suggestive example: two bipolar lobes of 4.5 \um\/ emission,
coincident with arcs of 44 GHz \methanol\/ masers, emanate from a 24 \um\/
point source coincident with a 6.7 GHz \methanol\/ maser.  

The results discussed above are strong evidence that the EGOs in our survey
sample are young MYSOs that are actively driving outflows, and hence actively
accreting.  While our sample was selected to cover a range of MIR properties,
it included only EGOs categorized as ``likely'' MYSO outflow candidates by
\citet{egocat}, and the sample of EGOs searched for 44 GHz \meth\/ maser
emission was, in essence, a 6.7 GHz \meth\/ maser-selected EGO subsample.
Further studies are required to establish whether \emph{all} EGOs are MYSOs
with active outflows, and particularly whether the distinction between
``possible'' and ``likely'' outflow candidates drawn by \citet{egocat} based on
MIR morphology is borne out by observations of other outflow tracers.

The kinematics of the \methanol\/ masers, while revealing for
individual sources, exhibit such diversity across our sample that they
allow few generalizations about EGOs as a population, beyond the
observation that 44 GHz \methanol\/ masers are generally at or near
the thermal gas \vlsr.  This is consistent with the excitation of maser
emission at the interface with surrounding dense ambient gas: for a
kinetic temperature of 80 K, \citet{Leurinithesis} found that maser
emission in the 44 GHz transition is strongest at densities of
\q10$^{5}$-10$^{6}$ \cc (n(H$_{2}$)).  It is also consistent with the
sketch of \citet{Kalenskii06} for Class I \methanol\/ masers excited
in a shell behind the shock front for outflows in the plane of the
sky--an orientation preferentially selected for by the \citet{egocat}
criterion of extended 4.5 \um\/ emission.



\subsection{\meth\/ masers in EGOs}\label{discussion_masersinmsfr}

As noted above, the sample of EGOs searched for 44 GHz \meth\/ maser emission
was essentially a 6.7 GHz \meth\/ maser-selected subsample.  Class I 44
GHz \meth\/ maser emission was detected towards \q 89\% (16/18) of EGOs with
6.7 GHz \meth\/ masers, as well as towards the one targeted EGO without an
associated 6.7 GHz \meth\/ maser.  \citet{Slysh94} noted that 82\% of new 44
GHz masers detected in their survey--which targeted \HII\/ regions, \water\/
masers, 6.7 GHz \meth\/ masers, and IRDCs--were associated with known 6.7 GHz
\meth\/ masers.  Twelve of the fields with UC\HII regions observed at 44 GHz
by \citet{Kurtz04} were also searched for 6.7 GHz \meth\/ masers by
\citet{Walsh98}.  All twelve have detected 6.7 GHz \meth\/ masers, though the
relative positions of the two types of masers vary widely, from coincident
within the errors to widely separated.  The high detection rate of 44 GHz
\meth\/ masers in our survey adds evidence that 44 GHz Class I and 6.7 GHz
Class II \meth\/ masers may be excited (via different mechanisms) by the same
MYSO.

The range of 6.7 GHz maser spectral properties in our EGO survey is
particularly striking because Class II \methanol\/ maser emission requires the
presence of warm dust (T$_{d}\gtrsim$125 K), which both releases \methanol\/
from icy grain mantles into the gas phase and emits the IR photons that pump
the population inversions \citep[][and references
therein]{Cragg92,Minier03,Cragg05}.  As a consequence, Class II masers may
only be excited relatively near the central protostar, where dust temperatures
are high.  This is reflected observationally in our data: the 6.7 GHz
\methanol\/ maser emission associated with EGOs is in all cases spatially
compact, and in most cases coincident with 24 \um\/ emission likely
attributable to an embedded YSO or YSOs.  The detection of thermal emission in
a warm transition of \methanol~ (5$_{2,3}$-4$_{1,3}$, \elow=44.3 K) toward
most EGOs in our JCMT survey also supports the presence of warm, dense
gas.  Precisely where in the circum(proto)stellar environment 6.7
GHz Class II \methanol\/ masers originate has been the subject of considerable
discussion in the literature \citep[c.f.][which includes a review of much of
the literature to date]{vanderWalt07}.  Circumstellar accretion disks
\citep[e.g.][]{Norris98}, outflows \citep[e.g.][]{debuizer03}, and shocks
propagating through rotating clouds \citep[e.g.][]{Dodson04} have all been
suggested to explain observed maser spatial and velocity distributions.  While
the evidence from our survey is insufficient to distinguish among these
possibilities for individual sources, the kinematic diversity of the ensemble
argues against any one physical origin (e.g. in a particular dynamical
structure) for 6.7 GHz \methanol\/ masers.


MSFRs may also exhibit considerable kinematic complexity: for example,
velocity dispersions of \q5 \kms\/ are observed among sources in
Cepheus A East \citep{Brogan07} and S255N \citep{s255n}.  The two 6.7
GHz \methanol\/ maser groups in G11.92$-$0.61 are separated by \q 1.6
\kms\/ in velocity and \q4\farcs4 ($\gtrsim$17000 AU at
3.8 kpc) and may be tracing two distinct protostars, a possibility
that exists for other sources in our survey.


\subsection{EGOs and Evolutionary Sequences of MSF}\label{discussion_evolution}

The evolutionary sequence of a forming massive star may be broadly described,
following the terminology of \citet{ZY07}, as a progression from a cold dense
massive core (CDMC) to a hot dense massive core (HDMC) to a disk-accreting
main-sequence star (DAMS) to the final main-sequence star (FIMS), where the
transition from a HDMC to a DAMS occurs when the central, accreting source
becomes powered more by hydrogen burning than by disk accretion.  Refining
this sequence and adding physical detail requires being able to establish the
relative ages of large samples of MYSOs on the basis of observational
evidence.  While \emph{Spitzer} surveys have provided large samples, the use
of broadband MIR data for this purpose is complicated by the fact that
multiple molecular, ionic, and PAH lines fall within each of the IRAC bands
\citep[c.f. Fig. 1 of][]{Reach06}, and evolutionary models that include all
these emission mechanisms in realistic geometries are not yet available.  At
longer wavelengths (e.g. MIPS 24 \um), the broadband flux is more likely to be
dominated by thermal dust emission.  \citet{debuizer06} has argued that such
thermal emission may emanate from heated dust in the walls of an outflow
cavity (as opposed to pinpointing the protostar itself), and the point in the
evolutionary sequence, relative to other indicators, at which 24 \um\/
emission becomes observable is not well characterized as a function of viewing
angle or protostellar mass.  For our EGO sample, the detection of thermal
emission in the \methanol~ (5$_{2,3}$-4$_{1,3}$) transition (\elow=44.3 K)
toward the majority of sources in our JCMT survey is suggestive of possible
hot core line emission \citep[e.g.][]{vDB98}, but followup with a more diverse
range of organic tracers is required.

Two more established observational indicators of evolutionary state are masers
and \HII\/ regions, although both remain imperfectly understood.
\citet{Ellingsen07conf} propose a ``straw man'' evolutionary sequence in which
a forming MYSO first excites Class I \methanol\/ masers (via a protostellar
outflow), then Class II \methanol\/ masers (radiatively pumped by photons
emitted by warm dust), then \water\/ masers (associated with outflows), and
finally OH masers (radiatively pumped by FIR photons) and UC \HII\/ regions in
quick succession.  Each of these stages overlaps with those before and after,
and \citet{Ellingsen07conf} note that there are sources known to be
inconsistent with this sequence.

Most EGOs in our sample are associated with both Class I and Class II
\methanol\/ masers, placing them early in the \citet{Ellingsen07conf}
evolutionary sequence.  Some Class I \methanol\/ masers detected in
our survey--which are apparently isolated from MIR emission or Class
II \methanol\/ masers--may be excited by outflows from even younger
protostars, or from YSOs not massive enough to excite 6.7 GHz Class II
\methanol\/ maser emission.  The EGOs in our sample have not been
systematically searched for \water\/ or OH masers at high angular
resolution.  Three sources (G11.92$-$0.62, G23.01$-$0.41, G35.03+0.35)
are known to also have associated \water\/ masers; of these, two
(G23.01$-$0.41, G35.03+0.35) also have OH masers coincident with the
EGO \citep{FC99}.  These EGOs may thus either be examples of sources
inconsistent with the \citet{Ellingsen07conf} sequence, or in fact be
protoclusters, with YSOs at different evolutionary stages responsible
for different types of maser emission.

While it is generally accepted that UC \HII\/ regions
represent a late stage of massive star formation \citep[the end state
  (FIMS) in the sequence of][]{ZY07}, the nature of the earliest
detectable free-free emission from forming massive stars is still the
subject of considerable discussion in the literature \citep[see for
  example][]{MvdT04,vdTM05,vdT05,Hoareppl,HoareFranco07,GibbHoare07,Lizano08}.
Theoretical calculations suggest that at the earliest stages of
massive star formation, high accretion rates can prevent luminous
(L$>10^{4}$ \lsun) MYSOs from ionizing their surroundings, either by
``quenching'' the \HII\/ region \citep[confining it very close to the
  stellar surface;][]{Walmsley95}, or by swelling the radius of the
MYSO, thus lowering its effective temperature \citep[below \q30000
  K;][and references therein]{Yorke02,HoareFranco07}.  Subsequently,
ionized gas may exist in a range of dynamical structures, including
accretion flows, outflows, jets, ionized/photoevaporating disks and
stellar winds, all of which may exist at the scale of hypercompact
(HC) \HII regions (size $\le$0.05 pc, density $\ge$10$^{6}$ \cc)
\citep[c.f.][]{Kurtz05,Hoareppl,HoareFranco07,Keto07,Lizano08}.  

As discussed in \S\ref{individualsources}, cm wavelength continuum emission
has been detected coincident with MIR emission towards two EGOs in our
sample (G35.03+0.35 and G49.27$-$0.34, see also
Table~\ref{samplelit}), but the existing data are insufficient to
establish the nature of this emission.  At 44 GHz, both
thermal emission from dust and free-free emission from ionized gas may
contribute to the continuum; without multiwavelength observations,
these cannot be disentangled.  The non-detections (18/19,\q95\%) of 44 GHz continuum emission
coincident with EGOs in our survey are consistent with other evidence
(masers, MIR properties, \S\ref{discussion_natureofegos}) that EGOs
are young MYSOs.  Given the depth of our survey,
however, the non-detections do not provide strong constraints on the
properties (luminosity, mass, ionizing photon flux, spectral type) of
the ``central'' sources responsible for powering the outflows traced
by 4.5 \um\/ and Class I \methanol\/ maser emission.  At a typical distance of 4 kpc, an
unresolved 5$\sigma$ detection would correspond to an ionizing photon
flux of \q9.4$\times$10$^{45}$ photons$^{-1}$ \citep[following][and
  assuming a typical electron temperature of 10$^{4}$ K]{KCW94}, equivalent to a
single star cooler than spectral type B1.5 \citep{Smith02}.  This
estimate gives a \emph{lower limit} for the ionizing flux that
corresponds to our detection threshold, as it assumes that every
ionizing photon emitted produces observable free-free radiation
(e.g. that quenching and absorption by dust are both unimportant).  It
also assumes optically thin free-free emission \citep{Mezger67,KCW94};
in the case of optically thick free-free emission, derived parameters
dependent on the electron density (including the ionizing photon flux)
will be underestimated \citep[c.f. discussion in][]{Keto08}.  Under
similar assumptions, the photon flux of 10$^{45}$ photons$^{-1}$
assumed by \citet{Keto07} in modeling very young ionized accretion
flows around forming MYSOs would be below our detection limit.
Extrapolating the peak 15 GHz (2 cm) intensities for UC \HII\/ regions
in the survey of \citet{KCW94} (KCW94) to 44 GHz assuming optically
thin free-free emission (S$_{\nu}\propto \nu^{\alpha}, \alpha=-$0.1),
only \q 50\% of the KCW94 UC \HII\/ regions would be detected at the
5$\sigma$ level (\q 5 \mjb) in our survey.  The 15 GHz observations
have approximately the same resolution as our 44 GHz
observations and KCW94 sources with S$_{\nu}$,peak(15
GHz)$>$S$_{\nu}$,peak(8.4 GHz) (inconsistent with optically thin
free-free emission) were excluded.

The overwhelming lack of 44 GHz continuum detections of EGOs in our
survey thus serves to rule out bright UC \HII\/ regions
as powering sources in most cases, but deeper observations are
required to constrain the presence of fainter UC or of HC \HII\/
regions.  Smaller and denser than UC \HII\/ regions, most HC \HII\/
regions have rising spectral indices ($\alpha$\q1) at cm-mm
wavelengths and are generally faint ($\lesssim$ 10 mJy) even at 44 GHz
\citep{Kurtz05,Garay07,Lizano08}.  Emission from MYSOs attributed to
ionized stellar winds is likewise faint at 44 GHz \citep[5/8 sources
  in the sample of][with peak intensity $<$5 \mjb]{GibbHoare07}.
Sensitive, high angular resolution observations over a range of cm
wavelengths are required to further constrain the presence, extent,
and physical properties of any ionized gas associated with EGOs.



\section{Conclusions}\label{conclusions}

The results of our VLA and JCMT surveys constitute a preponderance of evidence that EGOs are
young MYSOs with active outflows:
\begin{itemize}
\item{Class II 6.7 GHz \methanol\/ masers, which are associated exclusively with
    MYSOs, are detected towards $\gtrsim$64\% of EGOs surveyed (18/28)--nearly double
    the rate of surveys using other MYSO selection criteria.}
\item{Class I 44 GHz \methanol\/ masers, which trace molecular outflows, are
    detected towards \q90\% of EGOs surveyed (17/19) and \q89\% (16/18) of EGOs with
    associated 6.7 GHz \methanol\/ masers.}
\item{$\Delta$v$_{FWZI}$ in the \hco(3-2) line is $\gtrsim$20 \kms\/ for the
    majority of EGOs surveyed, consistent with the presence of outflows.}
\item{SiO(5-4) emission, which is predicted to persist for only \q10$^{4}$
    years after the passage of a shock, is detected towards 90\%  (9/10) of EGOs
    surveyed--a rate comparable to that of sources with narrowband \h\/
    emission believed to be excited by outflow shocks.}
\item{Thermal \methanol~ (5$_{2,3}$-4$_{1,3}$) emission, indicative of the
    presence of warm dense gas, is detected toward 83\% (15/18) of EGOs surveyed.} 
\item{No 44 GHz continuum emission is detected at the 5 mJy level (5$\sigma$)
toward \q95\% of EGOs surveyed (18/19), ruling out bright UC \HII\/ regions as
powering sources.}
\end{itemize}

Class II 6.7 GHz \methanol\/ maser emission is spatially concentrated in
compact maser groups (extent $\lesssim$1\pp\/), which are generally coincident
with 24 \um\/ emission.  Each EGO is associated with only one or two 6.7 GHz
\methanol\/ maser groups.  Class I 44 GHz \methanol\/ maser emission, in
contrast, is generally widely distributed over many tens of arcseconds,
coincident with extended 4.5 \um\/ emission.  \hco(3-2) emission is detected
towards all EGOs surveyed (19/19); the profiles are characterized by broad
line wings, consistent with the presence of outflows.  In addition, all
surveyed EGOs (19/19) are detected in \hisoco(3-2) emission, allowing the
determination of revised kinematic distances based on the new Galactic
rotation model of \citet{Reid09}.  Thermal \methanol(5$_{2,3}$-4$_{1,3}$)
emission is also detected towards the majority of EGOs in our sample (\q83\%,
15/18), at the same velocity as the \hisoco(3-2) emission (median velocity
offset $\lesssim$0.3 \kms).

These results verify that EGOs are a promising sample for studying
accretion and outflow at the early stages of massive star
formation. There is also evidence that two of the EGOs surveyed may in
fact be protoclusters of at least two MYSOs, providing opportunities
to study how MYSO outflows interact in the dense environments
characteristic of most MSF.  The next steps are to establish the
nature of the driving sources.  Deep cm continuum observations will
more strictly constrain the presence or absence of ionized gas (and
hence ionizing photons).  High angular resolution (sub)mm line and
continuum observations will constrain the physical properties
(temperature, density, mass) and numbers of compact cores.  High
angular resolution observations in direct molecular outflow tracers
are also required to better characterize outflow kinematics and
energetics: 44 GHz Class I \methanol\/ masers are a strong indicator
of the presence of a molecular outflow, but are preferentially excited
near the systemic velocity and provide no information about outflow
mass or momentum.  These projects are well matched with the
capabilities of the EVLA and ALMA, and EGOs will be promising targets
for these facilities.

\acknowledgments

This research has made use of NASA's Astrophysics Data System Bibliographic
Services and the SIMBAD database operated at CDS, Strasbourg, France.  C.J.C.
would like to thank J. Wouterloot for helpful discussions about JCMT data
reduction and R. Indebetouw for helpful discussions about MIPS saturated
source photometry.  We thank M. Reid for providing his new kinematic Galactic
distance estimate source code ahead of its publication.  Support for this work
was provided by NSF grant AST-0808119.  C.J.C. was partially supported during
this work by a National Science Foundation Graduate Research Fellowship and a
NRAO Graduate Summer Student Research Assistantship.

{\it Facilities:}  \facility{VLA ()}, \facility{JCMT ()}, \facility{Spitzer ()}


\begin{deluxetable}{lccccccc}
\tablewidth{0pt}
\tablecaption{Observational Parameters of VLA 6.7 GHz Observations \label{67obstable}} 
\tablehead{ 
\colhead{Source Name} & 
\colhead{V$_{range}$\tablenotemark{a}} &
\colhead{Phase Cal.} &
\colhead{Bandpass Cal.} &
\colhead{Flux Cal.} &
\colhead{Obs. Date} &
\colhead{Syn. Beam} &
\colhead{SAD cutoff\tablenotemark{b}} \\
\colhead{} &
\colhead{\kms} &
\colhead{} & 
\colhead{} & 
\colhead{} & 
\colhead{} & 
\colhead{\arcsec $\times$ \arcsec} & 
\colhead{Min,Max,Median}\\
\colhead{} &
\colhead{} &
\colhead{} &
\colhead{} & 
\colhead{} & 
\colhead{} & 
\colhead{} & 
\colhead{\jb}
}
\tablecolumns{8}
\tabletypesize{\scriptsize}
\setlength{\tabcolsep}{0.05in}
\startdata
G10.29$-$0.13 & -4.6 to 22.8 & J1733-130 &
3C286,J2136+006 & J2136+006\tablenotemark{c} & 2008 Feb 9,17 &
3.35$\times$2.61 & 0.16,0.75,0.20 \\
G10.34$-$0.14 & -1.6 to 25.8 & J1733-130 &
3C286,J2136+006 & J2136+006\tablenotemark{c} & 2008 Feb 9,17 &
3.49$\times$2.68 & 0.16,1.44,0.20 \\
G11.92$-$0.61 & 25.3 to 56.8 &  J1733-130 &
3C84,3C286,J2136+006 & 3C48,3C286 & 2007 Oct 31 & 1.94$\times$0.96 &
0.20,0.30,0.22 \\
G18.67+0.03 & 66.4 to 93.8 & J1733-130 &
3C286,J2136+006 & J2136+006\tablenotemark{c} & 2008 Feb 9,17 &
3.31$\times$2.42 & 0.16,0.35,0.16\\
G18.89$-$0.47 & 44.4 to 71.8 & J1733-130 &
3C286,J2136+006 & J2136+006\tablenotemark{c} & 2008 Feb 9,17 &
3.37$\times$2.46 & 0.15,0.16,0.16\\
G19.01$-$0.03 & 44.4 to 71.8 & J1733-130 &
3C286,J2136+006 & J2136+006\tablenotemark{c} & 2008 Feb 9,17 &
3.47$\times$2.27 & 0.16,0.25,0.16\\
G19.36$-$0.03 & 15.4 to 42.8 & J1733-130 &
3C286,J2136+006 & J2136+006\tablenotemark{c} & 2008 Feb 9,17 &
3.44$\times$2.22 & 0.16,0.36,0.16\\
G22.04+0.22 & 38.4 to 65.8 & J1733-130 &
3C286,J2136+006 & J2136+006\tablenotemark{c} & 2008 Feb 9,17 &
3.50$\times$2.21 & 0.16,0.22,0.17\\
G23.01$-$0.41 & 64.4 to 91.8 & J1733-130 &
3C286,J2136+006 & J2136+006\tablenotemark{c} & 2008 Feb 9,17 &
3.48$\times$2.13 & 0.16,12.3,0.19\\
G23.96$-$0.11 & 59.4 to 86.8 & J1733-130 &
3C286,J2136+006 & J2136+006\tablenotemark{c} & 2008 Feb 9,17 &
3.67$\times$2.12 & 0.15,0.21,0.15\\
G24.94+0.07 & 30.4 to 57.8 & J1743-038 & J2136+006
& J2136+006\tablenotemark{c} & 2008 Feb 1,19 & 1.82$\times$1.40 &
0.16,0.22,0.17\\
G25.27$-$0.43 & 48.4 to 75.8 & J1743-038 & J2136+006
& J2136+006\tablenotemark{c} & 2008 Feb 1,19 &  1.78$\times$1.24 &
0.14,0.20,0.15\\
G28.28$-$0.36 & 28.4 to 55.8 & J1743-038 & J2136+006
& J2136+006\tablenotemark{c} & 2008 Feb 1,19 &   2.31$\times$1.37 &
0.16,0.26,0.16\\
G28.83$-$0.25 & 75.4 to 105.6 & J1743-038 & J2136+006
& J2136+006\tablenotemark{c} & 2008 Feb 1,19 &  1.96$\times$1.49 & 0.16,0.40,0.18\\
G35.03+0.35 & 33.4 to 60.8 &  J1743-038 & J2136+006
& J2136+006\tablenotemark{c} & 2008 Feb 1,19 &   2.32$\times$1.34 &
0.16,0.25,0.16\\
G37.48$-$0.10 & 47.4 to 74.8 & J1743-038 & J2136+006
& J2136+006\tablenotemark{c} & 2008 Feb 1,19 & 1.82$\times$1.45 &
0.16,0.20,0.16\\
G39.10+0.49 & 3.4 to 30.8 & J1925+211 & J2136+006
& J2136+006\tablenotemark{c} & 2008 Feb 1,19 &  2.37$\times$1.30 &
0.15,0.33,0.17\\
G49.27$-$0.34 & 47.4 to 74.8 & J1925+211 & J2136+006
& J2136+006\tablenotemark{c} & 2008 Feb 1,19 &  1.99$\times$1.37  &
0.17,0.42,0.19\\
G49.42+0.33 & -5.2 to -32.6 & J1925+211 & J2136+006
& J2136+006\tablenotemark{c} & 2008 Feb 1,19 &  1.96$\times$1.33  &
0.12,0.20,0.20\\
\enddata
\tablenotetext{a}{Velocity range searched for maser emission.}
\tablenotetext{b}{Minimum, maximum, and median cutoff levels used in the
automated identification of masers with the AIPS task SAD.  Cutoff levels used
were $\sim$6 or 10 $\times$ $\sigma_{channel}$.}
\tablenotetext{c}{Assuming S(6.7 GHz)=9.03 Jy.} 
\end{deluxetable}

\begin{deluxetable}{lcccccccc}
\tablewidth{0pt}
\tablecaption{Observational Parameters of VLA 44 GHz Observations \label{44obstable}} 
\tablehead{ 
\colhead{~} &
\colhead{~} &
\colhead{~} &
\multicolumn{3}{c}{Line Observations} &
\multicolumn{3}{c}{Wideband Observations} \\
\colhead{Source Name} & 
\colhead{V$_{range}$\tablenotemark{a}} &
\colhead{Phase Cal.} &
\colhead{Obs. Date} &
\colhead{Syn. Beam} &
\colhead{SAD cutoff\tablenotemark{b}} & 
\colhead{Obs. Date} &
\colhead{Syn. Beam} &
\colhead{rms}\\
\colhead{} &
\colhead{\kms} &
\colhead{} & 
\colhead{} &
\colhead{\arcsec $\times$ \arcsec} & 
\colhead{\jb} &
\colhead{} &
\colhead{\arcsec $\times$ \arcsec} & 
\colhead{\mjb}
}
\tablecolumns{9}
\tabletypesize{\scriptsize}
\setlength{\tabcolsep}{0.05in}
\startdata
G10.29$-$0.13 & -3.2 to 15.4 & J1820-254 & 18 Feb 08 &
1.03$\times$0.54 & 0.15,0.18,0.17 & 18 Feb 08 &  0.98$\times$0.51 & 1.4 \\
G10.34$-$0.14 & 7.0 to 25.6 & J1820-254 & 18 Feb 08
&  0.87$\times$0.49 & 0.16,0.19,0.17 & 18 Feb 08 & 0.89$\times$0.48 & 1.5\\
G11.92$-$0.61 & 30.5 to 49.1 & J1820-254 & 18 Feb 08
&  0.75$\times$0.51 & 0.22,0.22,0.22 &  18 Feb 08 & 0.99$\times$0.44 & 1.4\\
G18.67+0.03 & 71.0 to 89.6 & J1832-105 & 18 Feb 08
&   0.76$\times$0.45 & 0.10,0.10,0.10 & 18 Feb 08 &  0.80$\times$0.48& 1.1\\
G18.89$-$0.47 & 48.0 to 66.5 & J1832-105 & 16 Feb 08
&   0.57$\times$0.43 & 0.32,0.32,0.32 & 18 Feb 08 & 0.76$\times$0.47 &1.1\\
G19.01$-$0.03 & 48.5 to 67.1 & J1832-105 & 18 Feb 08
&  0.62$\times$0.50 & 0.12,0.12,0.12 & 18 Feb 08 &   0.69$\times$0.51 & 1.0\\
G19.36$-$0.03 & 18.5 to 37.1 & J1832-105 & 18 Feb 08
&   0.63$\times$0.50& 0.12,0.27,0.12 & 18 Feb 08 &  0.66$\times$0.49& 1.0\\
G22.04+0.22 & 41.5 to 60.1 & J1832-105 & 18 Feb 08 &
0.58$\times$0.51 & 0.10,0.23,0.11 & 18 Feb 08 &  0.59$\times$0.51 & 0.9 \\
G23.01$-$0.41 & 69.5 to 88.1 & J1832-105 & 18 Feb 08 &
 0.58$\times$0.51 & 0.09,0.10,0.10 & 18 Feb 08 &  0.58$\times$0.49 & 0.9\\
G23.96$-$0.11 & 59.5 to 78.1 & J1832-105 & 18 Feb 08 &
 0.60$\times$0.51 & 0.08,0.11,0.11 & 18 Feb 08 &  0.61$\times$0.48 & 1.0\\
G24.94+0.07 & 38.5 to 57.1 & J1832-105 & 18 Feb 08 &
0.64$\times$0.46 & 0.10,0.10,0.10 & 18 Feb 08 &  0.59$\times$0.44 & 1.0 \\
G25.27$-$0.43 & 53.5 to 72.1 & J1832-105 & 18 Feb 08 &
 0.63$\times$0.48 & 0.11,0.11,0.11 & 18 Feb 08 & 0.59$\times$0.48 &1.0 \\
G28.28$-$0.36 & 32.5 to 51.1 & J1851+005 & 18 Feb 08 &
 0.59$\times$0.50 & 0.08,0.08,0.08 & 18 Feb 08 &  0.60$\times$0.46 & 1.2\\
G28.83$-$0.25 & 80.5 to 99.1 &  J1851+005 & 18 Feb 08
&  0.61$\times$0.41 & 0.08,0.08,0.08 & 18 Feb 08 &  0.62$\times$0.44 & 0.9 \\
G35.03+0.35 & 35.5 to 54.5 & J1851+005 & 18 Feb 08 &
 0.58$\times$0.46 & 0.08,0.08,0.08 & 18 Feb 08 & 0.57$\times$0.45 & 0.8\\
G37.48$-$0.10 & 50.5 to 69.5 & J1851+005 & 18 Feb 08 &
 0.55$\times$0.46 & 0.08,0.08,0.08 & 18 Feb 08 &  0.61$\times$0.41 & 0.9\\
G39.10+0.49 & 11.0 to 29.3 & J1851+005 & 16 Feb 08 &
 0.66$\times$0.43 & 0.26,0.26,0.26 & 18 Feb 08 & 0.61$\times$0.38& 1.0 \\
G49.27$-$0.34 & 50.5 to 69.5 & J1924+156 & 18 Feb 08
& 0.47$\times$0.43 & 0.10,0.10,0.10 & 18 Feb 08 &  0.49$\times$0.40 & 0.9\\
G49.42+0.33 & -32.5 to -1.9\tablenotemark{c} &
J1924+156 & 18 Feb 08 & 0.50$\times$0.40 & 0.08,0.08,0.08 & 18 Feb 08 &
0.53$\times$0.40 & 0.9 \\
\enddata
\tablenotetext{a}{Velocity range searched for maser emission.}
\tablenotetext{b}{Minimum, maximum, and median cutoff levels used in the
automated identification of masers with the AIPS task SAD.  Cutoff levels used
were $\sim$6 or 10 $\times$ $\sigma_{channel}$.}
\tablenotetext{c}{This source was observed with two different tunings.} 
\end{deluxetable}

\begin{deluxetable}{llcccccccc}
\tablewidth{0pt}
\tablecaption{JCMT Fitted Line Properties \label{jcmtfitstable}} 
\tablehead{ 
\colhead{Source Name} & 
\colhead{Transition} &
\colhead{Obs. Date} & 
\colhead{rms} & 
\colhead{T$_{MB}$\tablenotemark{a}} & 
\colhead{v$_{center}$} & 
\colhead{FWHM} & 
\colhead{$\int T_{MB} dv$} & 
\colhead{Distance\tablenotemark{b}} \\
\colhead{} &
\colhead{} &
\colhead{} &
\colhead{(K)} &
\colhead{(K)} &
\colhead{(\kms)} &
\colhead{(\kms)} &
\colhead{(K * \kms)} &
\colhead{(kpc)} 
}
\tablecolumns{10}
\tabletypesize{\scriptsize}
\setlength{\tabcolsep}{0.05in}
\startdata
G10.29$-$0.13 & \hisoco(3-2) & 28 May 08 & 0.07 & 0.55(0.04) & 13.6(0.1) & 3.3(0.3) & 1.9(0.2) &  2.19(+0.73,-0.92)\\
 & \methanol(5$_{2,3}$-4$_{1,3}$) & 28 May 08 & 0.07 & 0.41(0.04) & 14.2(0.1) & 3.2(0.3) & 1.4(0.2) & \\
G10.34$-$0.14\tablenotemark{c} & \hisoco(3-2) & 1 May 02 & 0.12 & 0.50(0.07)& 12.7(0.2) &2.7(0.4) & 1.4(0.3)& 2.08(+0.75,-0.95) \\
G11.92$-$0.61 & \hisoco(3-2) & 28 May 08 & 0.06 & 1.01(0.03) & 36.0(0.1) & 4.1(0.2) & 4.4(0.2) & 3.80(+0.40,-0.46) \\
& \methanol(5$_{2,3}$-4$_{1,3}$) & 28 May 08 & 0.06 & 0.72(0.03) & 36.3(0.1) & 6.5(0.3) & 5.0(0.3) & \\
& SiO(5-4) & 29 May 08 & 0.02 & 0.23(0.02) & 36.1(0.3) & 9.6(0.8) & 2.4(0.22) & \\
G18.67+0.03 & \hisoco(3-2) & 29 May 08 & 0.06 & 0.30(0.03) & 79.6(0.2) & 3.4(0.5)& 1.1(0.2) & 4.98(+0.24,-0.25) \\
& \methanol(5$_{2,3}$-4$_{1,3}$) & 29 May 08 & 0.06 & 0.18(0.03) & 79.7(0.4) & 4.7(0.9) & 0.9(0.2)\\
G18.89$-$0.47 & \hisoco(3-2) & 30 May 08 & 0.07 & 0.61(0.05)& 66.6(0.2) & 4.5(0.4)& 2.9(0.3) &  4.49(+0.27,-0.29) \\
& \methanol(5$_{2,3}$-4$_{1,3}$) & 30 May 08 & 0.07 & 0.46(0.04) & 66.6(0.2) & 4.8(0.5) & 2.3(0.3) & \\
G19.01$-$0.03 & \hisoco(3-2) & 29 May 08 & 0.06 & 0.79(0.04) & 59.9(0.1) & 2.7(0.2) & 2.3(0.2) & 4.20(+0.29,-0.31)\\
& \methanol(5$_{2,3}$-4$_{1,3}$) & 29 May 08 & 0.06 & 0.34(0.03) & 59.7(0.2) & 4.8(0.5) & 1.8(0.2) & \\
& SiO(5-4) & 29 May 08 & 0.02 & 0.08(0.01) & 57.1(0.6) & 9.8(1.6) & 0.9(0.2) & \\
G19.36$-$0.03 & \hisoco(3-2) & 29 May 08 & 0.06 & 1.03(0.04) & 26.6(0.1) & 2.7(0.1) & 2.9(0.2) & 2.43(+0.44,-0.49) \\
&  \methanol(5$_{2,3}$-4$_{1,3}$) & 29 May 08 & 0.05 & 0.82(0.03) & 26.5(0.1) & 4.4(0.2) & 3.8(0.2) & \\
& SiO(5-4) & 29 May 08 & 0.03 & 0.09(0.01) & 24.2(1.3) & 20.6(3.1) & 1.9(0.4) & \\
G22.04+0.22 & \hisoco(3-2) & 29 May 08 & 0.04 & 1.34(0.05) & 51.3(0.1) & 2.5(0.1) & 3.5(0.2) &  3.62(+0.32,-0.34) \\
& \methanol(5$_{2,3}$-4$_{1,3}$) & 29 May 08 & 0.06 & 1.05(0.04) & 51.3(0.1) & 4.9(0.2) & 5.5(0.3) & \\
& SiO(5-4) & 30 May 08 & 0.03 & 0.13(0.01) & 53.1(1.0) & 18.3(2.4) & 2.5(0.4) & \\
G23.01$-$0.41 & \hisoco(3-2) & 29 May 08 & 0.06 & 1.16(0.03) & 77.4(0.1) & 3.9(0.1) & 4.8(0.2) & 4.59(+0.38,-0.33)\tablenotemark{d} \\
& \methanol(5$_{2,3}$-4$_{1,3}$) & 29 May 08 & 0.06 & 0.54(0.03) & 77.7(0.2) & 6.6(0.4) & 3.8(0.3) & \\
& SiO(5-4) & 12 June 08 & 0.04 & 0.16(0.02) & 79.7(0.8) & 14.9(1.9) & 2.5(0.4)& \\
G23.96$-$0.11 & \hisoco(3-2) & 29 May 08 & 0.05 & 0.67(0.03) & 72.5(0.1) & 4.0(0.2) & 2.8(0.2) &   4.45(+0.28,-0.29)\\
& \methanol(5$_{2,3}$-4$_{1,3}$) & 29 May 08 & 0.05 & 0.60(0.03) & 72.7(0.1) & 4.6(0.2) & 3.0(0.2) & \\
G24.94+0.07 &  \hisoco(3-2) & 29 May 08 & 0.05 & 0.45(0.05) & 41.5(0.2) & 3.3(0.4) & 1.5(0.2) &  2.99(+0.35,-0.38) \\
& \methanol(5$_{2,3}$-4$_{1,3}$) &  29 May 08 & 0.05 & 0.23(0.03) & 41.8(0.3) & 4.8(0.8) & 1.2(0.3) & \\
& SiO(5-4) & 29 May 08 & 0.03 & 0.11(0.02) & 42.1(0.7) & 7.2(1.6) & 0.9(0.2) & \\
G25.27$-$0.43 & \hisoco(3-2) & 29 May 08 & 0.04 & 0.31(0.06) & 59.8(0.1) & 1.0(0.2) & 0.3(0.1) &3.86(+0.31,-0.32) \\
& \methanol(5$_{2,3}$-4$_{1,3}$) &  29 May 08 & 0.05 & $<$0.15 & \nodata &  \nodata &\nodata & \\
G28.28$-$0.36 & \hisoco(3-2) & 29 May 08 & 0.05 & 0.32(0.06) & 49.5(0.1) & 1.3(0.3) & 0.5(0.1) & 3.29(+0.34,-0.35)\\
& \methanol(5$_{2,3}$-4$_{1,3}$) &  29 May 08 & 0.05 & $<$0.15 & \nodata &\nodata & \nodata & \\
G28.83$-$0.25 & \hisoco(3-2) & 29 May 08 & 0.05 & 0.94(0.03) & 87.3(0.1) & 3.3(0.1) & 3.3(0.2) & 5.03(+0.33,-0.32)\\
& \methanol(5$_{2,3}$-4$_{1,3}$) &  29 May 08 & 0.06 & 0.46(0.03) & 87.7(0.1) & 4.2(0.3) & 2.1(0.2) & \\
& SiO(5-4) & 29 May 08 & 0.02 & 0.10(0.02) & 88.1(0.5) & 5.9(1.1) & 0.6(0.2) & \\
G35.03+0.35 & \hisoco(3-2) & 29 May 08 & 0.05 & 1.04(0.03) & 53.1(0.1) & 5.4(0.2) & 6.0(0.2) & 3.43(+0.38,-0.38)\\
& \methanol(5$_{2,3}$-4$_{1,3}$) &  29 May 08 & 0.06 & 0.56(0.03) & 52.4(0.1) & 5.1(0.3) & 3.0(0.2) & \\
& SiO(5-4) & 29 May 08 & 0.03 & 0.21(0.02) & 52.5(0.5) & 11.1(1.1) & 2.4(0.3)& \\
G37.48$-$0.10 & \hisoco(3-2) & 30 May 08 & 0.07 & 0.41(0.05) & 59.2(0.2) & 2.6(0.4) & 1.1(0.2) & 3.83(+0.42,-0.41) \\
& \methanol(5$_{2,3}$-4$_{1,3}$) &  30 May 08 & 0.07 & $<$0.21 & \nodata &\nodata & \nodata & \\
G39.10+0.49 & \hisoco(3-2) & 29 May 08 & 0.05 & 0.31(0.03) & 23.0(0.2) & 4.2(0.4) & 1.4(0.2) & 1.73(+0.43,-0.44)\\
& \methanol(5$_{2,3}$-4$_{1,3}$) &  29 May 08 & 0.05 & 0.24(0.02) & 23.3(0.3) & 8.8(0.7) & 2.3(0.3) & \\
& SiO(5-4) & 13 June 08 & 0.04 & $<$0.12 & \nodata & \nodata & \nodata & \\
G49.27$-$0.34 & \hisoco(3-2) & 30 May 08 & 0.07 & 1.01(0.04) & 67.9(0.1) & 3.3(0.2) & 3.6(0.2) & 5.55(1.66,-1.66) \\
& \methanol(5$_{2,3}$-4$_{1,3}$) &  30 May 08 & 0.07 & 0.22(0.06) & 67.4(0.2) & 1.8(0.6) & 0.4(0.2) & \\
& SiO(5-4) &  30 May 08 & 0.04 & 0.15(0.02) & 65.8(0.5) & 6.3(1.2) & 1.0(0.3)& \\
G49.42+0.33 & \hisoco(3-2) & 30 May 08 & 0.06 & 0.17(0.05) & -19.8(0.3) &
2.3(0.8) & 0.4(0.2) & 12.29(0.57,-0.54)\tablenotemark{e} \\ 
& \methanol(5$_{2,3}$-4$_{1,3}$) &  30 May 08 & 0.07 & 0.20(0.06) &-22.0(0.2) & 1.3(0.4) & 0.3(0.1) & \\
\enddata
\tablenotetext{a}{Where a transition was observed and no line detected, the
  3$\sigma$ upper limit is given.  \hco\/ was detected towards all sources,
  but the profiles are characterized by
  self-absorption dips and were not fit for this reason.}
\tablenotetext{b}{Near kinematic distance based on \hisoco\/ velocity and the 
  revised kinematic distance prescription and code of \citet{Reid09}, unless
  otherwise noted.}
\tablenotetext{c}{The \methanol(5$_{2,3}$-4$_{1,3}$) line was not included
  in the tuning of the archival data used for this source (\S\ref{jcmtobs})}.
\tablenotetext{d}{Trigonometric parallax distance from \citet{Brunthaler09}.}
\tablenotetext{e}{Outer galaxy object.}
\end{deluxetable}

\begin{deluxetable}{lcccccccc}
\tablewidth{0pt}
\tablecaption{EGO Sample: Properties from the Literature \label{samplelit}} 
\tablehead{ 
\colhead{Source Name} & 
\multicolumn{2}{c}{J2000 Coordinates\tablenotemark{a}} &
\colhead{EGO Catalog\tablenotemark{b}} &
\colhead{IRDC?\tablenotemark{c}} &
\colhead{Maser?\tablenotemark{d}} & 
\colhead{cm continuum?} & 
\colhead{Outflow?\tablenotemark{e}} & 
\colhead{(Sub)mm core?\tablenotemark{f}}\\
\colhead{} &
\colhead{$\alpha$ ($^{\rm h}~~^{\rm m}~~^{\rm s}$)} &
\colhead{$\delta$ ($^{\circ}~~{\arcmin}~~{\arcsec}$)} &
\colhead{} &
\colhead{} & 
\colhead{} &
\colhead{} & 
\colhead{} &
\colhead{} 
}
\tablecolumns{9}
\tabletypesize{\scriptsize}
\setlength{\tabcolsep}{0.05in}
\startdata
G10.29$-$0.13 & 18 08 49.3 & -20 05 57.3 & 2 & Y & 6.7$^{1}$ & \nodata & \nodata$^{2}$ & Y \\
G10.34$-$0.14 & 18 09 00.0 & -20 03 35.0 & 2 & Y & 6.7$^{1}$ & \nodata & \hco(1-0)$^{2}$ & Y \\
G11.92$-$0.61 & 18 13 58.1 & -18 54 16.7 & 1 & Y & 6.7?, \water\/$^{3}$ & \nodata & $^{12}$CO(1-0)$^{4}$ & Y \\
G18.67+0.03 & 18 24 53.7 & -12 39 20.0 & 1	& N & 6.7? & \nodata & \nodata & \nodata \\
G18.89$-$0.47 & 18 27 07.9 & -12 41 35.5 & 1 & Y & \nodata & \nodata & \nodata & Y \\
G19.01$-$0.03 & 18 25 44.8 & -12 22 45.8 & 1 & Y & 6.7? & \nodata & \nodata & Y \\
G19.36$-$0.03 & 18 26 25.8 & -12 03 56.9 & 2 & Y & 6.7$^{1}$, \water\/? & \nodata & \nodata$^{2}$ & Y \\
G22.04+0.22 & 18 30 34.7 & -09 34 47.0 & 1	& Y & 6.7?, \water\/? & \nodata & \nodata & \nodata \\
G23.01$-$0.41 & 18 34 40.2 & -09 00 38.0 & 1 & N & 6.7?, 44?$^{5}$, \water\/$^{6}$ & \nodata & SiO$^{7}$,$^{12}$CO,$^{13}$CO,HNCO$^{8}$ &Y \\
G23.96$-$0.11 & 18 35 22.3 & -08 01 28.0 & 1 & N & 6.7$^{9,10}$, \water\/? & \nodata & \hco,\hisoco,CS$^{10}$ & Y \\
G24.94+0.07 & 18 36 31.5 & -07 04 16.0 & 1	& N & 6.7?, \water\/?? & \nodata & \nodata & \nodata \\
G25.27$-$0.43 & 18 38 56.9 & -07 00 48.0 & 1 & Y & \nodata & \nodata & \nodata & \nodata \\
G28.28$-$0.36 & 18 44 13.2 & -04 18 04.0 & 2 & N & 6.7$^{1}$ & \nodata & \nodata$^{2}$ & Y \\
G28.83$-$0.25 & 18 44 51.0 & -03 45 49.0 & 1 & Y & 6.7$^{1}$, \water\/? & \nodata & \hco(1-0)$^{2}$ & Y \\
G35.03+0.35 & 18 54 01.0 & +02 01 20.0 & 1 & N& 6.7?, \water\/$^{6}$ & Y$^{11}$ & $^{12}$CO(1-0)$^{4}$ & Y\\
G37.48$-$0.10 & 19 00 07.0 & +03 59 53.0 & 1 & N & 6.7$^{10}$, \water\/? & \nodata & $^{13}$CO?$^{10}$ &  Y \\
G39.10+0.49 & 19 00 58.1 & +05 42 44.0 & 1	& N & 6.7$^{9,10}$, \water\/? & \nodata & \hco,CS$^{10}$ & \nodata \\
G49.27$-$0.34 & 19 23 06.7 & +14 20 13.0 & 1 & Y & \nodata & Y$^{12}$ & \nodata & \nodata \\
G49.42+0.33 & 19 20 59.3 & +14 46 48.0 & 2	& N & 6.7? & \nodata & \nodata & \nodata \\
\hline
G24.00-0.10\tablenotemark{g} & 18 35 23.5\tablenotemark{g} & -07 59 32\tablenotemark{g} & 1 & Y & 6.7? & \nodata & \nodata & \nodata \\
G28.85-0.23\tablenotemark{g} & 18 44 47.5\tablenotemark{g} & -03 44 15\tablenotemark{g} & 4 & N & 6.7$^{1}$ & \nodata & \nodata & Y? \\
G49.27-0.32\tablenotemark{g} & 19 23 02.2\tablenotemark{g} & 14 20 52\tablenotemark{g} & 3 & N & \nodata & \nodata & \nodata & \nodata \\
\enddata
\tablenotetext{a}{Phase tracking centers of the VLA observations (\S\ref{vlaobs}).  In some cases these differ slightly from the positions in \citet{egocat}.} 
\tablenotetext{b}{EGO catalog table number from \citet{egocat}.}
\tablenotetext{c}{Based on 8 \um\/ GLIMPSE images, from \citet{egocat}.}
\tablenotetext{d}{6.7 indicates a 6.7 GHz \meth\/ maser with an
  interferometric position as referenced.  6.7? indicates a 6.7 GHz \meth\/ maser detection from a
single-dish survey with a reported or targeted position within one beam of the
EGO position.  All 6.7? are from \citet{Caswell95, Sz00, Sz02, Pandian07,
Ellingsen07}.  \water\/ indicates a \water\/ maser with an interferometric
position, as referenced.  \water\/?
indicates a water maser detection from a single-dish survey with a reported or
targeted position within one beam of the EGO position, \water\/?? within 1.5
beams.  All \water\/? and \water\/?? are from \citet{Wouterloot93,arcetri,Sz05}.}  
\tablenotetext{e}{All from single dish, single-pointing surveys as referenced,
  except for G23.01-0.41, for which \citet{Furuya08} data are
  interferometric.  ? indicates line wings reported as questionable.  If only
  a reference is listed, the source was included in the cited survey, but no
  evidence of line wings was reported.}
\tablenotetext{f}{All submm and mm core counterpart identifications are from
single-dish maps at 450 \um\/, 850 \um\/, 870 \um\/ or 1.2 mm 
\citep{Walsh03,Faundez04,Hill05,Hill06,Thompson06,DiFran08,atlasgal}, except for G23.01-0.41,
which is from interferometric 3mm data presented in \citet{Furuya08}.}
\tablenotetext{g}{EGO not targeted in our survey but within VLA primary beam
  (6\farcm7 FHWP) at 6.7 GHz.   The positions given are the EGO positions from
\citet{egocat}, not the VLA pointing center.}
References: $^{1}$\citet{Walsh98} $^{2}$ \citet{Purcell06} $^{3}$ \citet{HC96} $^{4}$
  \citet{ShC96}  $^{5}$\citet{Slysh94} 
  $^{6}$ \citet{FC99} $^{7}$ \citet{Harju98} $^{8}$ \citet{Furuya08} $^{9}$
  \citet{Niez05} $^{10}$ \citet{Sz07}  $^{11}$ \citet{KCW94} $^{12}$ \citet{Mehringer94}
\end{deluxetable}

\begin{deluxetable}{lcccl}
\tablewidth{0pt}
\tablecaption{EGO Sample: MIPs 24\um\/ Counterparts \label{mipstable}} 
\tablehead{ 
\colhead{EGO Name} & 
\multicolumn{2}{c}{J2000 24 \um\/ Source Position\tablenotemark{a}} & 
\colhead{Flux Density\tablenotemark{b}} &
\colhead{Contour Levels (Figure~\ref{3color})} \\
\colhead{} &
\colhead{$\alpha$ ($^{\rm h}~~^{\rm m}~~^{\rm s}$)} &   
\colhead{$\delta$ ($^{\circ}~~{'}~~{''}$)} &
\colhead{(Jy)} &
\colhead{(MJy sr$^{-1}$)} 
}
\tablecolumns{5}
\tabletypesize{\scriptsize}
\setlength{\tabcolsep}{0.05in}
\startdata
G10.29$-$0.13 & \nodata\tablenotemark{c}    & \nodata\tablenotemark{c}   & \nodata\tablenotemark{c} &  900,1200,1500,1800,2100,2400\\
G10.34$-$0.14 & 18 09 00.0 & -20 03 35 & \nodata\tablenotemark{d} & 900,1200,1500,1800,2100,2400\\
G11.92$-$0.61 & 18 13 58.0 & -18 54 18 & 10.5(2.1)\tablenotemark{e} & 600,900,1200,1500,1800,2100\\
G18.67+0.03   & 18 24 53.8 & -12 39 21 & 3.9($<$0.1) &  600,1200,1800\\
G18.89$-$0.47 & 18 27 07.8 & -12 41 35 & 0.5($<$0.1) & 300,450,600\\
G19.01$-$0.03 & 18 25 44.7 & -12 22 46 & 3.5(0.3) & 600,900,1200,1500,1800\\
G19.36$-$0.03 & 18 26 25.6 & -12 03 51 & \nodata\tablenotemark{d} & 600,900,1200,1500,1800\\
G22.04+0.22   & 18 30 34.6 & -09 34 46 & 3.3(0.2) & 600,900,1500 \\
G23.01$-$0.41 & 18 34 40.2 & -09 00 38 & 10.0(2.0)\tablenotemark{e} & 600,1200,1800\\
G23.96$-$0.11 & 18 35 22.2\tablenotemark{f} & -08 01 22\tablenotemark{f} & \nodata\tablenotemark{d} & 150,300,600\\
G24.94+0.07   & 18 36 31.5 & -07 04 16 & 1.4(0.1) & 300,450,600,750\\
G25.27$-$0.43 & 18 38 56.9 & -07 00 48 & 0.2($<$0.1) & 75,150,300,450,600\\
G28.28$-$0.36 & \nodata\tablenotemark{c} & \nodata\tablenotemark{c} & \nodata\tablenotemark{c} & 900,1200,1500,1800\\
G28.83$-$0.25 & 18 44 51.1 & -03 45 47 & 3.6($<$0.1) & 600,900,1200,1500,1800\\
G35.03+0.35   & 18 54 00.6 & +02 01 19 & 18.0(3.6)\tablenotemark{e} & 900,1200,1800\\
G37.48$-$0.10 & 19 00 07.1 & +03 59 53 & 1.1(0.4) & 150,300,450,600,750\\
G39.10+0.49   & 19 00 58.0 & +05 42 45 & 1.4($<$0.1) & 150,450,750\\
G49.27$-$0.34 & 19 23 06.8 & +14 20 15 & 11(2.2)\tablenotemark{e} & 600,900,1200,1500,1800\\ 
G49.42+0.33   & 19 20 59.8\tablenotemark{g} & +14 46 50\tablenotemark{g} & \nodata & 150,300,450\\
\hline
G24.00-0.10 & 18 35 23.5 & -07 59 29 & 0.3($<$0.1) & \nodata \\
G28.85-0.23 & 18 44 47.5 & -03 44 17 & \nodata\tablenotemark{d} &
\nodata \\
G49.27-0.32 & 19 23 01.9 & +14 20 53 & 0.9(0.2) & \nodata \\
\enddata
\tablenotetext{a}{Estimated, using ds9 regions, as the center of a circle around the most
  prominent Airy ring for each 24 \um\/ source.  We employ this method because
  it can be applied consistently for all sources.  Estimated uncertainties are \q1\pp\/
  for isolated, unsaturated sources, up to \q2-4\pp\/  for blended and
  saturated sources (based on comparison with centroids
  generated by the modifed ``iracworks'' program, see note (e)).} 
\tablenotetext{b}{From \citet{egocat}, except where otherwise noted.}
\tablenotetext{c}{No discrete 24 \um\/ counterpart.}
\tablenotetext{d}{Probable EGO counterpart is blended with other 24 \um\/
  source(s) and/or diffuse emission that
  may or may not be associated with the EGO, precluding a reliable flux estimate.}
\tablenotetext{e}{Source is saturated.  Flux was estimated using a modified
version of the "iracworks" saturated source software contributed by Tom
Jarrett and available at
http://spider.ipac.caltech.edu/staff/jarrett/irac/tools/ The software fits the
PSF wings of a source whose core is saturated, and was modified by Katherine
Johnston and Remy Indebetouw to operate on MIPS 24 $\mu$m data.  Uncertainty in
the reported flux is estimated to be 20\%.}
\tablenotetext{f}{Position of the northern of two blended 24 \um\/ sources
  (see Figure~\ref{3color}j), the probable EGO counterpart.} 
\tablenotetext{g}{Nearest 24 \um\/ source, see Figure~\ref{3color}s.  It is
unclear whether or not this MIPS source is associated with the EGO to its west.}
\end{deluxetable}

\begin{deluxetable}{lccccccccc}
\tablewidth{0pt}
\tablecaption{Observational Properties of EGOs \tablenotemark{a} \label{summarytable}} 
\tablehead{ 
\colhead{Source Name} & 
\multicolumn{2}{c}{Maser?} & 
\colhead{Cont.?} &
\multicolumn{4}{c}{Thermal line emission?} & 
\multicolumn{2}{c}{$V_{6.7~GHz~maser}$\tablenotemark{c}} \\ 
\colhead{} &
\colhead{6.7 GHz} &
\colhead{44 GHz} &
\colhead{44 GHz} & 
\colhead{\hco \tablenotemark{b}} & 
\colhead{\hisoco} & 
\colhead{\methanol} &
\colhead{SiO} &
\colhead{\vlsr \tablenotemark{d}} & 
\colhead{line wings\tablenotemark{e}}\\
\colhead{} &
\colhead{} &
\colhead{} &
\colhead{} &
\colhead{(3-2)} &
\colhead{(3-2)} & 
\colhead{(5$_{2,3}$-4$_{1,3}$)} & 
\colhead{(5-4)} & 
\colhead{} &
\colhead{} 
}
\tablecolumns{10}
\tabletypesize{\scriptsize}
\setlength{\tabcolsep}{0.05in}
\startdata
G10.29$-$0.13 & Y & Y & N & SA-R & Y & Y & \nodata & Y & Y-B,R \\
G10.34$-$0.14 & Y & Y & N & SA-B & Y &  \nodata & \nodata & Y?\tablenotemark{f} & Y-B,R \\ 
G11.92$-$0.61 & Y & Y & N & SA-R & Y & Y & Y & Y & N \\
G18.67+0.03 & Y & Y & N & As-R & Y & Y & \nodata & Y & N \\
G18.89$-$0.47 & Y & Y & N & SA-B & Y & Y & \nodata & N & Y-B\tablenotemark{g} \\
G19.01$-$0.03 & Y & Y & N & SA-R & Y & Y & Y & Y & Y-B \\
G19.36$-$0.03 & Y & Y & N & SA-B & Y & Y & Y & Y & N \\
G22.04+0.22 & Y & Y & N & SA-R & Y & Y & Y & Y & Y-B \\
G23.01$-$0.41 & Y & Y & N & SA-R & Y & Y & Y & Y & Y-B,R \\
G23.96$-$0.11 & Y & Y & N & SA & Y & Y & \nodata & Y & Y-B \\
G24.94+0.07 & Y & Y & N & SA-B & Y & Y & Y & N & Y-R \\
G25.27$-$0.43 & Y & Y & N & SA-B & Y & N & \nodata & N & Y-B,R\\
G28.28$-$0.36 & Y & N & N & SA-R & Y & N & \nodata & N & Y-B \\
G28.83$-$0.25 & Y & Y & N & SA-R & Y & Y & Y & Y & Y-B,R \\
G35.03+0.35 & Y & Y & Y & SA-B & Y & Y & Y & N & Y-B \\
G37.48$-$0.10 & Y & Y & N & SP & Y & N & \nodata & Y & Y-B,R \\
G39.10+0.49 & Y & Y & N & SA-R & Y & Y & N & N & Y-B\tablenotemark{g} \\
G49.27$-$0.34 & N & Y & N & SA-B & Y & Y & Y & \nodata & \nodata \\
G49.42+0.33 & Y & N & N & SP & Y & Y & \nodata & N & Y-B,R\tablenotemark{h} \\ 
\hline     
G24.00$-$0.10 & Y & \nodata & \nodata & \nodata & \nodata & \nodata & \nodata& \nodata
 & \nodata \\
G28.85$-$0.23 & Y & \nodata & \nodata & \nodata & \nodata & \nodata & \nodata& \nodata
 & \nodata \\
G49.27$-$0.32 & N & \nodata & \nodata & \nodata & \nodata & \nodata & \nodata& \nodata
 & \nodata \\
\enddata
\tablenotetext{a}{For each source and each transition, Y indicates a
  detection, N a nondetection.  A detection is defined as $\ge$6$\sigma$ for
  the VLA maser data (\S\ref{vlaobs}, Tables~\ref{67obstable} and \ref{44obstable}), $\ge$5$\sigma$ for
  the 44 GHz continuum (VLA, \S\ref{44cont}), and $\ge$3$\sigma$ for the
  thermal molecular line data (JCMT, \S\ref{mollines}).  \nodata indicates that a source was not observed in that
  transition.}
\tablenotetext{b}{All observed sources were detected in the \hco(3-2) line.
  The coding indicates the profile type: SA=Self-absorbed. -R indicates red peak is stronger, -B
  indicates blue peak is stronger.  As-R=asymmetric-red, without a clear
  self-absorption dip.  SP=single-peaked.}
\tablenotetext{c}{If the source was not included in the JCMT survey or if no
  6.7 GHz maser was detected, this is indicated by \nodata.}
\tablenotetext{d}{Indicated as Y=yes if any 6.7 GHz maser velocity component falls
  within the \hisoco\/ profile.}
\tablenotetext{e}{Indicated as Y=yes if any 6.7 GHz maser velocity component
  falls on or beyond the wings of the \hco\/ profile.  R=red, B=blue}
\tablenotetext{f}{Questionable because of low S/N in \hisoco\/ spectrum.}
\tablenotetext{g}{Red side of \hco profile falls outside velocity range
  searched for maser emission.}
\tablenotetext{h}{Two spatially and kinematically distinct 6.7 GHz masers
  within JCMT beam.}
\end{deluxetable}

\begin{deluxetable}{cccccccc}
\tablewidth{0pt}
\tablecaption{Fitted Properties: 6.7 GHz Masers \label{maserfitparams_67}} 
\tablehead{ 
\colhead{Field Name} & 
\multicolumn{4}{c}{J2000.0 Coordinates} & 
\colhead{Intensity\tablenotemark{a}} &
\colhead{dI\tablenotemark{b}} &
\colhead{Velocity} \\
\colhead{(Target EGO)} &
\colhead{$\alpha$ ($^{\rm h}~~^{\rm m}~~^{\rm s}$)} &   
\colhead{dx (\pp)} & 
\colhead{$\delta$ ($^{\circ}~~{'}~~{''}$)} &
\colhead{dy (\pp)} & 
\colhead{(Jy beam$^{-1}$)} &
\colhead{(Jy beam$^{-1}$)} &
\colhead{(\kms)} 
}
\tablecolumns{8}
\tabletypesize{\scriptsize}
\setlength{\tabcolsep}{0.05in}
\startdata
G10.29-0.13 & 18 08 49.359 & 0.073 & -20 05 58.989 & 0.068 & 0.52 &0.03 & 1.70\\
Full table is 27 pages, online only\\
\enddata
\tablenotetext{a}{T$_{B}$(K)=7794 * I(\jyb) * $\frac{3.5313}{\theta_{maj}\times\theta_{min}}$ where $\theta_{maj}$
  and $\theta_{min}$ are the dimensions of the synthesized beam in arcseconds
  from Table~\ref{67obstable} (3.5313 square arcseconds is the median value of
  $\theta_{maj}\times\theta_{min}$ for the 6.7 GHz observations).}    
\tablenotetext{b}{Error in fitted intensity.}
\end{deluxetable}

\begin{deluxetable}{lccccccc}
\tablewidth{0pt}
\tablecaption{Fitted Properties: 44 GHz Masers \label{maserfitparams_44}} 
\tablehead{ 
\colhead{Field Name} & 
\multicolumn{4}{c}{J2000.0 Coordinates} & 
\colhead{Intensity\tablenotemark{a}} &
\colhead{dI\tablenotemark{b}} &
\colhead{Velocity} \\
\colhead{(Target EGO)} &
\colhead{$\alpha$ ($^{\rm h}~~^{\rm m}~~^{\rm s}$)} &   
\colhead{dx (\pp)} & 
\colhead{$\delta$ ($^{\circ}~~{'}~~{''}$)} &
\colhead{dy (\pp)} & 
\colhead{(Jy beam$^{-1}$)} &
\colhead{(Jy beam$^{-1}$)} &
\colhead{(\kms)} 
}
\tablecolumns{8}
\tabletypesize{\scriptsize}
\setlength{\tabcolsep}{0.05in}
\startdata
G10.29-0.13 & 18 08 49.113 & 0.040 & -20 05 58.185 &  0.041 &  0.24 &0.03 &10.43\\
Full table is 34 pages, online only
\enddata
\tablenotetext{a}{T$_{B}$(K)=2198 * I(\jyb) * $\frac{0.2867}{\theta_{maj}\times\theta_{min}}$ where $\theta_{maj}$
  and $\theta_{min}$ are the dimensions of the synthesized beam in arcseconds
  from Table~\ref{44obstable} (0.2867 square arcseconds is the median value of
  $\theta_{maj}\times\theta_{min}$ for the 44 GHz observations).}   
\tablenotetext{b}{Error in fitted intensity.}
\end{deluxetable}

\begin{deluxetable}{lcccccccc}
\tablewidth{0pt}
\tablecaption{EGO Sample: Intensity-weighted 6.7 GHz Maser Group Positions\label{67iweighted_egos}} 
\tablehead{ 
\colhead{EGO} & 
\colhead{Gname} & 
\multicolumn{2}{c}{Maser Spot Coord., J2000} & 
\colhead{I$_{peak}$} &
\colhead{v$_{peak}$} &
\colhead{v$_{min}$} &
\colhead{v$_{max}$} &
\colhead{Ang. Sep.\tablenotemark{a}} \\
\colhead{} & 
\colhead{(Maser Spot)} &
\colhead{$\alpha$ ($^{\rm h}~~^{\rm m}~~^{\rm s}$)} &   
\colhead{$\delta$ ($^{\circ}~~{'}~~{''}$)} &
\colhead{(\jyb)} &
\colhead{(\kms)} &
\colhead{(\kms)} &
\colhead{(\kms)} &
\colhead{(\pp)} 
}
\tablecolumns{9}
\tabletypesize{\scriptsize}
\setlength{\tabcolsep}{0.05in}
\startdata
G10.29$-$0.13 & G10.287$-$0.125 & 18 08 49.37 & -20 05 59.0 & 5.92 & 4.58 & 1.70& 19.94 & \nodata \\
G10.34$-$0.14 & G10.342$-$0.142 & 18 08 59.98 & -20 03 35.4 & 23.38 & 10.87 & 3.60 & 18.0 & $<$1 \\
G11.92$-$0.61 & G11.919$-$0.613 & 18 13 58.14 & -18 54 16.3 & 1.12 & 38.99 & 38.58 & 39.54 & 2\\
              & G11.918$-$0.613 & 18 13 58.11 & -18 54 20.7 & 0.50 & 37.34 & 37.07 & 37.62 & 3\\
G18.67+0.03   & G18.667+0.025 & 18 24 53.78 & -12 39 20.8 & 11.83 & 78.73 & 76.13 & 82.16 & $<$1 \\
G18.89$-$0.47 & G18.888$-$0.474 & 18 27 07.83 & -12 41 35.9 & 4.59 & 56.46 & 53.30 & 57.56 & $<$1\\
G19.01$-$0.03 & G19.009$-$0.029 & 18 25 44.77 & -12 22 46.0 & 12.45 & 55.36 & 53.72 & 61.12 & $<$1\\
G19.36$-$0.03 & G19.365$-$0.030 & 18 26 25.78 & -12 03 53.3 & 23.30 & 25.26 & 24.30 & 30.07 & 3\\
G22.04+0.22   & G22.039+0.222 & 18 30 34.70 & -09 34 47.0 & 5.07 & 49.50 & 45.25& 54.85 & 2\\
G23.01$-$0.41 & G23.010$-$0.411 & 18 34 40.28 & -09 00 38.2 & 469.36 & 74.81 &69.46 & 83.73 & 1\\
G23.96$-$0.11 & G23.966$-$0.109 & 18 35 22.21 & -08 01 22.5 & 17.49 & 70.91 &66.93 & 75.03 & $<$1\\
G24.94+0.07   & G24.943+0.074 & 18 36 31.55 & -07 04 16.8 & 2.14 & 46.57 &45.89 & 47.67 & $<$1\\
G25.27$-$0.43 & G25.270$-$0.434 & 18 38 56.96 & -07 00 49.2 & 2.31 & 65.81 &51.95 & 68.28 &$<$1\\
G28.28$-$0.36 & G28.282$-$0.359 & 18 44 13.26 & -04 18 04.8 & 2.28 & 41.42 &40.60 & 41.97 & \nodata \\
G28.83$-$0.25 & G28.832$-$0.253 & 18 44 51.08 & -03 45 48.5 & 56.66 & 83.48 &79.5 & 92.67 & 1\\
G35.03+0.35   & G35.025+0.350 & 18 54 00.66 & +02 01 19.3 & 20.61 & 44.36 &41.07 & 46.97 & 2\\
G37.48$-$0.10 & G37.479$-$0.105 & 19 00 07.14 & +03 59 53.0 & 5.71 & 62.61 &53.83 & 63.16 & $<$1\\
G39.10+0.49   & G39.100+0.491 & 19 00 58.04 & +05 42 45.1 & 17.28 & 15.87 &12.72 & 29.04 &$<$1\\
G49.27$-$0.34 & \nodata & \nodata & \nodata & \nodata & \nodata & \nodata &\nodata & \nodata \\
G49.42+0.33   & G49.416+0.326 & 19 20 59.21 & +14 46 49.7 & 6.35 & -12.04 &-25.89 & -9.70 & 8 \\
              & G49.417+0.324\tablenotemark{b} & 19 20 59.82 & +14 46 49.1 & 2.27 & -26.58 &-26.99 & -23.83 & $<$1\\
\hline     
G24.00$-$0.10 & G23.996-0.100 & 18 35 23.49 & -07 59 29.8 & 2.26 & 68.3 & 65.97 &68.58 & $<$1\\
G28.85$-$0.23 & G28.848-0.228 & 18 44 47.46 & -03:44:17.2 & 1.54 & 102.82 & 99.67& 103.37 & $<$1 \\
G49.27$-$0.32 & \nodata & \nodata & \nodata & \nodata & \nodata & \nodata &\nodata & \nodata \\
\enddata
\tablenotetext{a}{Between intensity-weighted 6.7 GHz maser position and MIPS
24 \um\/ source position (Table~\ref{mipstable}).  See text for discussion.}
\tablenotetext{b}{It is unclear whether this maser associated with the
  targeted EGO, or a distinct, serendipitously detected, source.} 
\end{deluxetable}

\begin{deluxetable}{llccccccccc}
\tablewidth{0pt}
\tablecaption{Serendiptious 6.7 GHz Maser Detections: Intensity-weighted Group Positions\label{67iweighted_serendip}} 
\tablehead{ 
\colhead{Target EGO} & 
\colhead{Gname} & 
\multicolumn{2}{c}{Maser Spot Coord., J2000} & 
\colhead{I$_{peak}$} &
\colhead{v$_{peak}$} &
\colhead{v$_{min}$} &
\colhead{v$_{max}$} &
\colhead{Nearest \emph{IRAS}\tablenotemark{a}} & 
\colhead{Ang. Sep.\tablenotemark{b}} & 
\colhead{Maser?\tablenotemark{c}} \\
\colhead{} & 
\colhead{(Maser Spot)} & 
\colhead{$\alpha$ ($^{\rm h}~~^{\rm m}~~^{\rm s}$)} &   
\colhead{$\delta$ ($^{\circ}~~{'}~~{''}$)} &
\colhead{(\jyb)} &
\colhead{(\kms)} &
\colhead{(\kms)} &
\colhead{(\kms)} &
\colhead{} & 
\colhead{(\pp)} &  
\colhead{(6.7 GHz)}
}
\tablecolumns{11}
\tabletypesize{\scriptsize}
\setlength{\tabcolsep}{0.05in}
\startdata
G10.29$-$0.13 & G10.299$-$0.146 & 18 08 55.54 & -20 05 57.5 & 2.74 & 10.48 & 7.60 & 20.22 &  18060-2005 & 54 & 6.7\\
G10.34$-$0.14 & G10.322$-$0.160 & 18 09 01.45 & -20 05 07.7 & 72.53 & 11.56 & 4.15 & 14.30 & 18060-2005 & 45 & 6.7 \\
G11.92$-$0.61 & G11.936$-$0.616 & 18 14 00.90 & -18 53 26.1 &    42.77 & 32.26 & 30.07 & 44.2 & 18110-1854 & 9 & 6.7 \\
G18.67+0.03   & G18.662+0.035 & 18 24 51.10 & -12 39 22.0 & 3.90 & 79.01 & 77.64 & 82.71 & 18220-1241 & 5& 6.7?\\
G23.96$-$0.11 & G23.986$-$0.089 & 18 35 20.09 & -07 59 45.0 & 2.62 & 65.15 & 64.05 & 67.07 & 18326-0802 & 5 & 6.7? \\
G24.94+0.07   & G24.920+0.088 & 18 36 25.94 & -07 05 07.8 & 4.19 & 53.16 & 45.34 & 53.71 & 18337-0707 & 30 & 6.7? \\
G37.48$-$0.10\tablenotemark{d} & G37.546$-$0.112& 19 00 16.00 & +04 03 16.2 & 3.98 & 49.99 & 49.31 & 52.87 & 18577+0358 & 10 & 6.7? \\
\enddata 
\tablenotetext{a}{B1950 coordinates.}  
\tablenotetext{b}{From \emph{IRAS} point source.}
\tablenotetext{c}{As in Table~\ref{samplelit}, 6.7 indicates a 6.7 GHz
  \meth\/ maser with an interferometric position.  All 
  ``6.7'' designations are from \citet{Walsh98}.  6.7? indicates a 6.7
  GHz \meth\/ maser detection from a single-dish survey with a
  reported or targeted position within one single-dish beam of our
  interferometric position.  All 6.7?  are from \citet{Sz00,Sz02,Pandian07}.}
\tablenotetext{d}{This source is located at \q30\% level of the VLA primary beam.  We
  include it in this table and Table~\ref{maserfitparams_67} because,
  to the best of our knowledge, an interferometric position has not
  been previously published.}

\end{deluxetable}

\clearpage

\begin{figure}
\plotone{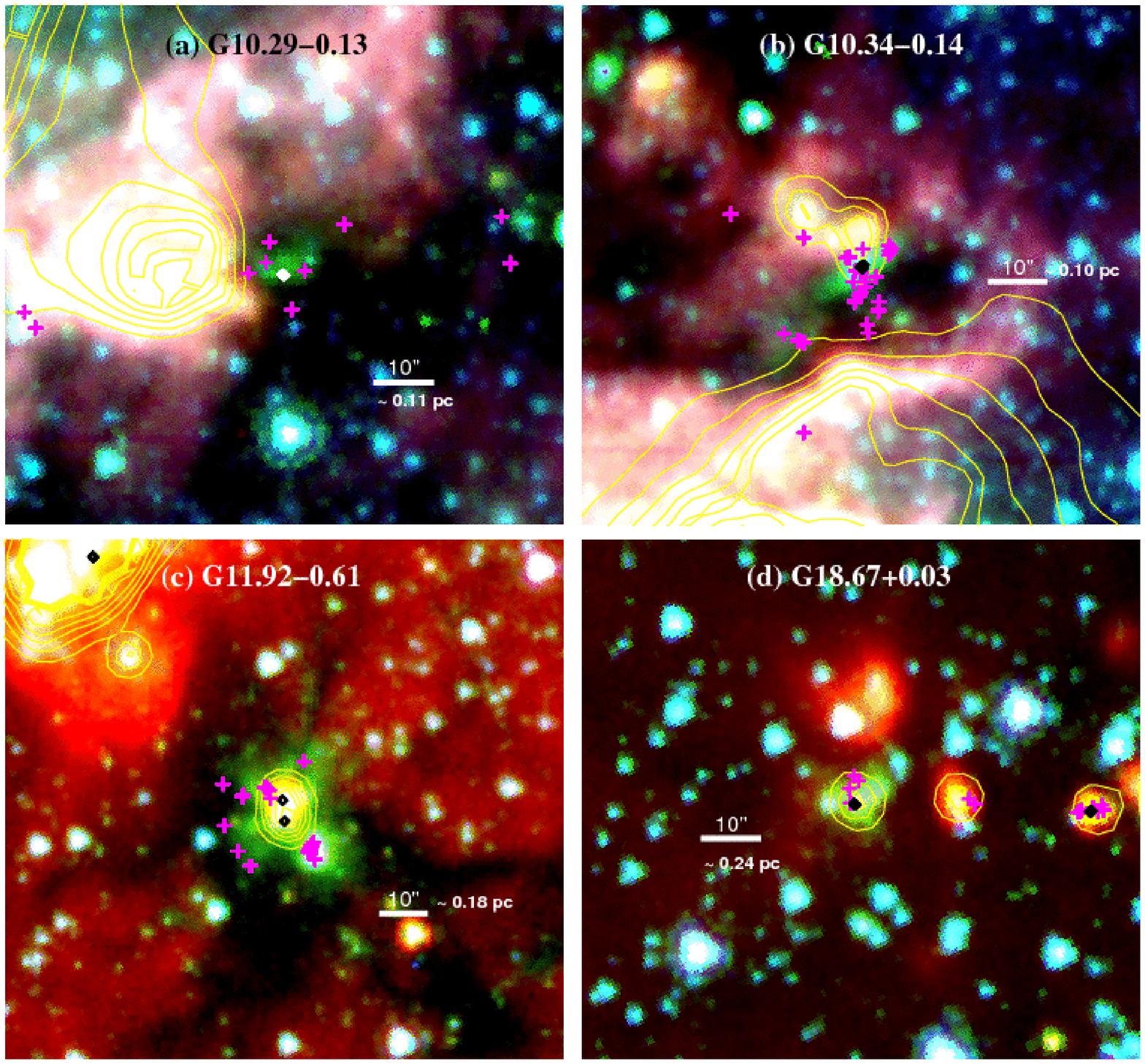}
\caption{Three-color GLIMPSE IRAC images showing 8.0 \um\/ (red), 4.5 \um\/
(green), and 3.6 \um\/ (blue).  Yellow 24 \um\/ MIPSGAL contours \citep{Carey09} are
superposed; contour levels for each source are given in Table~\ref{mipstable}.
Each panel is centered on the targeted EGO; north is up and east to
the left in all images.  The physical scale label on each scalebar (in pc)
assumes the distance to the source listed in Table~\ref{jcmtfitstable}.  Positions of 6.7 GHz \methanol\/
masers from Table~\ref{maserfitparams_67} are marked with diamonds.  Positions of
44 GHz \methanol\/ masers from Table~\ref{maserfitparams_44} are marked with
magenta crosses.  The primary beam (FWHP) of the VLA is 6\farcm7 at 6.7 GHz and
1\farcm0 at 44 GHz.}
\label{3color}
\end{figure}

\begin{figure}
\plotone{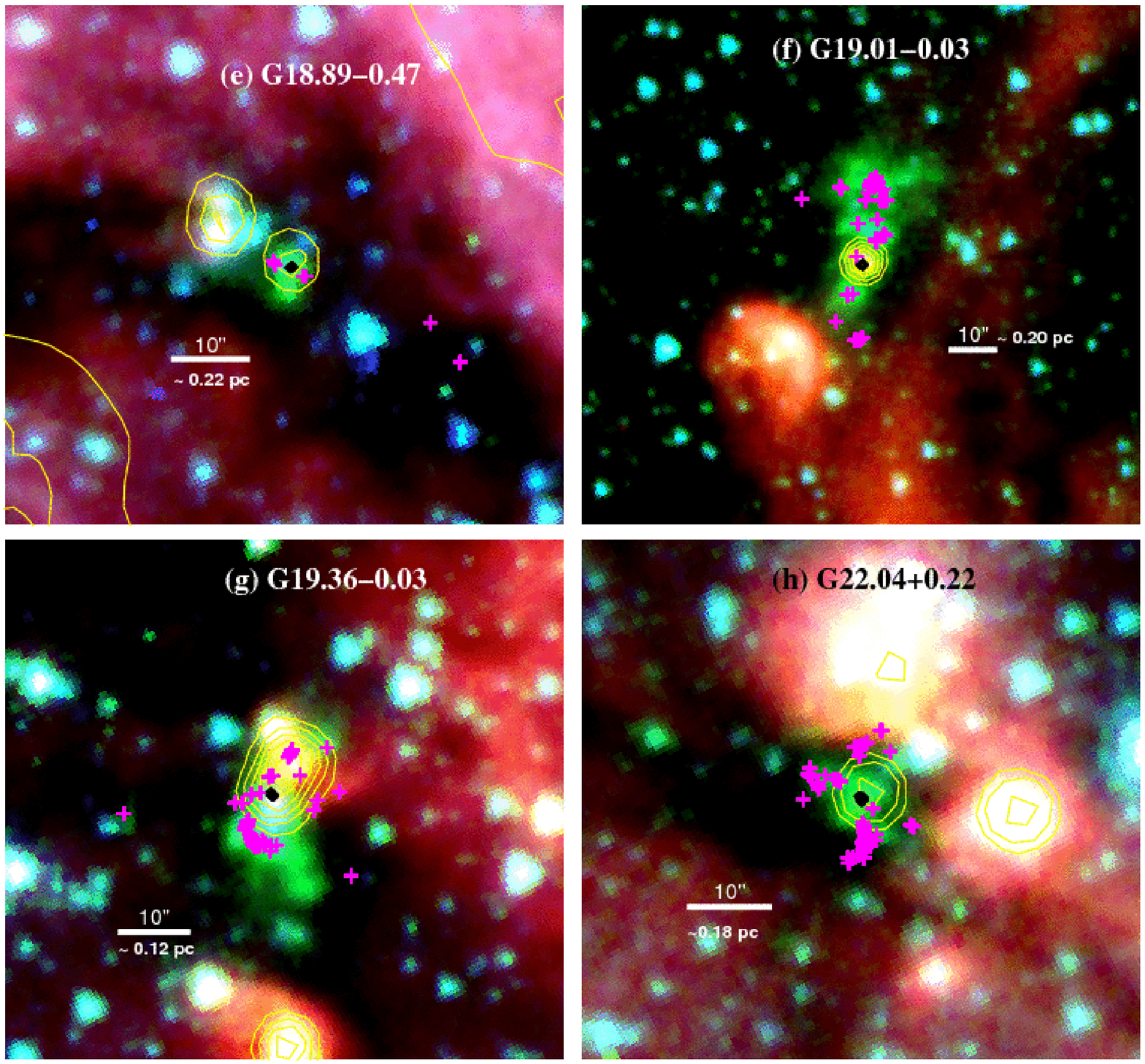}
\addtocounter{figure}{-1}
\caption{(continued)}
\end{figure}

\begin{figure}
\plotone{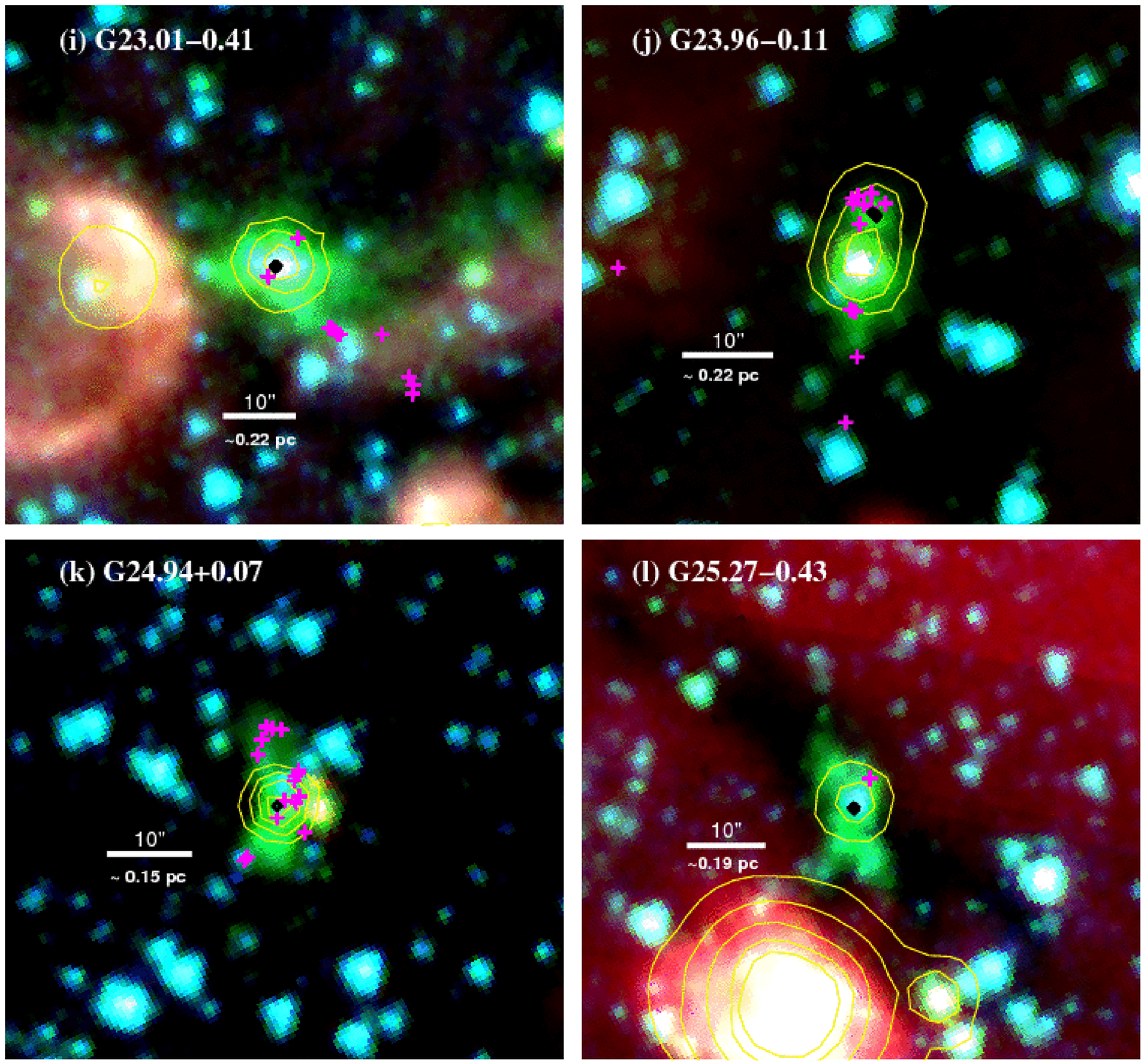}
\addtocounter{figure}{-1}
\caption{(continued)}
\end{figure}

\begin{figure}
\plotone{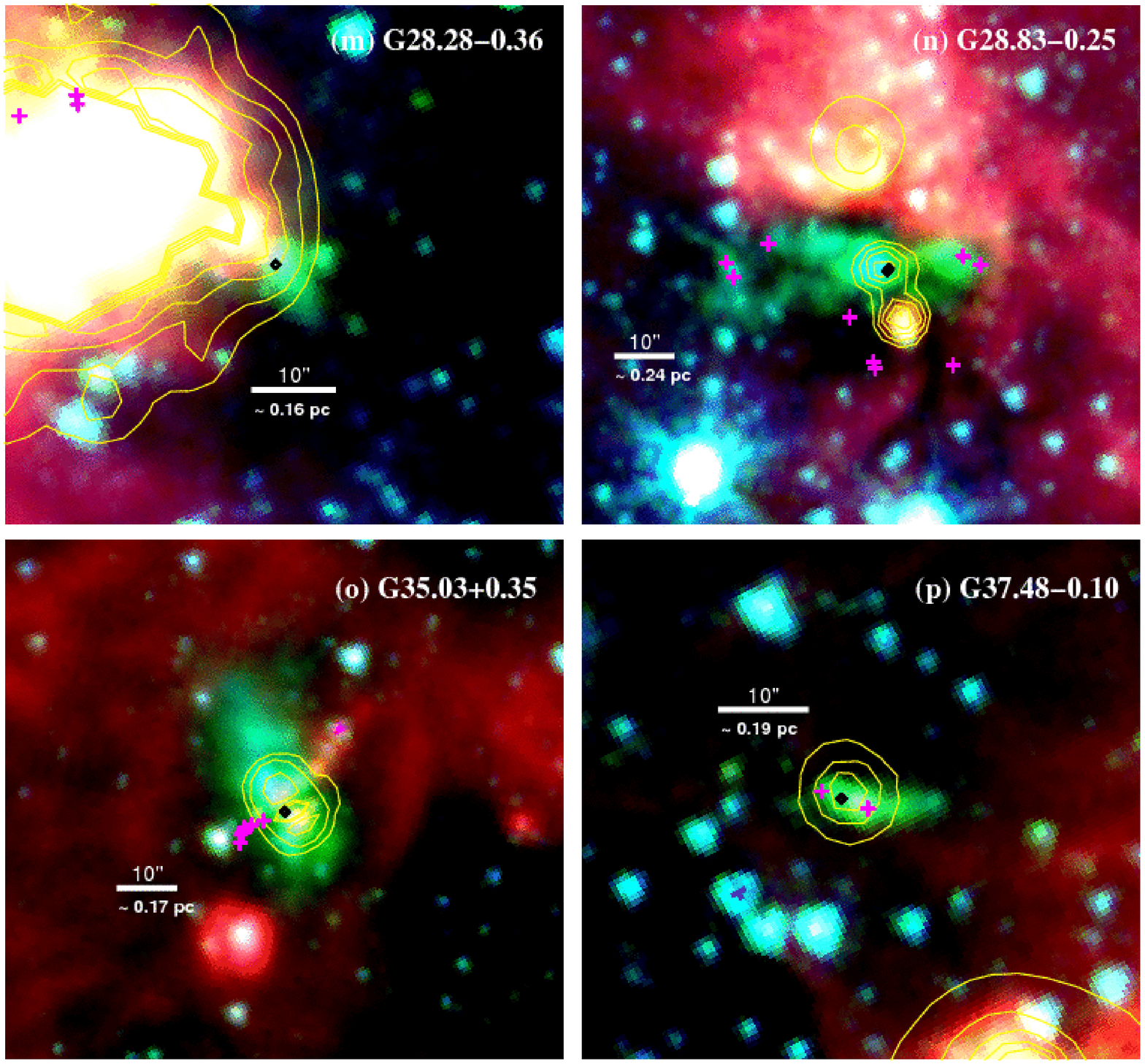}
\addtocounter{figure}{-1}
\caption{(continued)}
\end{figure}

\begin{figure}
\plotone{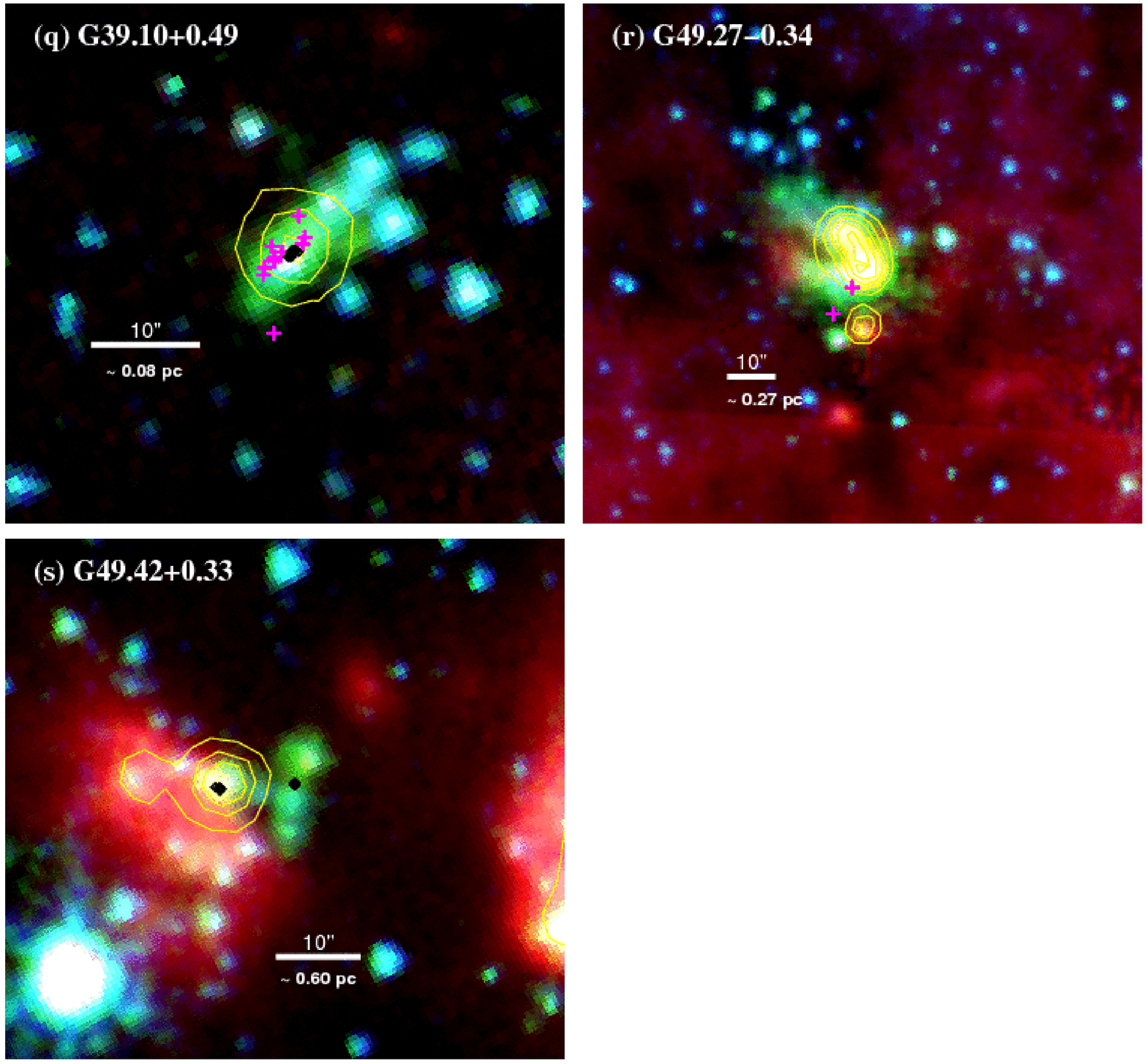}
\addtocounter{figure}{-1}
\caption{(continued)}
\end{figure}

\clearpage

\begin{figure}
\plotone{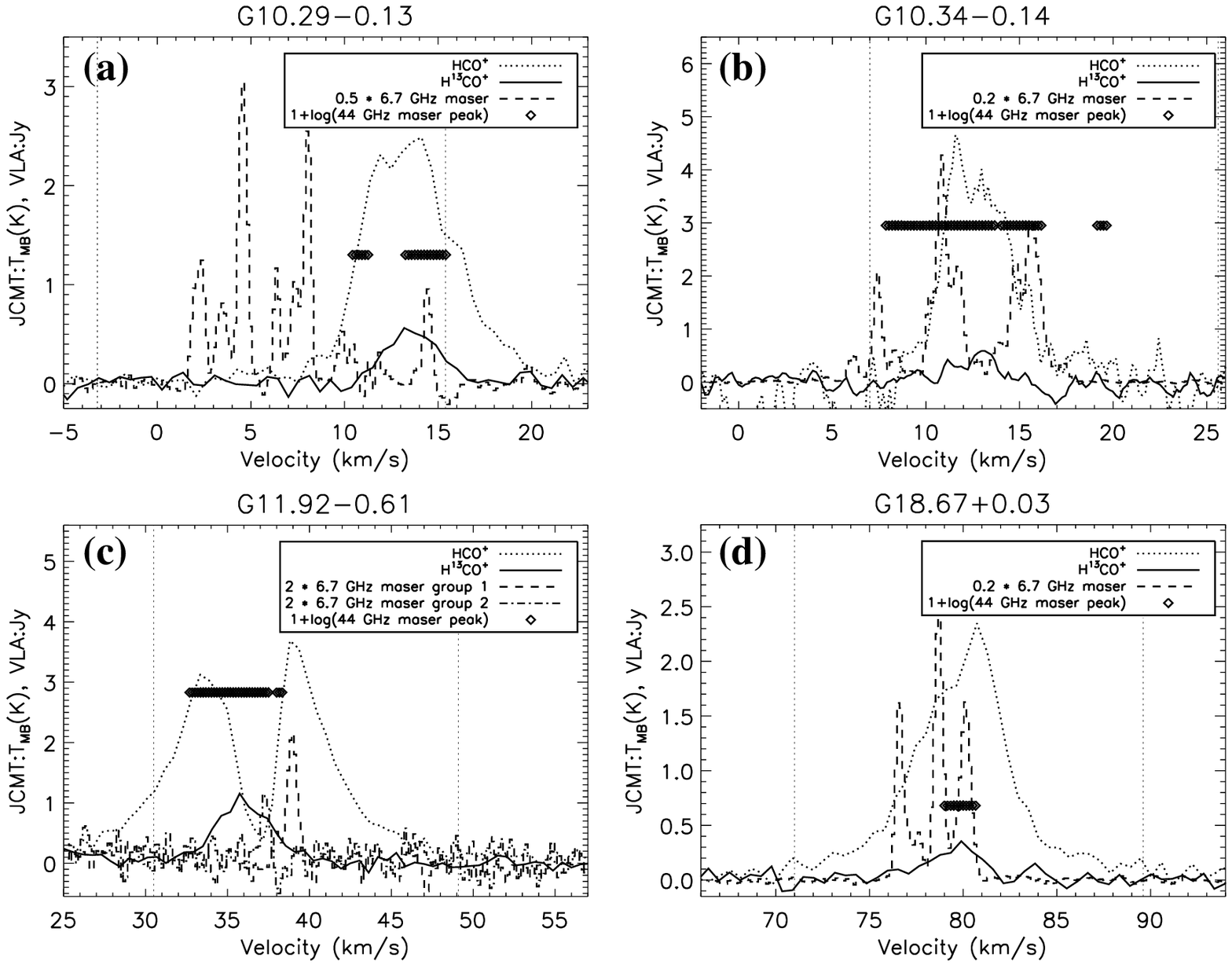}
\caption{JCMT \hco(3-2) (dotted line) and \hisoco(3-2) (solid line) spectra
towards each EGO, overplotted with the (scaled) integrated 6.7 GHz \methanol\/
maser spectrum of the associated maser group(s) (dashed and dot-dashed lines).
(See \S\ref{spatial67} and Table~\ref{67iweighted_egos}).  Velocities of 44
GHz \methanol\/ masers (Table~\ref{maserfitparams_44}) are marked by diamonds.
Scaling factors are given in the legend of each panel.  The velocity range
shown in each panel is that searched for 6.7 GHz \methanol\/ masers
(Table~\ref{67obstable}), except for (s), in which it is denoted by vertical
dashed lines.  The velocity range searched for 44 GHz \methanol\/ masers
(Table~\ref{44obstable}) is denoted by dotted vertical lines in all panels.}
\label{jcmtlineplots}
\end{figure}

\begin{figure}
\plotone{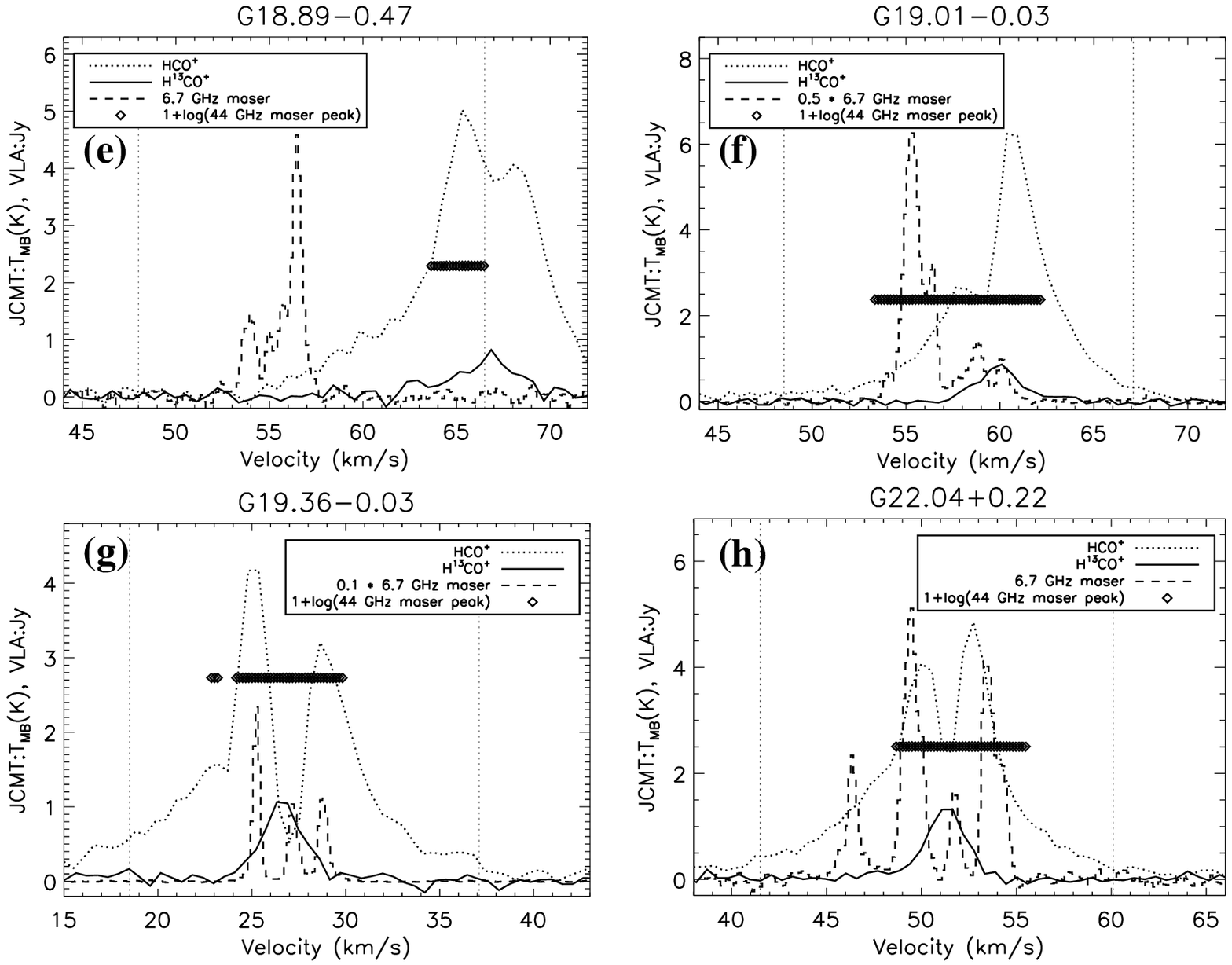}
\addtocounter{figure}{-1}
\caption{(continued)}
\end{figure}

\begin{figure}
\plotone{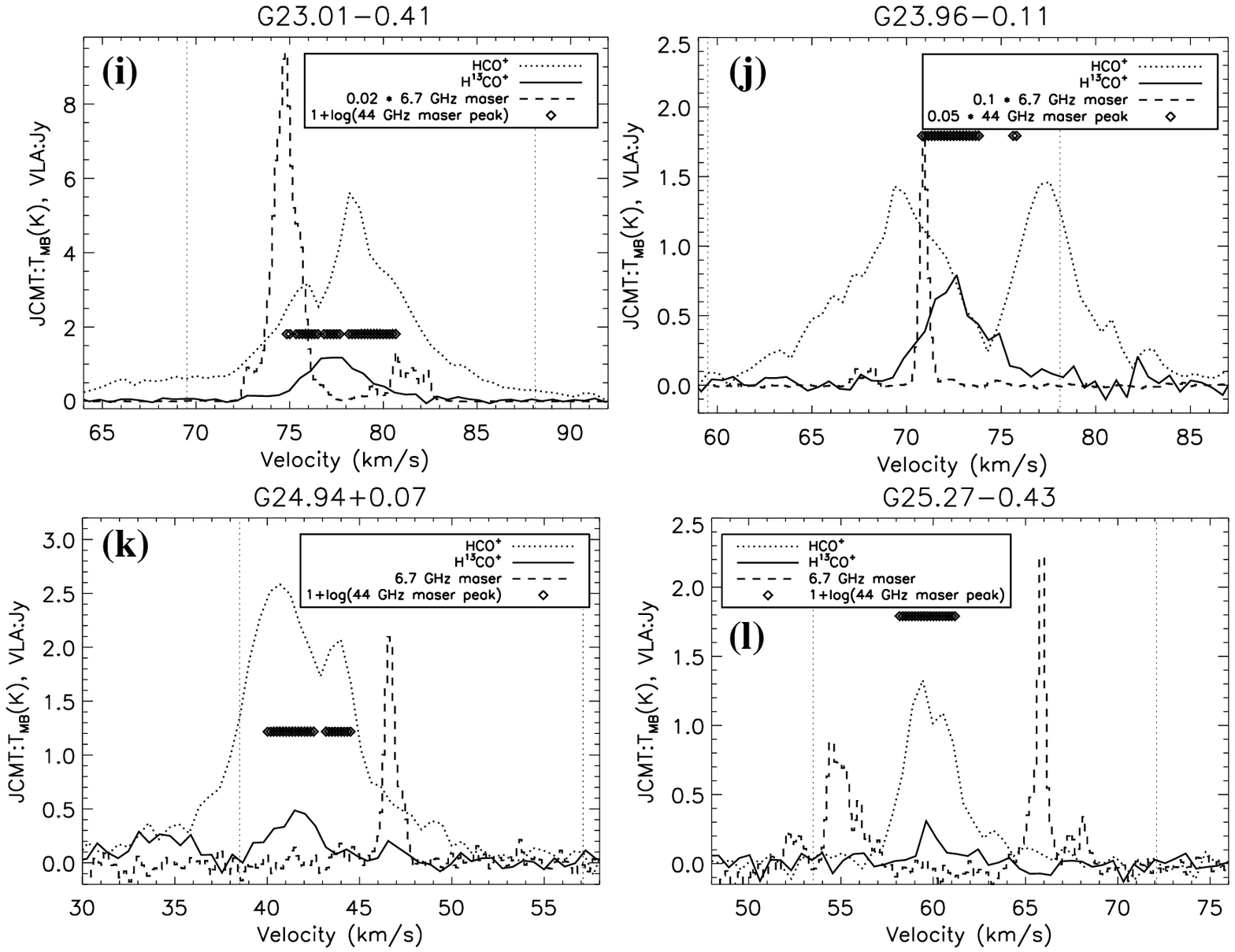}
\addtocounter{figure}{-1}
\caption{(continued)}
\end{figure}

\begin{figure}
\plotone{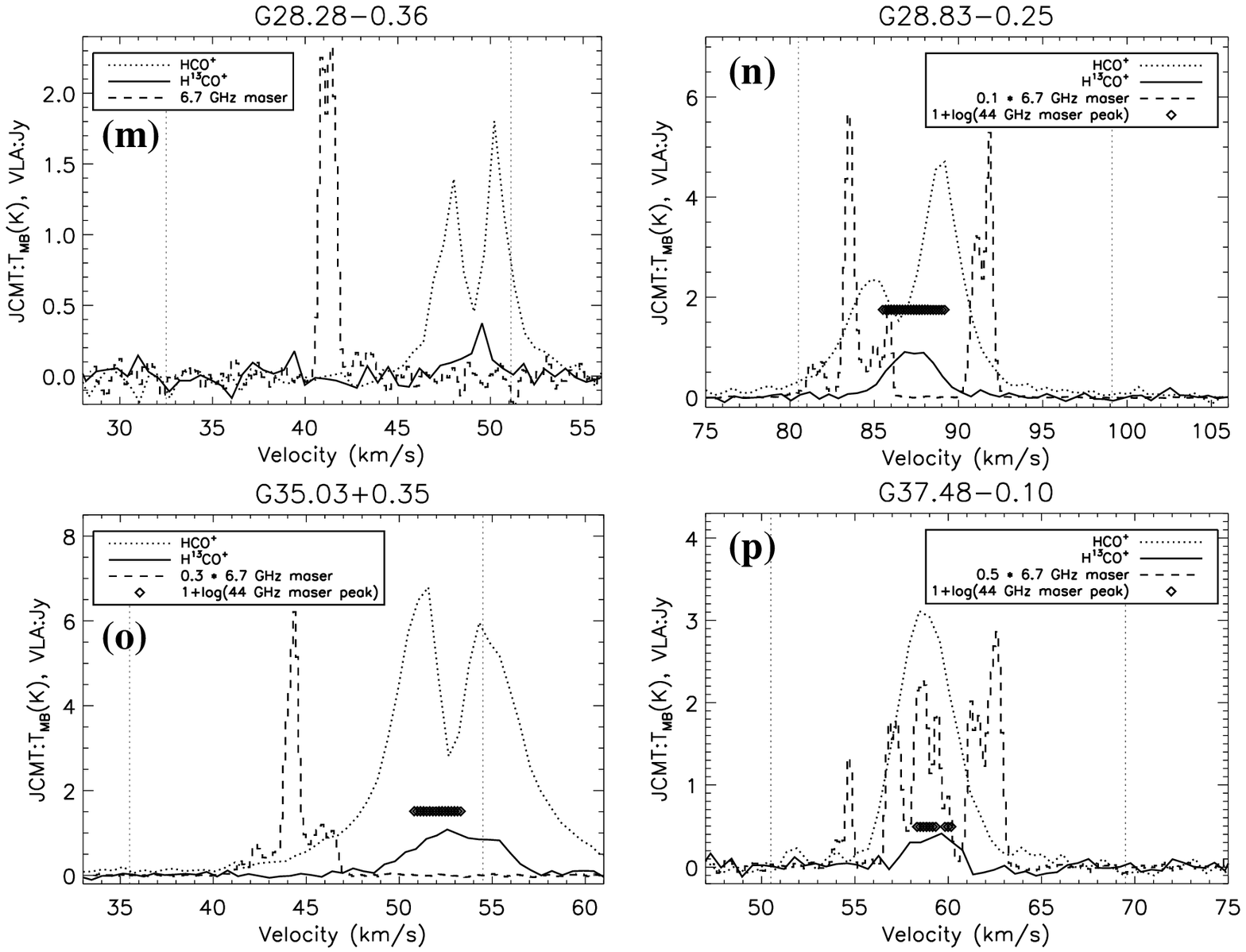}
\addtocounter{figure}{-1}
\caption{(continued)}
\end{figure}

\begin{figure}
\plotone{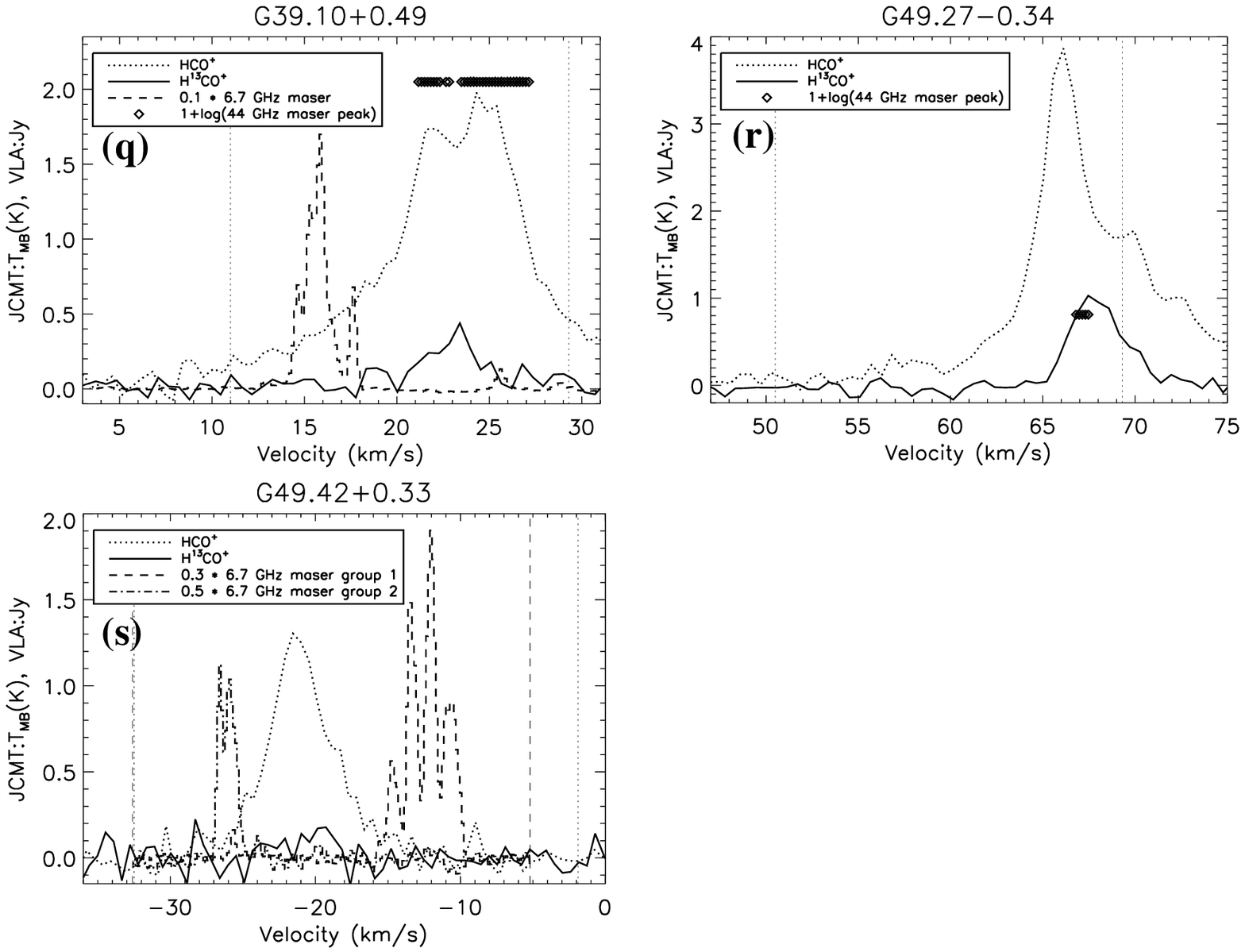}
\addtocounter{figure}{-1}
\caption{(continued)}
\end{figure}

\begin{figure}
\plotone{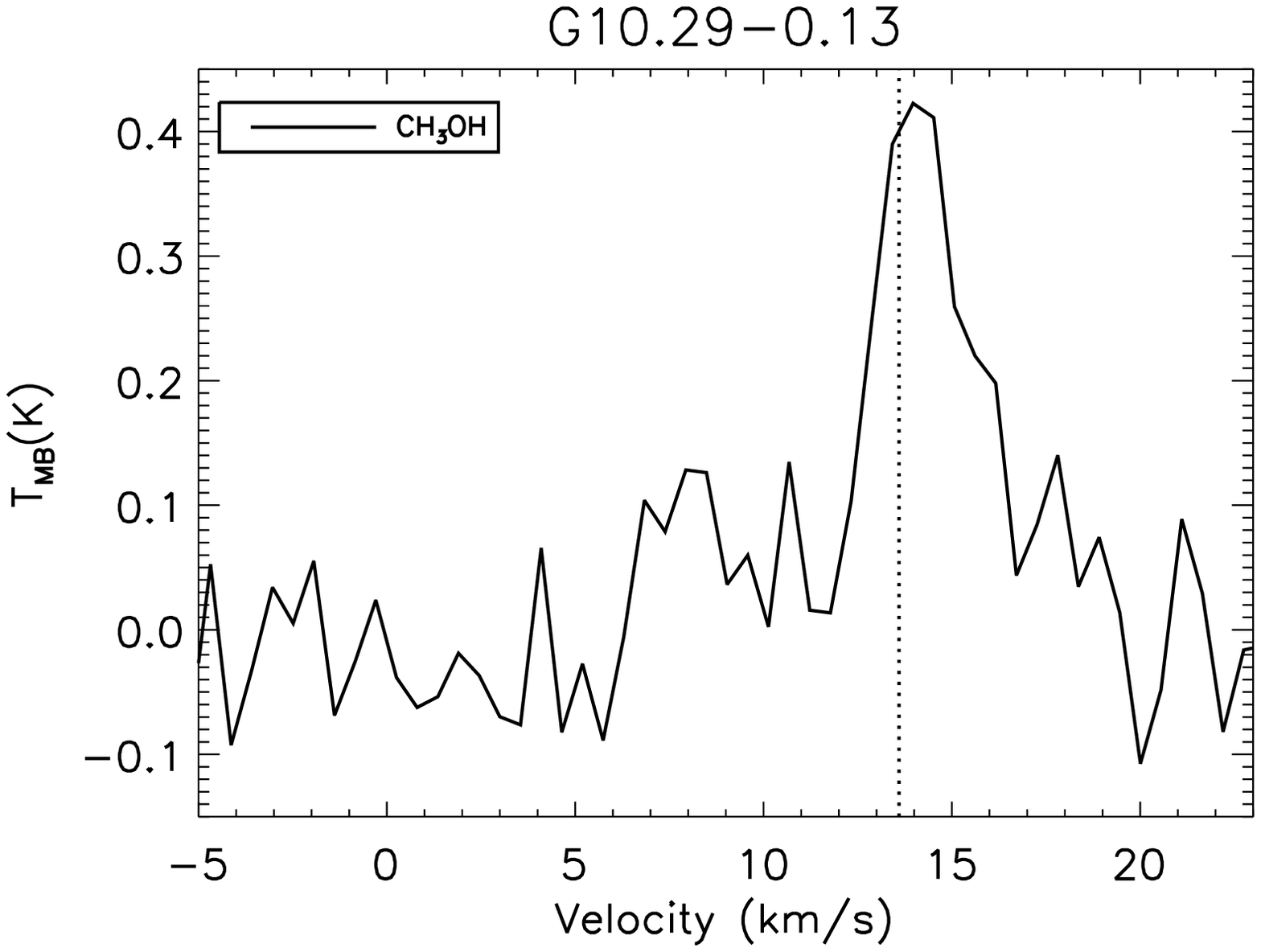}
\caption{JCMT thermal CH$_{3}$OH(5$_{2,3}$-4$_{1,3}$,
\elow=44.3 K) spectrum.  Spectra towards all EGOs observed in this transition 
with the JCMT are available online, including nondetections.  
(See \S\ref{mollines} and Table~\ref{jcmtfitstable}.)
The velocity range shown for each EGO is the same as that 
in Figure~\ref{jcmtlineplots}.  The dotted vertical line marks the \hisoco\/
velocity from Table~\ref{jcmtfitstable}.}
\label{thermalmethspectra}
\end{figure}

\begin{figure}
\plotone{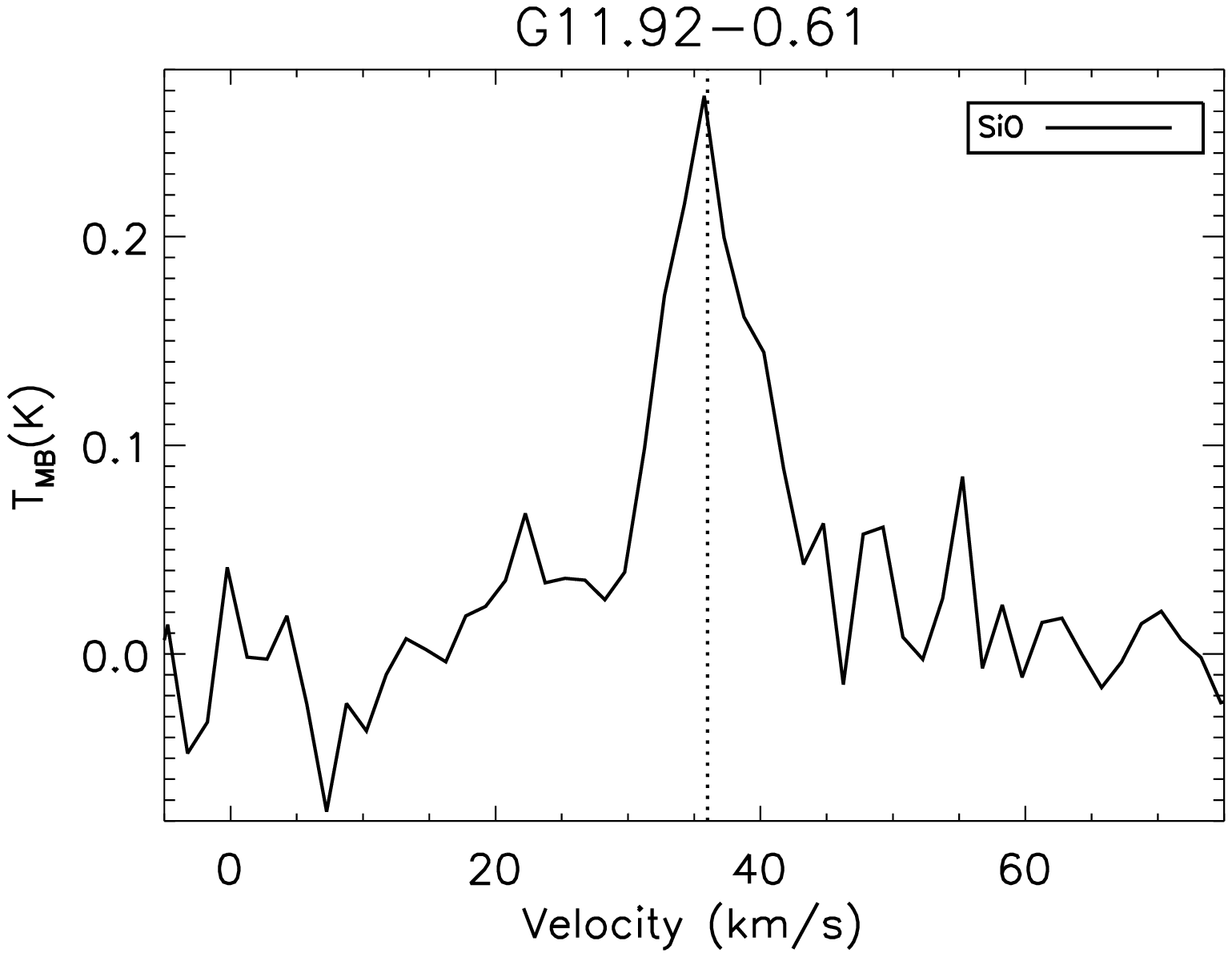}
\caption{JCMT thermal SiO (5-4, \elow=20.8 K) spectrum.  
(See \S\ref{mollines} and Table~\ref{jcmtfitstable}.) 
Spectra towards all EGOs observed in this transition 
with the JCMT are available online, including nondetections.  
A velocity range of \q80\kms\/ is shown, centered on
the velocity of the SiO line.  The dotted vertical line marks the \hisoco\/
velocity from Table~\ref{jcmtfitstable}.}
\label{siospectra}
\end{figure}

\clearpage
\begin{figure}
\plotone{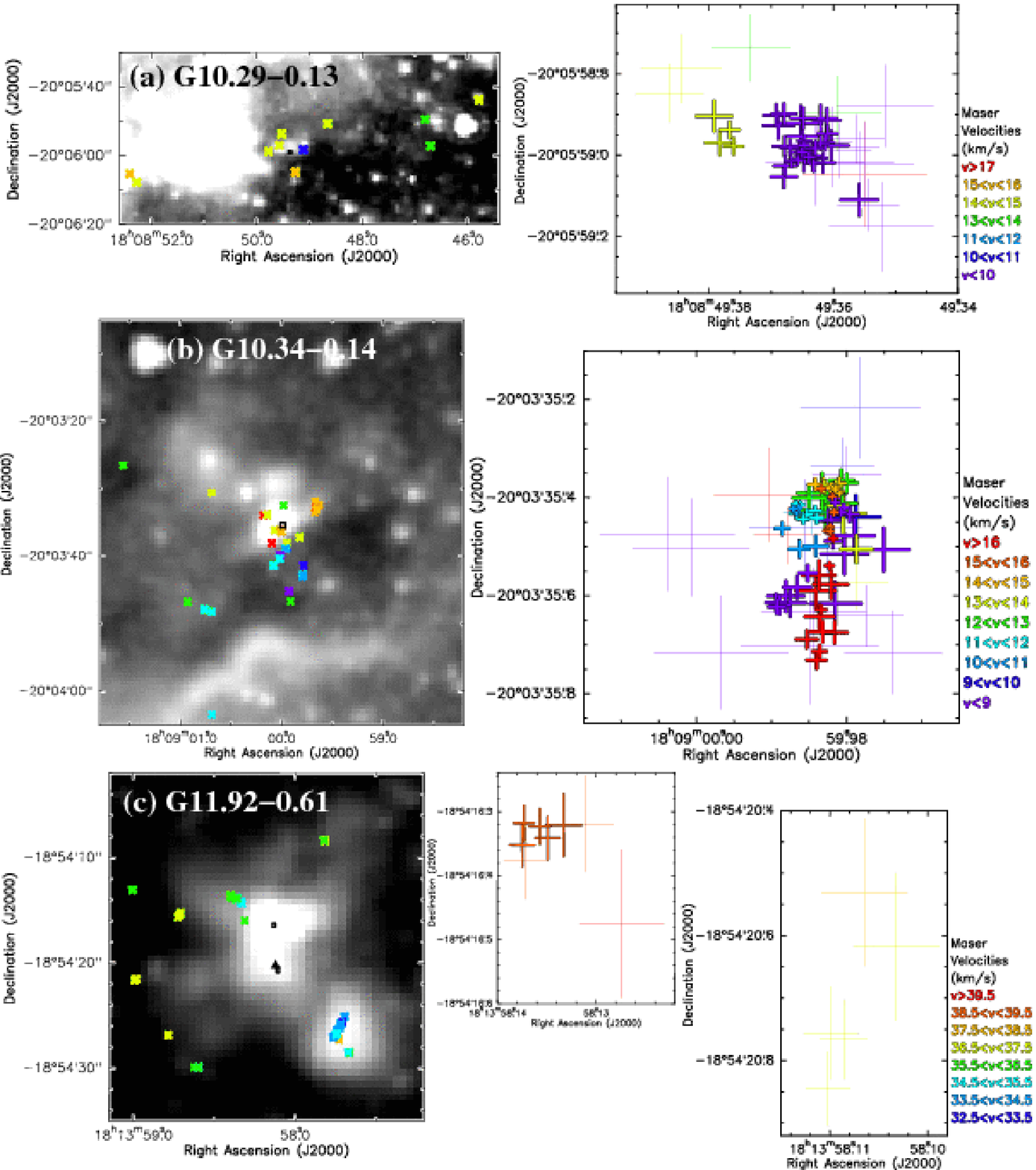}
\caption{}
\label{kinplots}
\end{figure}

\begin{figure}
\plotone{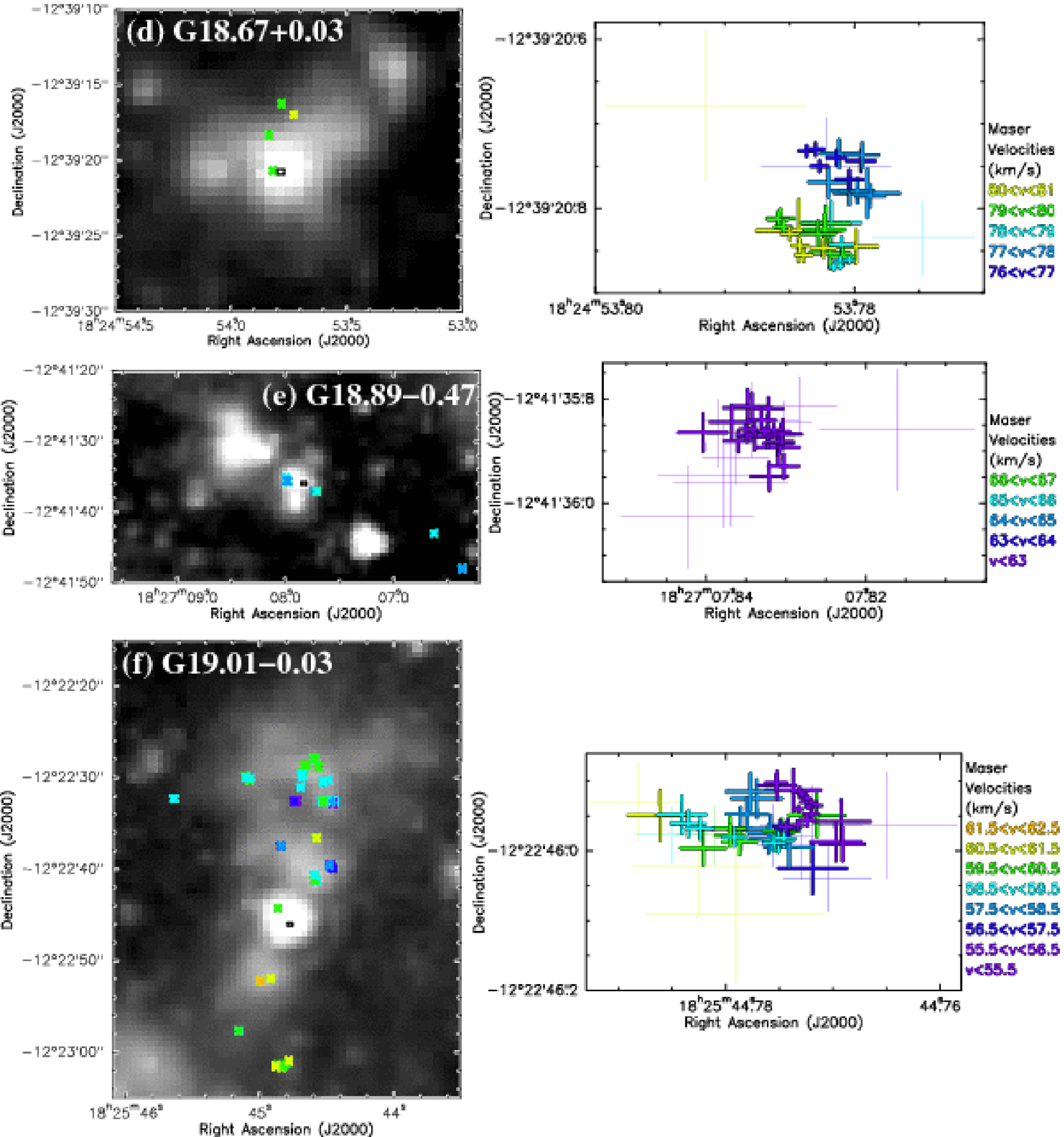}
\addtocounter{figure}{-1}
\caption{}
\end{figure}

\begin{figure}
\plotone{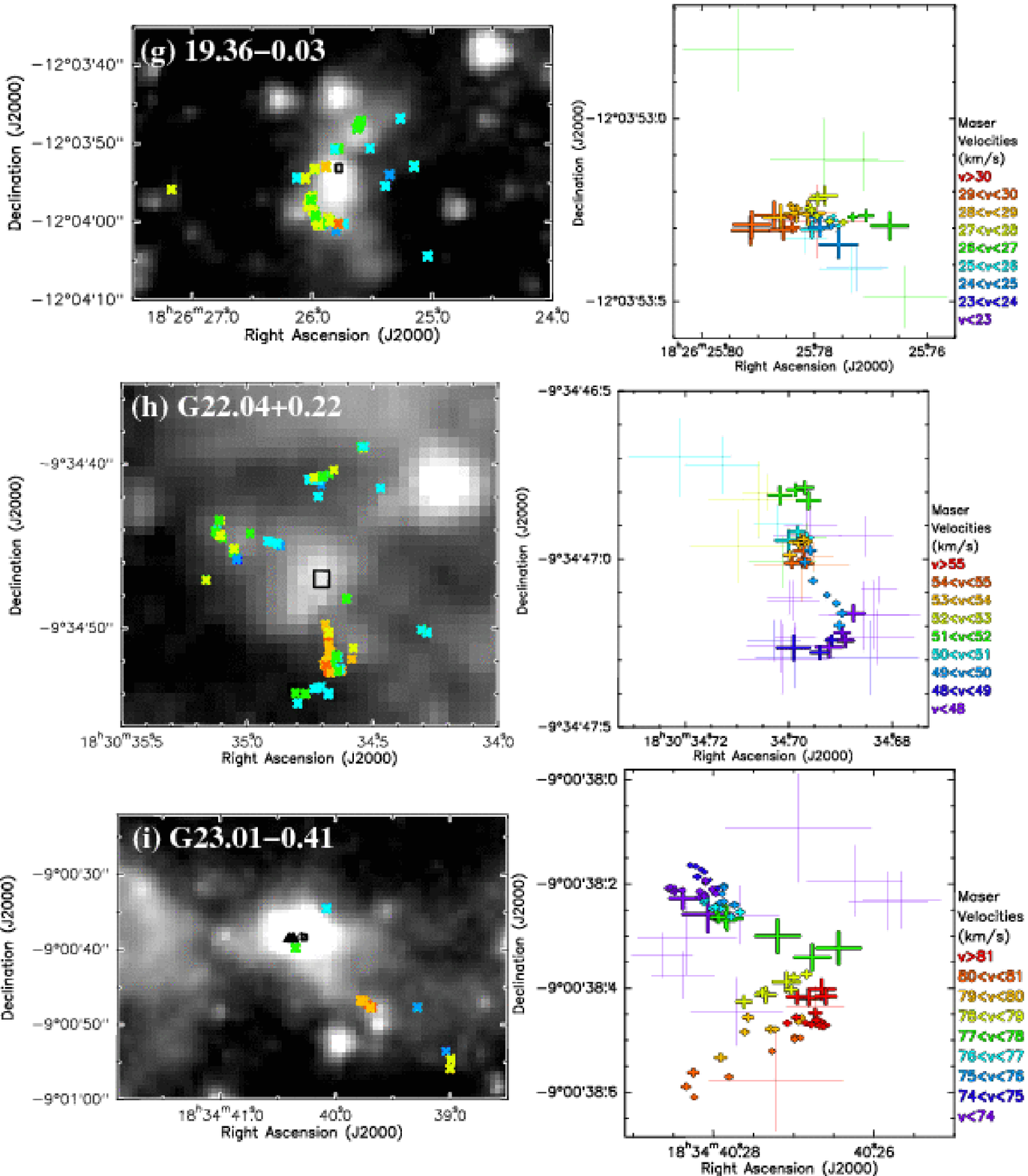}
\addtocounter{figure}{-1}
\caption{}
\end{figure}

\begin{figure}
\plotone{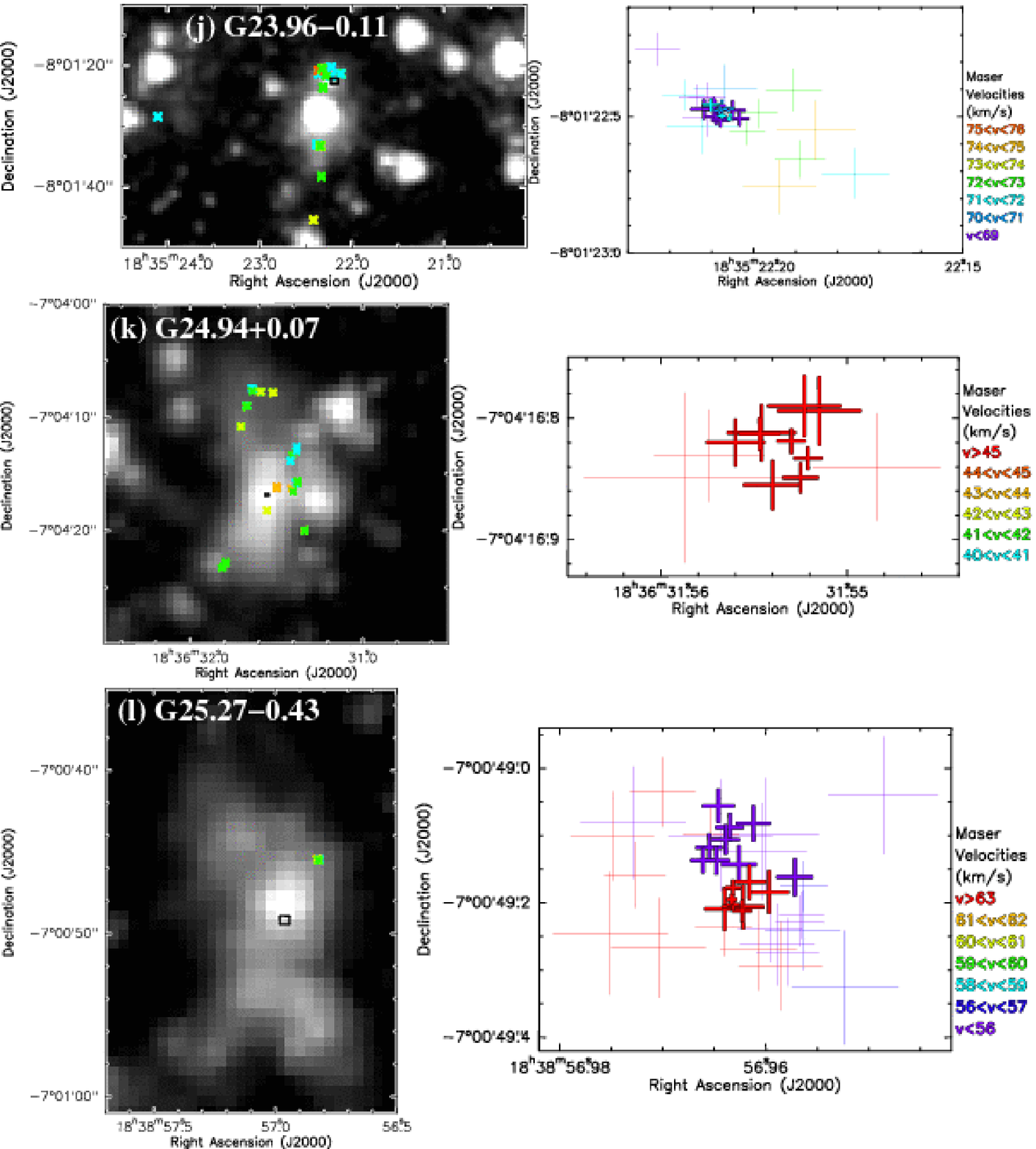}
\addtocounter{figure}{-1}
\caption{}
\end{figure}

\begin{figure}
\plotone{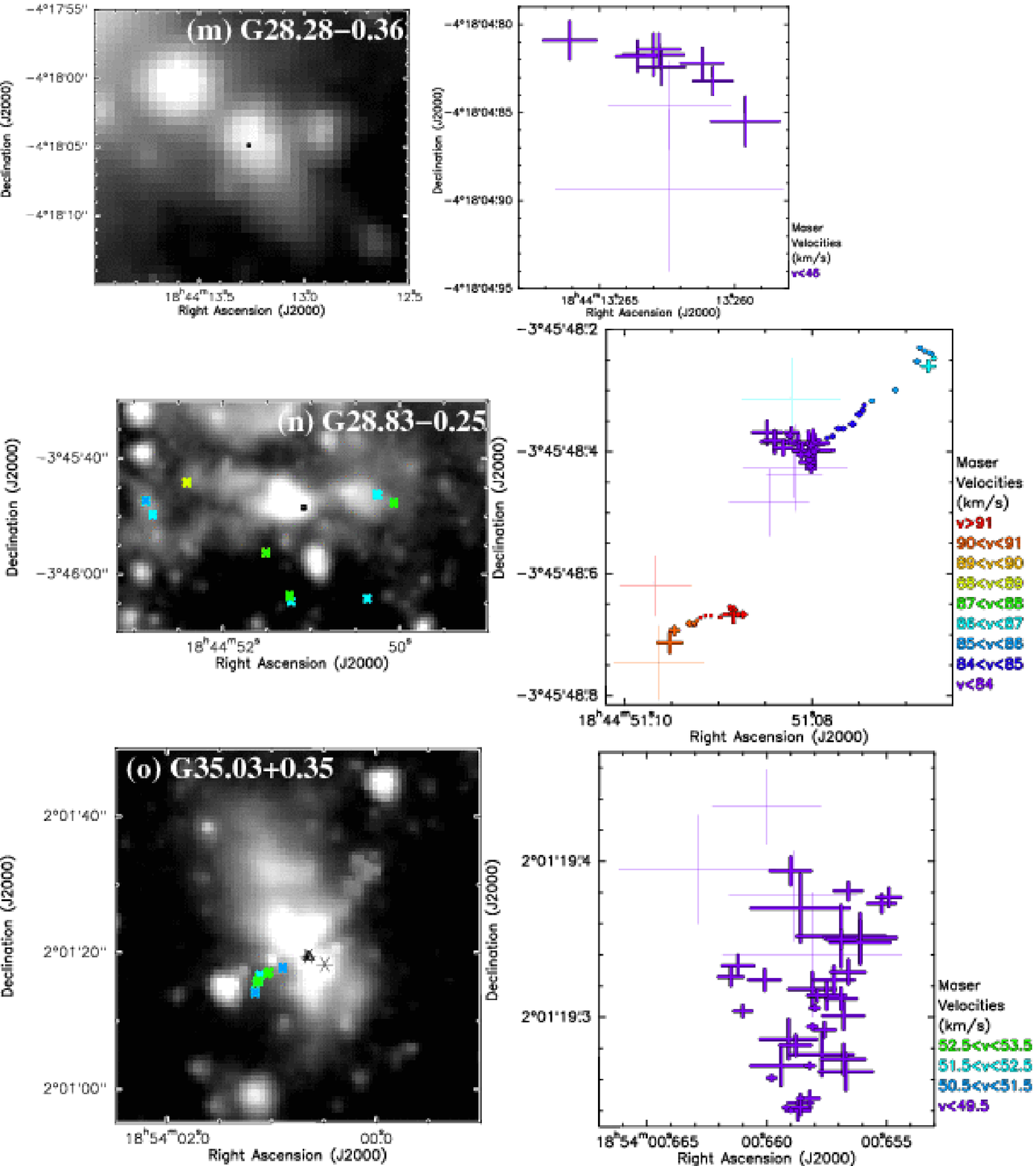}
\addtocounter{figure}{-1}
\caption{}
\end{figure}

\begin{figure}
\plotone{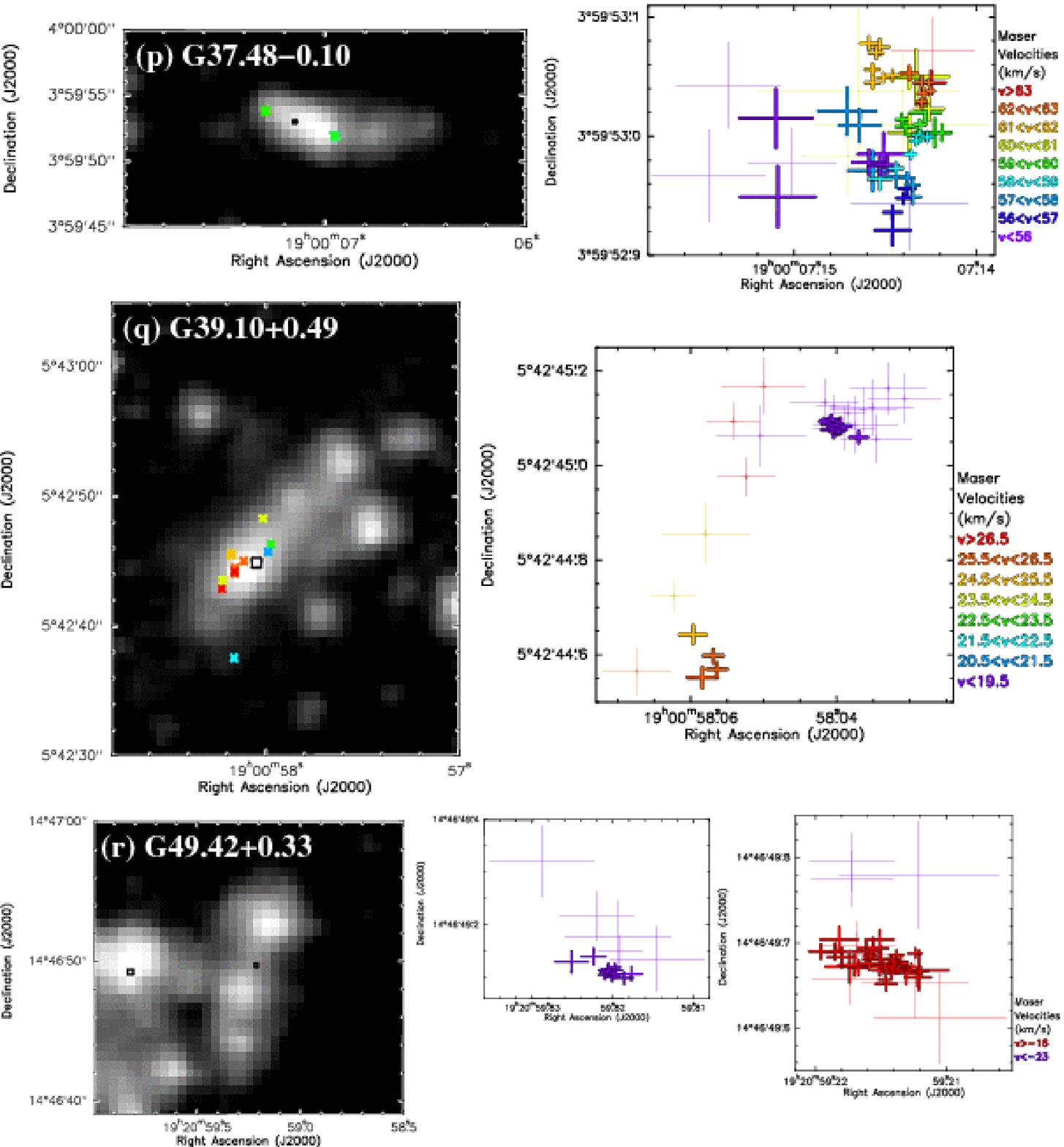}
\addtocounter{figure}{-1}
\caption{}
\end{figure}

\begin{figure}
\addtocounter{figure}{-1}
\caption{\textbf{Left:} Greyscale: 4.5 \um\/ (GLIMPSE), X: 44 GHz
  \methanol\/ masers, color-coded by velocity.  Other masers discussed
  in \S\ref{individualsources} are plotted as triangles (\water\/
  masers) or diamond (\formald\/ maser).  Peak positions for 44 GHz
  continuum sources detected in our survey (G35.03+0.35,
  \S\ref{g3503}) are plotted as asterisks.  For each source, the black
  rectangle(s) overplotted in the left panel is the field of view shown
  in the right panel(s). \textbf{Right:} Fitted positions, with error
  bars, of 6.7 GHz \methanol\/ masers from
  Table~\ref{maserfitparams_67}, color-coded by velocity.  For
  low-declination sources (a-i), fitted positions with errors
  (Table~\ref{maserfitparams_67}) $\delta \geq$0\farcs15 are not
  displayed (SNR$\lesssim$10$\sigma$), and fitted positions with
  errors 0\farcs05$\geq\delta<$0\farcs15
  (10$\sigma<$SNR$\lesssim$30$\sigma$) are drawn as light lines (see
  \S\ref{maserkin}).  For high-declination sources (k-r), the
  corresponding cutoffs are 0\farcs10 and 0\farcs033.  For each
  source, a legend (at right) lists the absolute limits in \kms\/ of
  the velocity bins for that source: the bin color-coded green is
  approximately centered on the thermal gas \vlsr\/
  (Table~\ref{jcmtfitstable}).  To increase the range of
  distinguishable colors, purple is used to represent the most
  blueshifted masers.  Velocity bins in which no 44 or 6.7 GHz
  \methanol\/ masers are detected are excluded from the legend.}
\end{figure}

\clearpage
\begin{figure}
\plotone{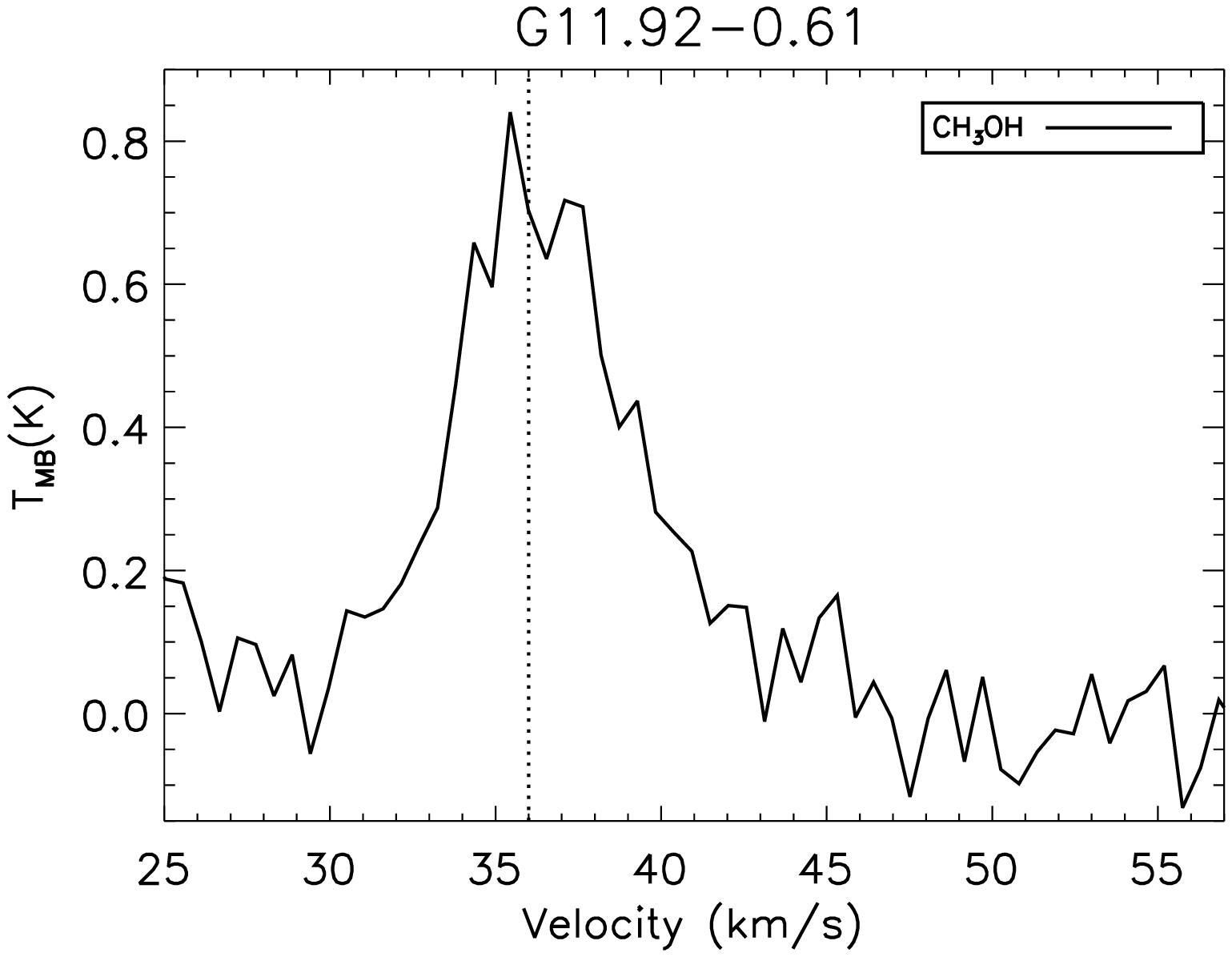}
\addtocounter{figure}{-3}
\caption{Figure Set 3: JCMT thermal CH$_{3}$OH(5$_{2,3}$-4$_{1,3}$,
\elow=44.3 K) spectrum.  
(See \S\ref{mollines} and Table~\ref{jcmtfitstable}.)
The velocity range shown for each EGO is the same as that 
in Figure~\ref{jcmtlineplots}.  The dotted vertical line marks the \hisoco\/
velocity from Table~\ref{jcmtfitstable}.}
\end{figure}

\begin{figure}
\plotone{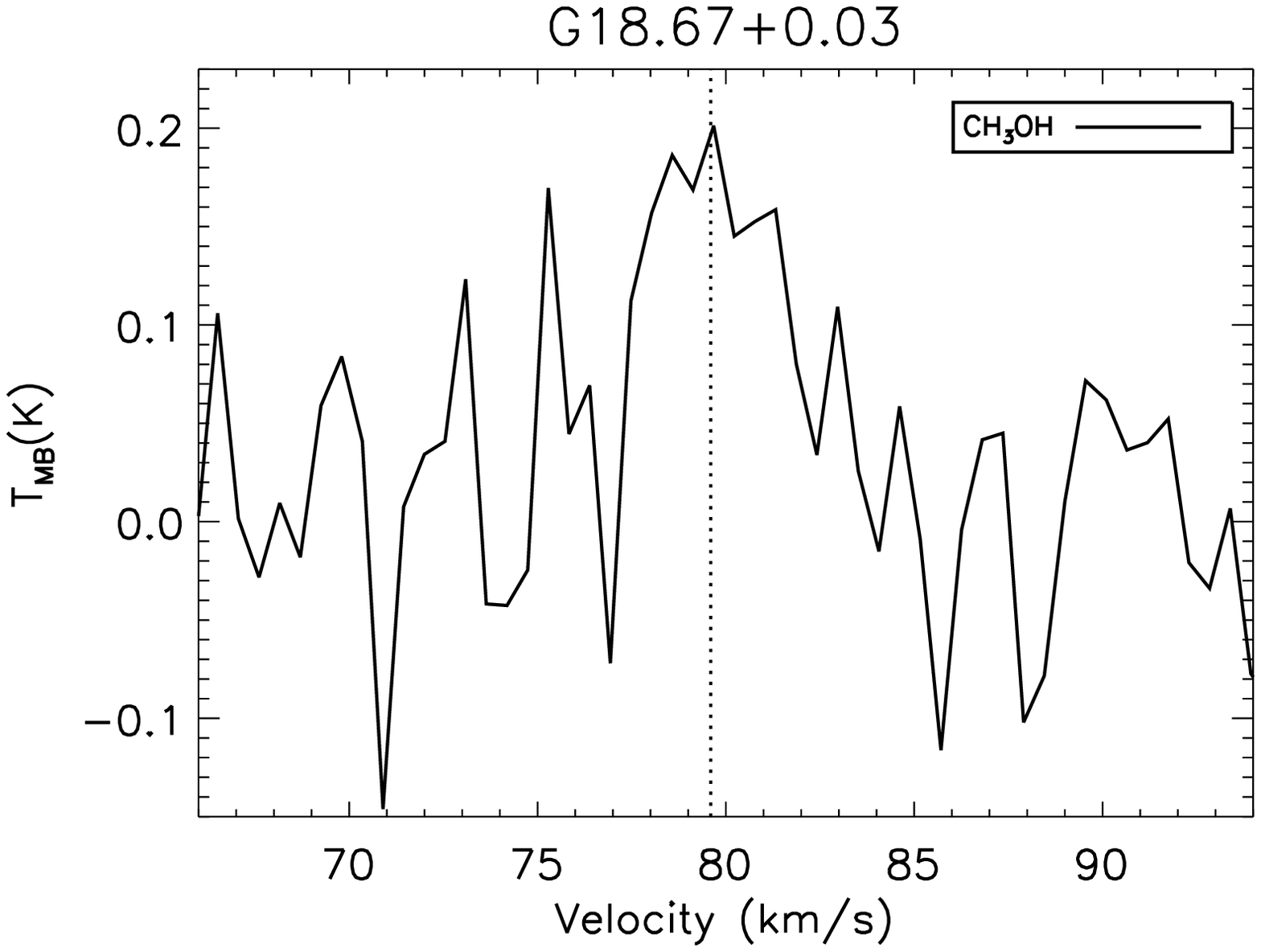}
\addtocounter{figure}{-1}
\caption{Figure Set 3: JCMT thermal CH$_{3}$OH(5$_{2,3}$-4$_{1,3}$,
\elow=44.3 K) spectrum.  
(See \S\ref{mollines} and Table~\ref{jcmtfitstable}.)
The velocity range shown for each EGO is the same as that 
in Figure~\ref{jcmtlineplots}.  The dotted vertical line marks the \hisoco\/
velocity from Table~\ref{jcmtfitstable}.}
\end{figure}

\begin{figure}
\plotone{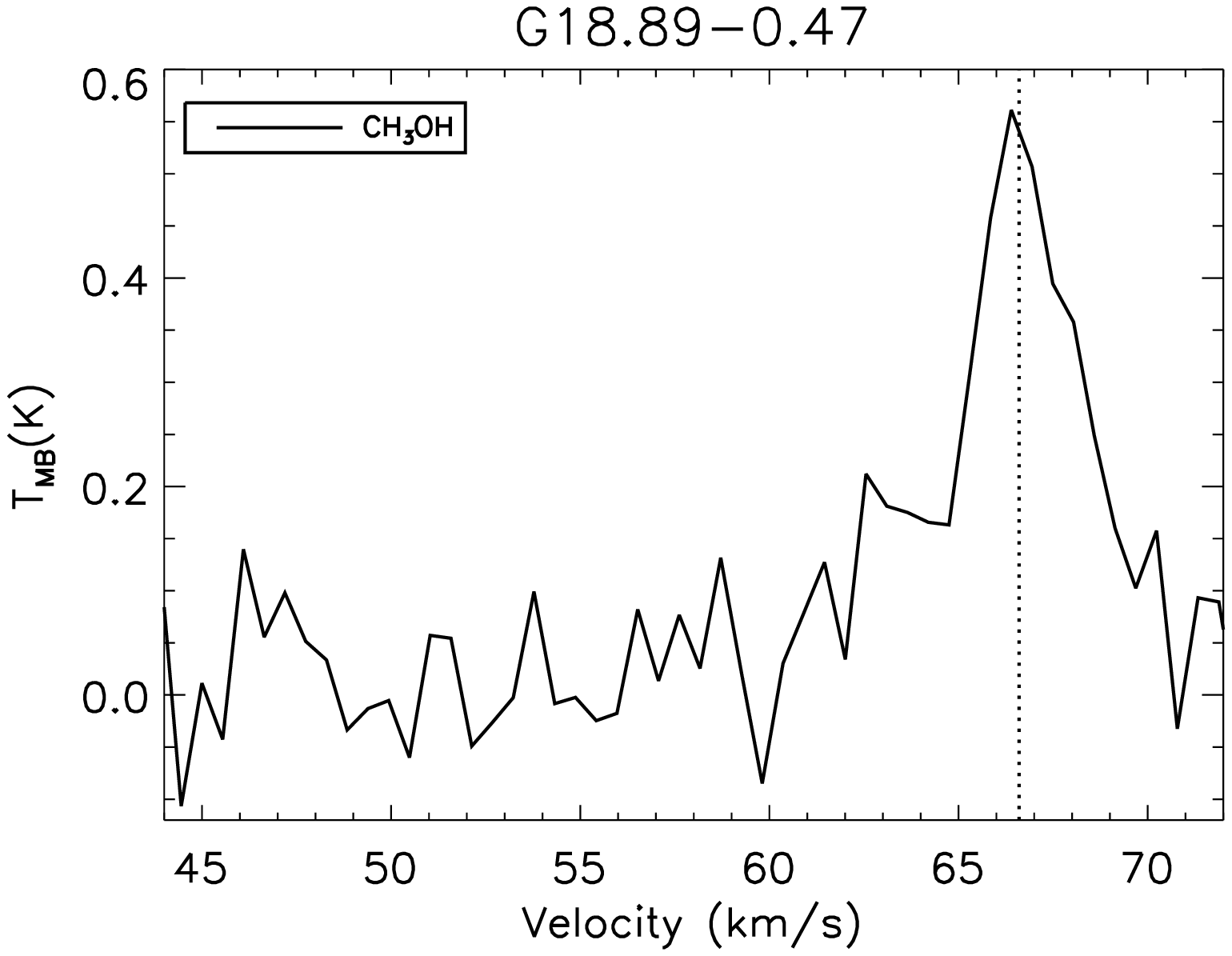}
\addtocounter{figure}{-1}
\caption{Figure Set 3: JCMT thermal CH$_{3}$OH(5$_{2,3}$-4$_{1,3}$,
\elow=44.3 K) spectrum.  
(See \S\ref{mollines} and Table~\ref{jcmtfitstable}.)
The velocity range shown for each EGO is the same as that 
in Figure~\ref{jcmtlineplots}.  The dotted vertical line marks the \hisoco\/
velocity from Table~\ref{jcmtfitstable}.}
\end{figure}

\begin{figure}
\plotone{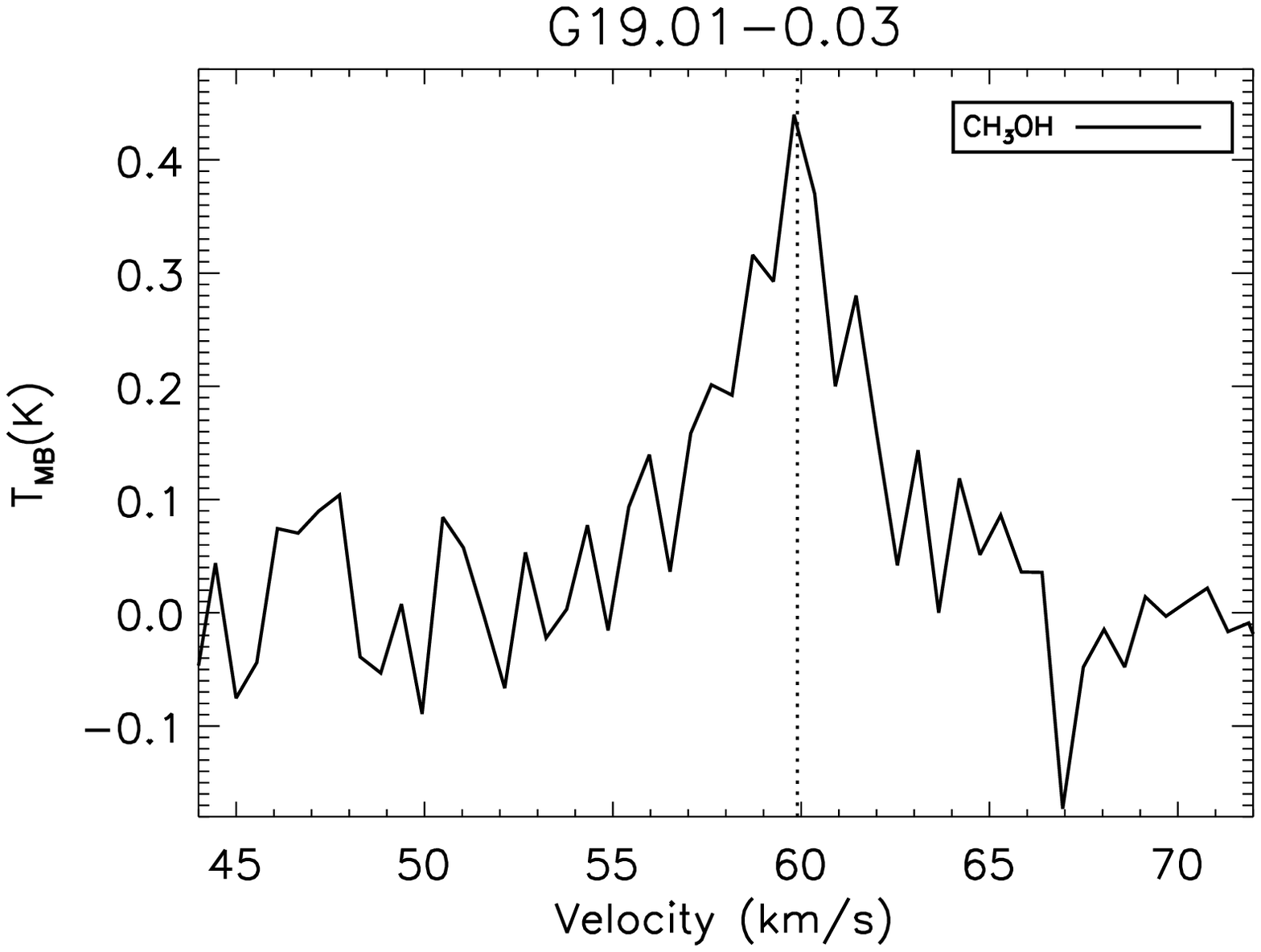}
\addtocounter{figure}{-1}
\caption{Figure Set 3: JCMT thermal CH$_{3}$OH(5$_{2,3}$-4$_{1,3}$,
\elow=44.3 K) spectrum.  
(See \S\ref{mollines} and Table~\ref{jcmtfitstable}.)
The velocity range shown for each EGO is the same as that 
in Figure~\ref{jcmtlineplots}.  The dotted vertical line marks the \hisoco\/
velocity from Table~\ref{jcmtfitstable}.}
\end{figure}

\begin{figure}
\plotone{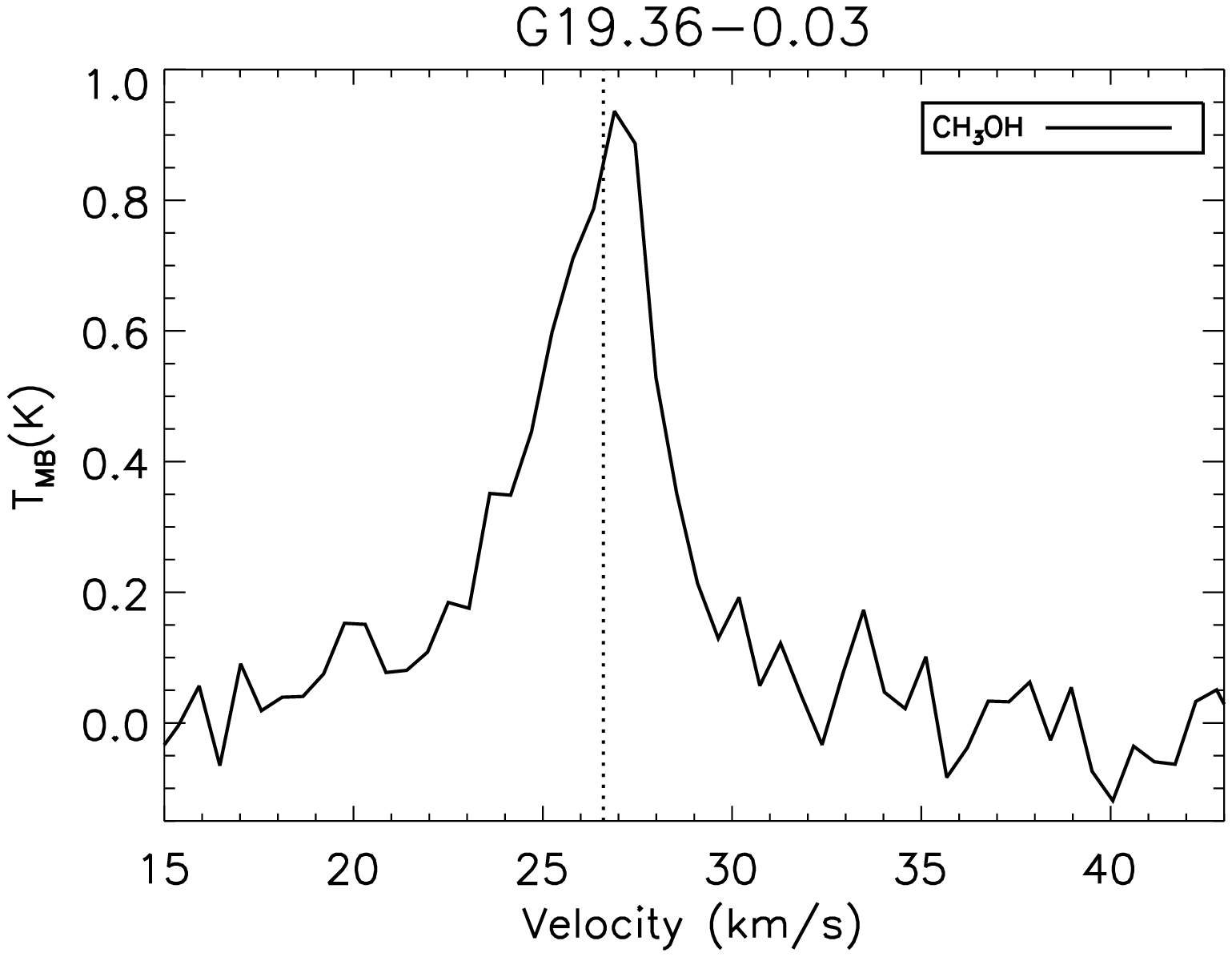}
\addtocounter{figure}{-1}
\caption{Figure Set 3: JCMT thermal CH$_{3}$OH(5$_{2,3}$-4$_{1,3}$,
\elow=44.3 K) spectrum.  
(See \S\ref{mollines} and Table~\ref{jcmtfitstable}.)
The velocity range shown for each EGO is the same as that 
in Figure~\ref{jcmtlineplots}.  The dotted vertical line marks the \hisoco\/
velocity from Table~\ref{jcmtfitstable}.}
\end{figure}

\begin{figure}
\plotone{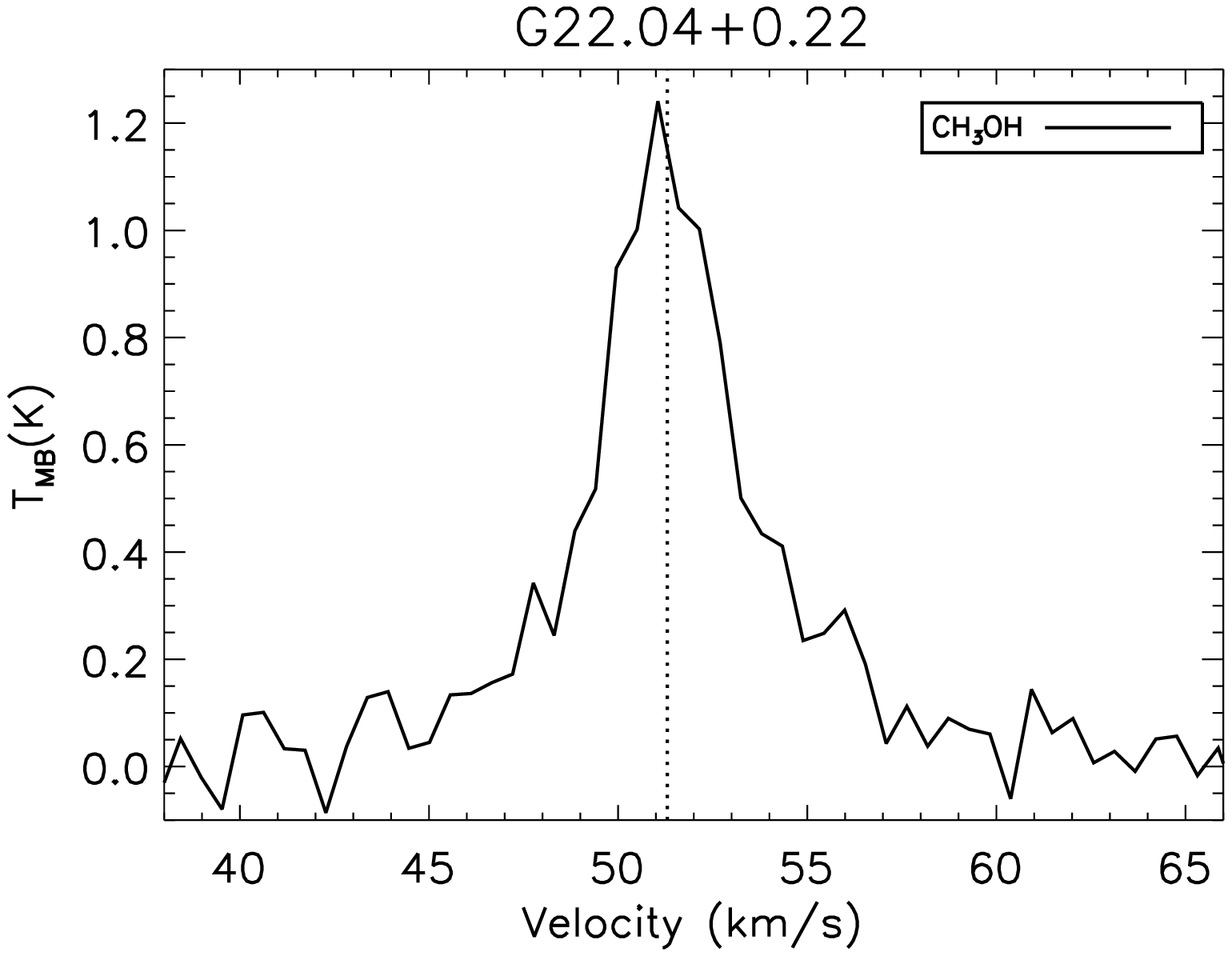}
\addtocounter{figure}{-1}
\caption{Figure Set 3: JCMT thermal CH$_{3}$OH(5$_{2,3}$-4$_{1,3}$,
\elow=44.3 K) spectrum.  
(See \S\ref{mollines} and Table~\ref{jcmtfitstable}.)
The velocity range shown for each EGO is the same as that 
in Figure~\ref{jcmtlineplots}.  The dotted vertical line marks the \hisoco\/
velocity from Table~\ref{jcmtfitstable}.}
\end{figure}

\begin{figure}
\plotone{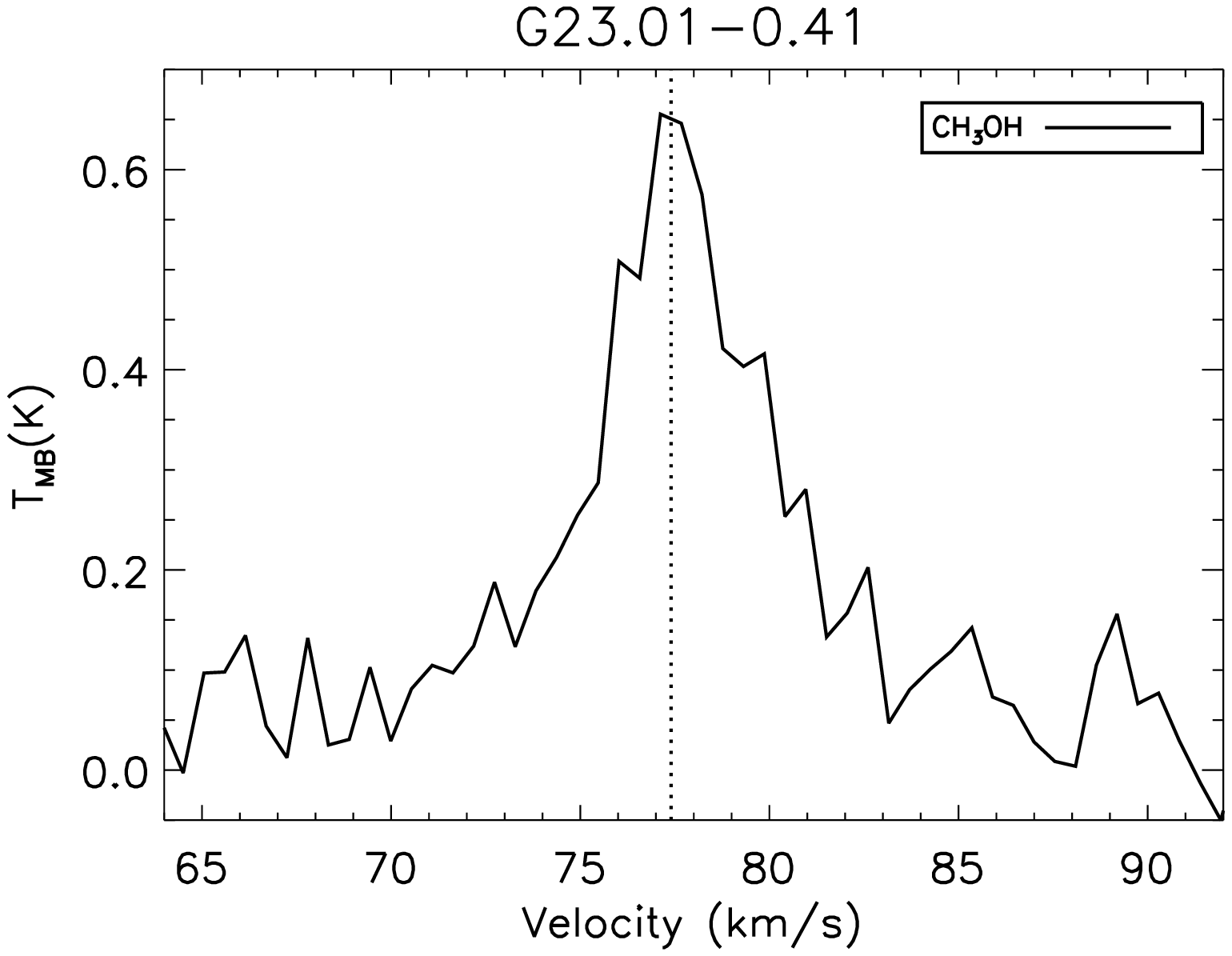}
\addtocounter{figure}{-1}
\caption{Figure Set 3: JCMT thermal CH$_{3}$OH(5$_{2,3}$-4$_{1,3}$,
\elow=44.3 K) spectrum.  
(See \S\ref{mollines} and Table~\ref{jcmtfitstable}.)
The velocity range shown for each EGO is the same as that 
in Figure~\ref{jcmtlineplots}.  The dotted vertical line marks the \hisoco\/
velocity from Table~\ref{jcmtfitstable}.}
\end{figure}

\begin{figure}
\plotone{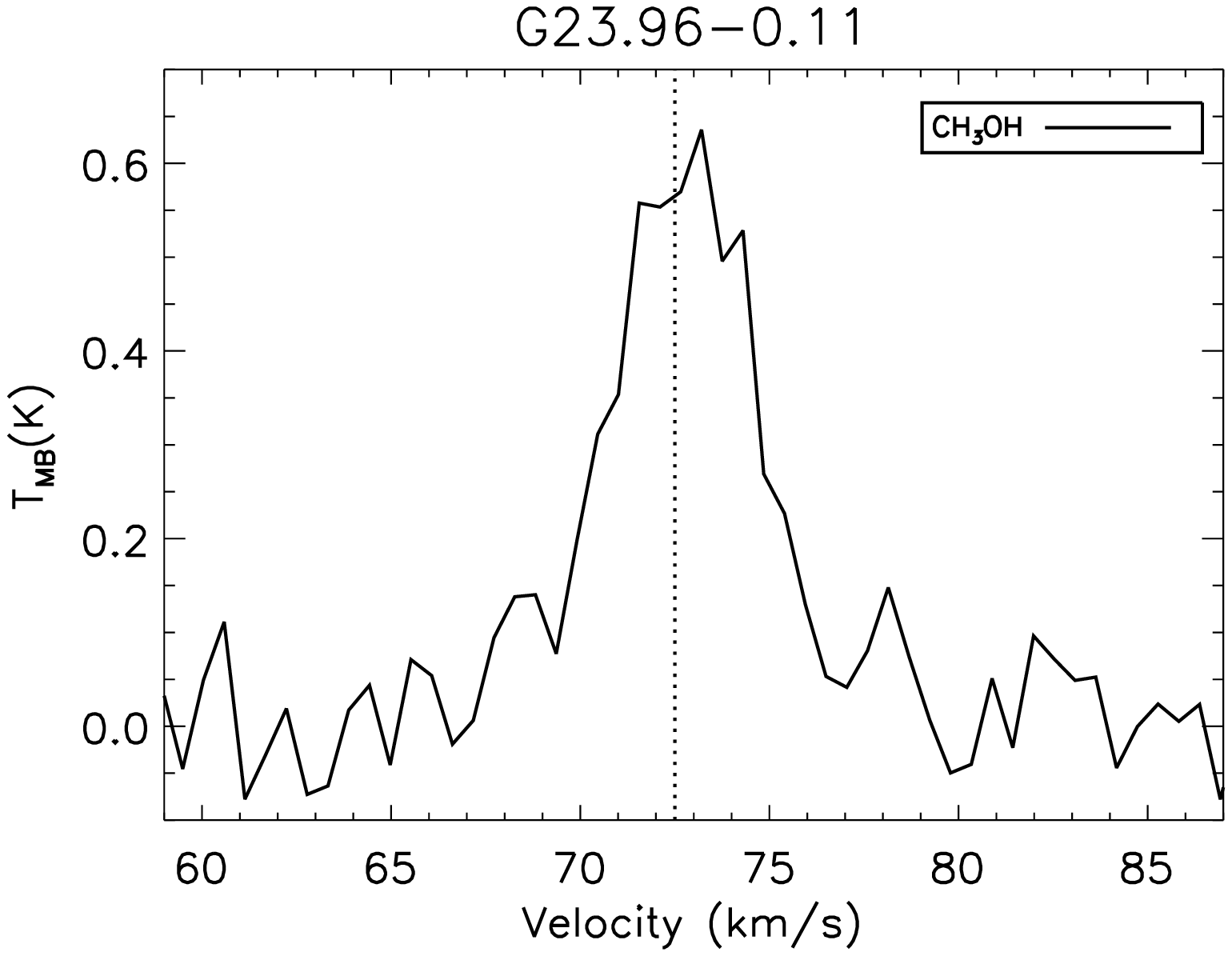}
\addtocounter{figure}{-1}
\caption{Figure Set 3: JCMT thermal CH$_{3}$OH(5$_{2,3}$-4$_{1,3}$,
\elow=44.3 K) spectrum.  
(See \S\ref{mollines} and Table~\ref{jcmtfitstable}.)
The velocity range shown for each EGO is the same as that 
in Figure~\ref{jcmtlineplots}.  The dotted vertical line marks the \hisoco\/
velocity from Table~\ref{jcmtfitstable}.}
\end{figure}

\begin{figure}
\plotone{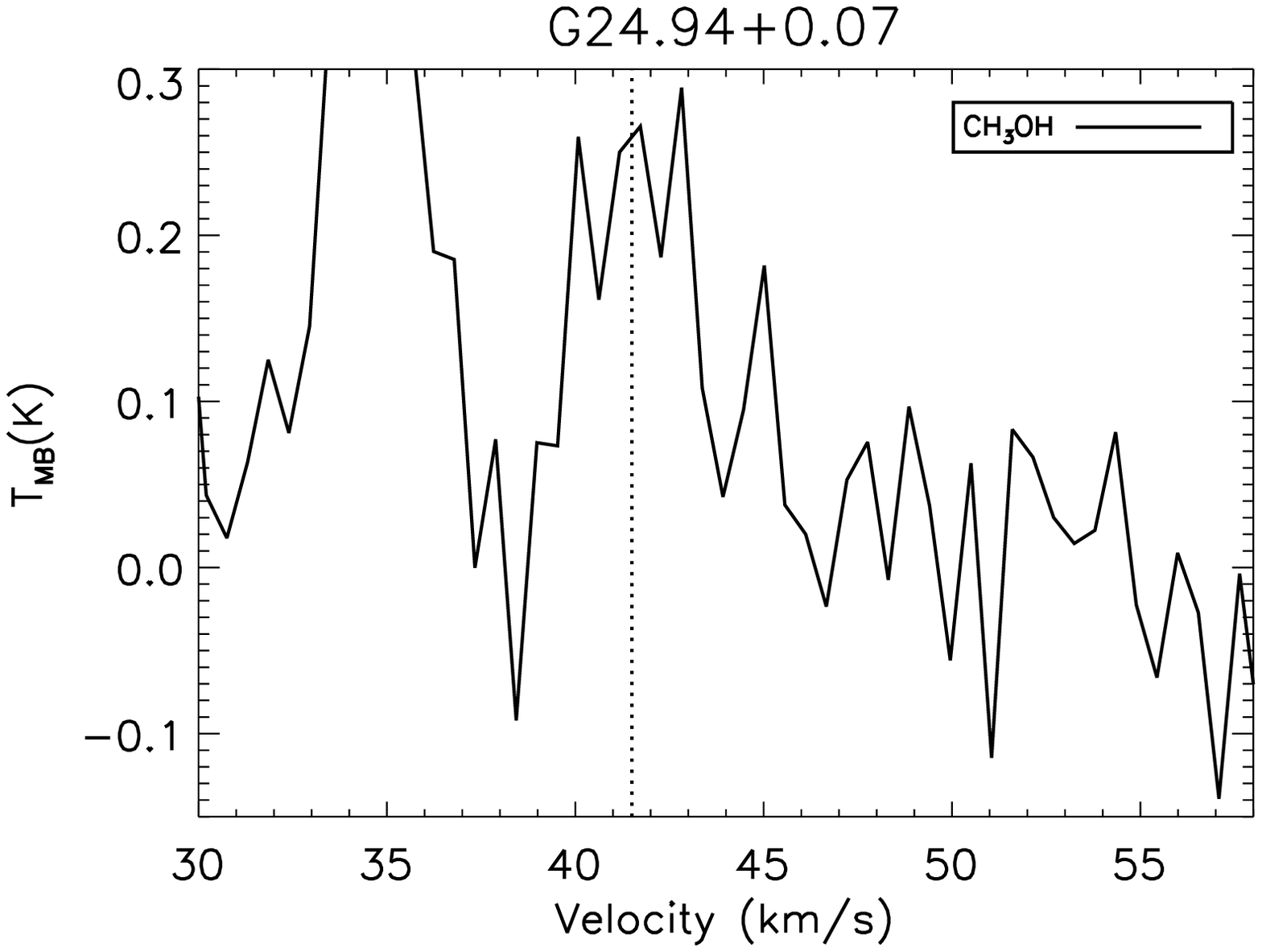}
\addtocounter{figure}{-1}
\caption{Figure Set 3: JCMT thermal CH$_{3}$OH(5$_{2,3}$-4$_{1,3}$,
\elow=44.3 K) spectrum.  
(See \S\ref{mollines} and Table~\ref{jcmtfitstable}.)
The velocity range shown for each EGO is the same as that 
in Figure~\ref{jcmtlineplots}.  The \hisoco\/ line falls within the velocity
range shown for this source; the T$_{MB}$ range has been adjusted to highlight 
the \methanol\/ line (\vlsr\q42\kms).  The dotted vertical line marks the \hisoco\/
velocity from Table~\ref{jcmtfitstable}.}
\end{figure}

\begin{figure}
\plotone{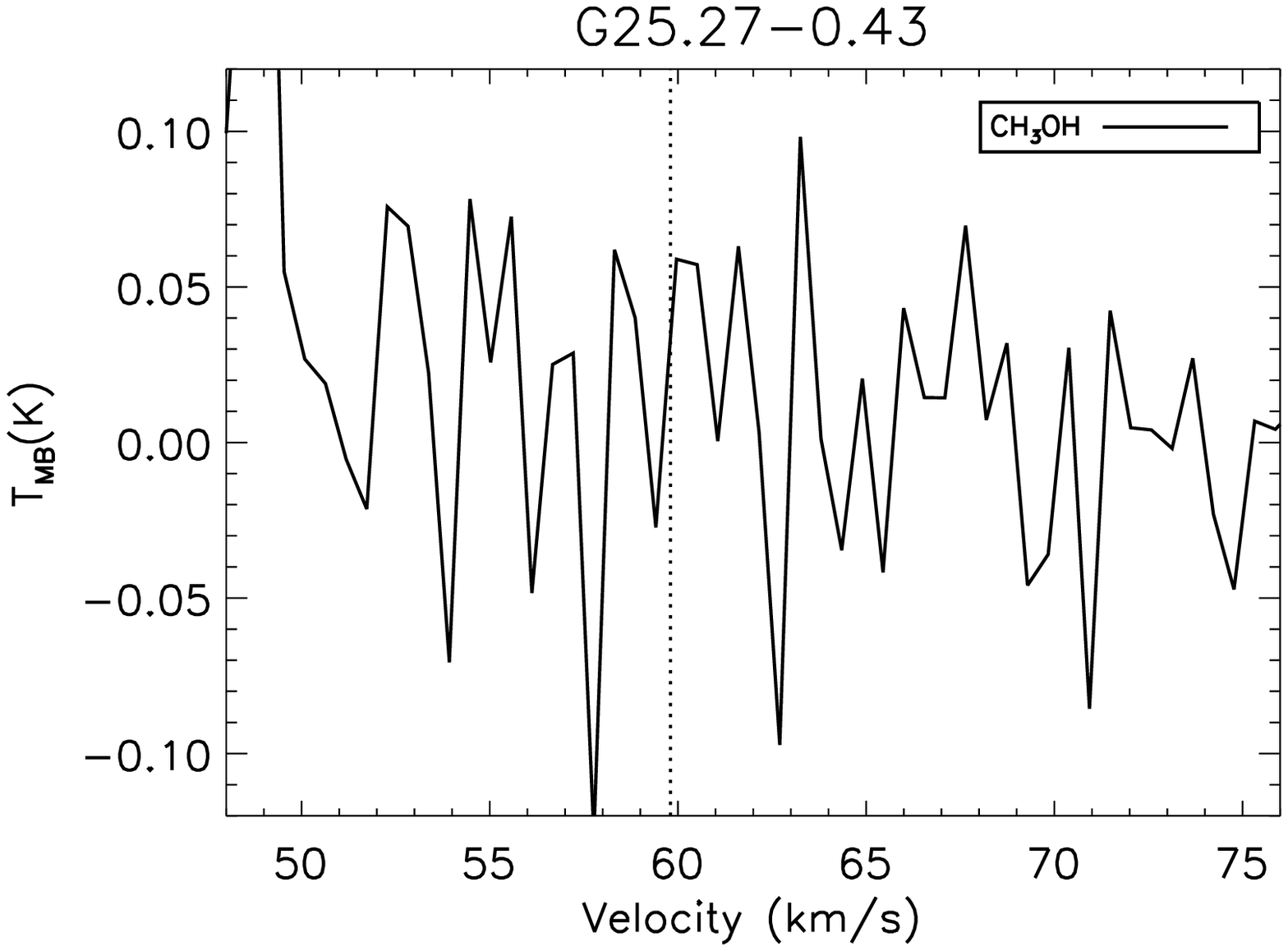}
\addtocounter{figure}{-1}
\caption{Figure Set 3: JCMT thermal CH$_{3}$OH(5$_{2,3}$-4$_{1,3}$,
\elow=44.3 K) spectrum.  
(See \S\ref{mollines} and Table~\ref{jcmtfitstable}.)
The velocity range shown for each EGO is the same as that 
in Figure~\ref{jcmtlineplots}.  The \hisoco\/ line falls within the velocity
range shown for this source;
the T$_{MB}$ range has been adjusted to highlight 
the \methanol\/ nondetection (\vlsr\q60\kms).  The dotted vertical line marks the \hisoco\/
velocity from Table~\ref{jcmtfitstable}.}
\end{figure}

\begin{figure}
\plotone{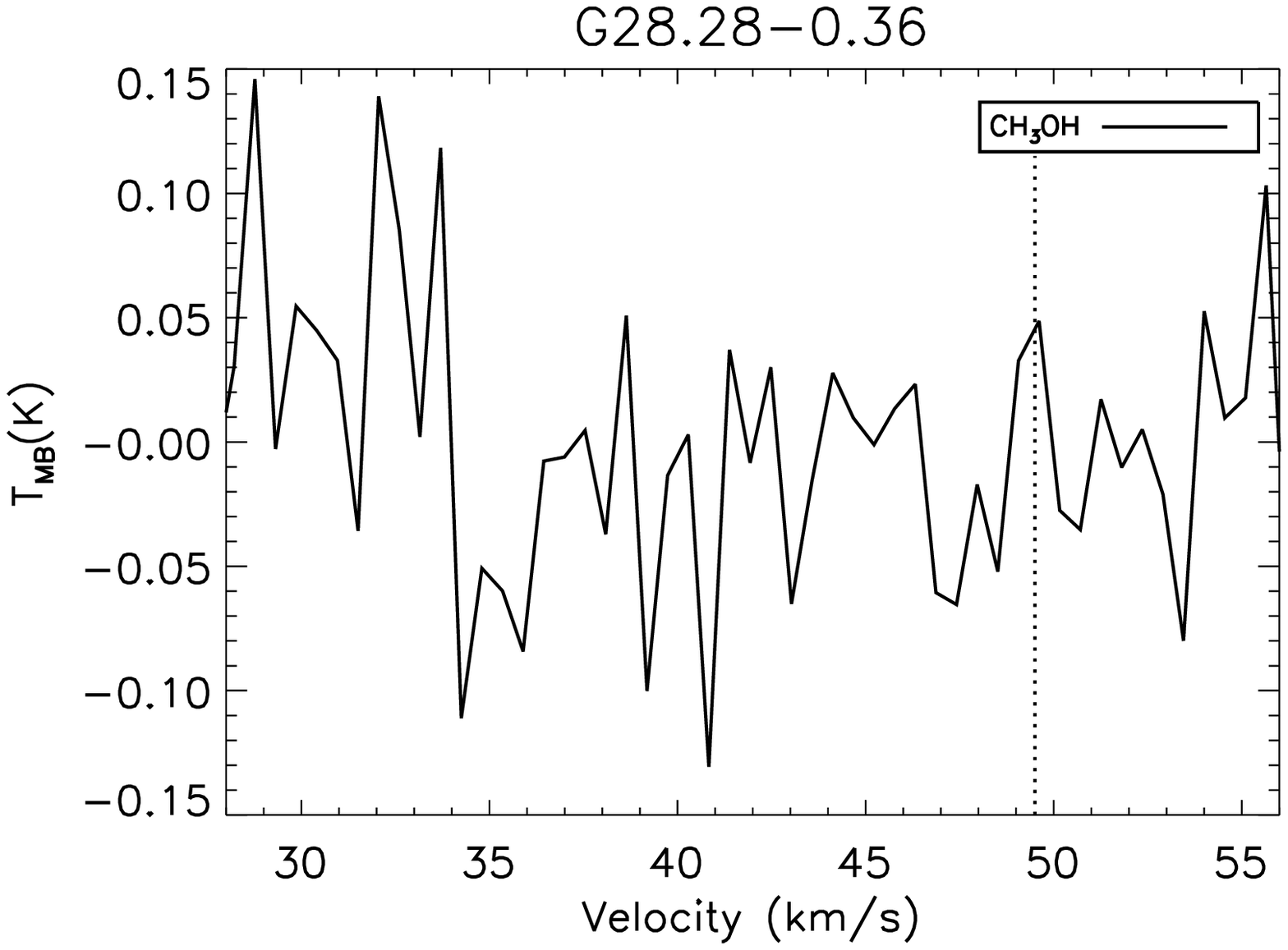}
\addtocounter{figure}{-1}
\caption{Figure Set 3: JCMT thermal CH$_{3}$OH(5$_{2,3}$-4$_{1,3}$,
\elow=44.3 K) spectrum.  
(See \S\ref{mollines} and Table~\ref{jcmtfitstable}.)
The velocity range shown for each EGO is the same as that 
in Figure~\ref{jcmtlineplots}.  The dotted vertical line marks the \hisoco\/
velocity from Table~\ref{jcmtfitstable}.}
\end{figure}

\begin{figure}
\plotone{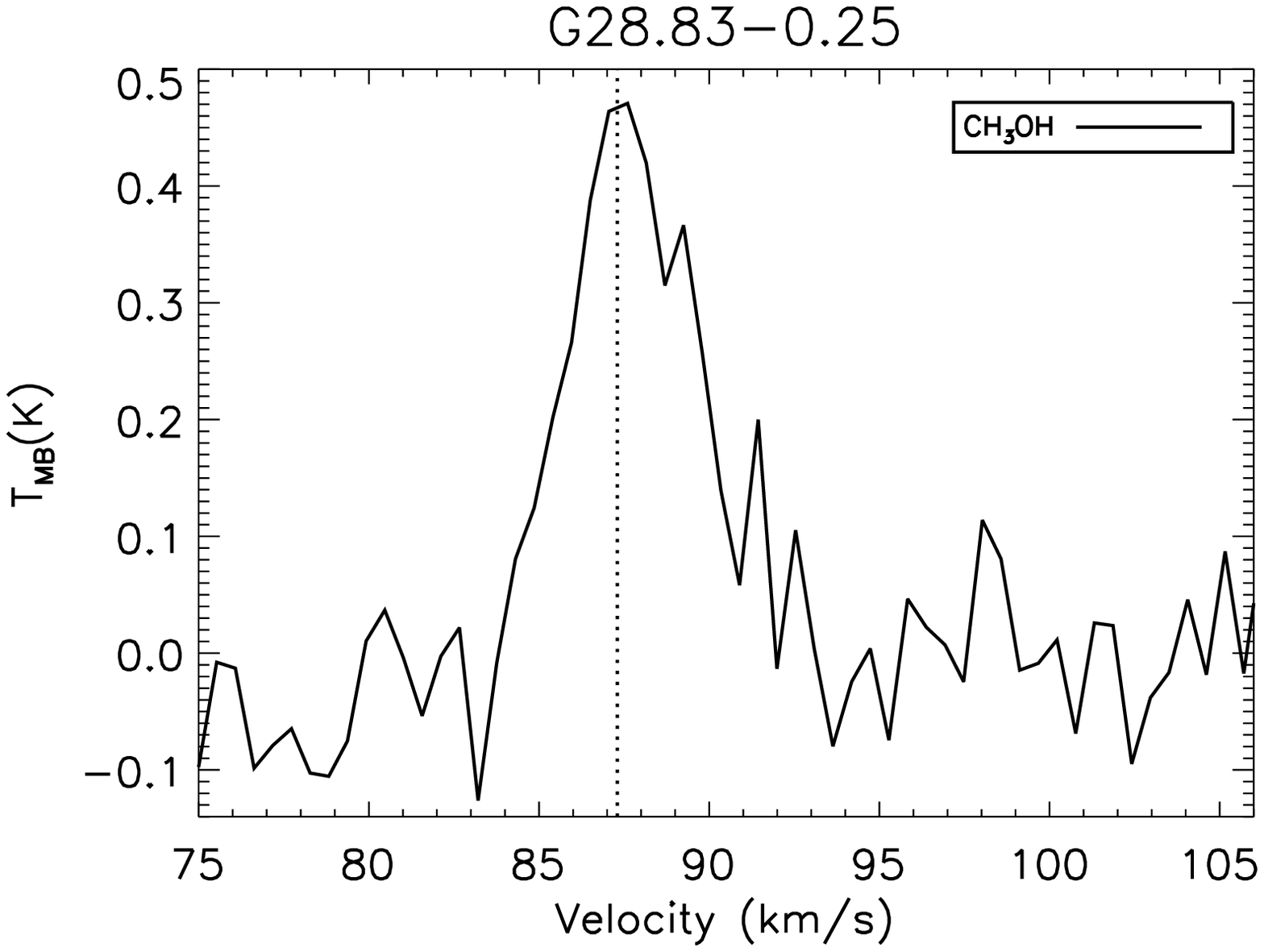}
\addtocounter{figure}{-1}
\caption{Figure Set 3: JCMT thermal CH$_{3}$OH(5$_{2,3}$-4$_{1,3}$,
\elow=44.3 K) spectrum.  
(See \S\ref{mollines} and Table~\ref{jcmtfitstable}.)
The velocity range shown for each EGO is the same as that 
in Figure~\ref{jcmtlineplots}.  The dotted vertical line marks the \hisoco\/
velocity from Table~\ref{jcmtfitstable}.}
\end{figure}

\begin{figure}
\plotone{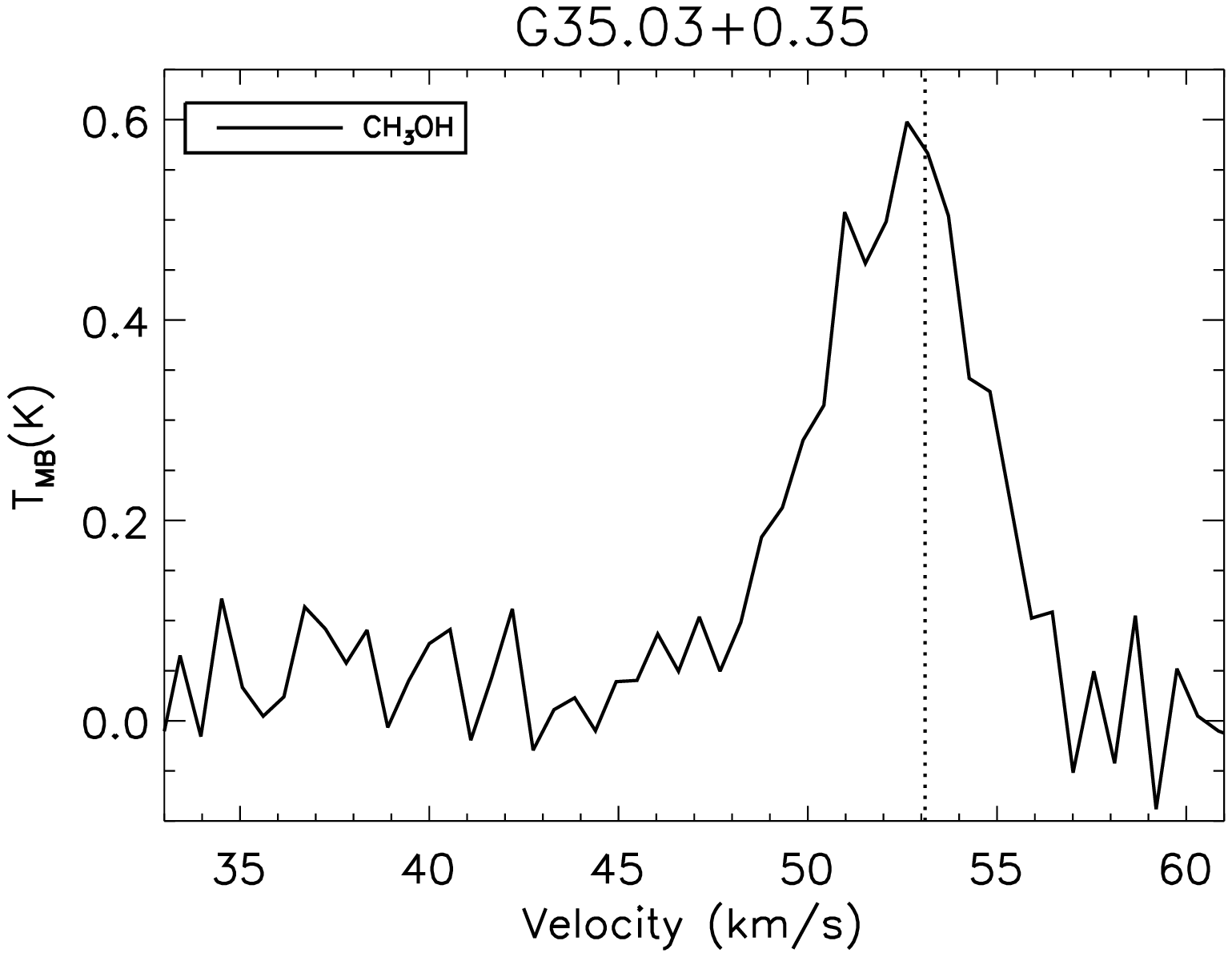}
\addtocounter{figure}{-1}
\caption{Figure Set 3: JCMT thermal CH$_{3}$OH(5$_{2,3}$-4$_{1,3}$,
\elow=44.3 K) spectrum.  
(See \S\ref{mollines} and Table~\ref{jcmtfitstable}.)
The velocity range shown for each EGO is the same as that 
in Figure~\ref{jcmtlineplots}.  The dotted vertical line marks the \hisoco\/
velocity from Table~\ref{jcmtfitstable}.}
\end{figure}

\begin{figure}
\plotone{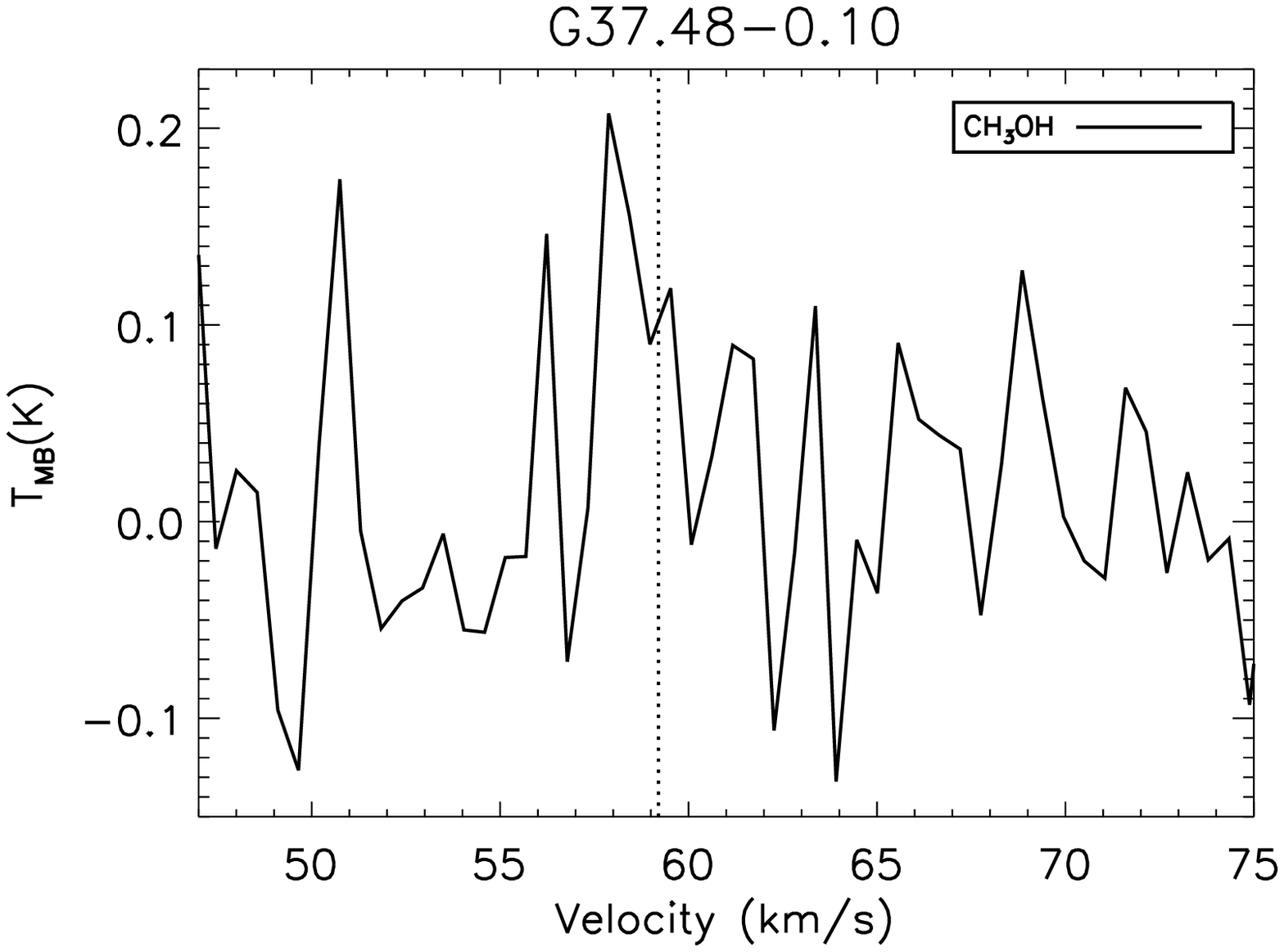}
\addtocounter{figure}{-1}
\caption{Figure Set 3: JCMT thermal CH$_{3}$OH(5$_{2,3}$-4$_{1,3}$,
\elow=44.3 K) spectrum.  
(See \S\ref{mollines} and Table~\ref{jcmtfitstable}.)
The velocity range shown for each EGO is the same as that 
in Figure~\ref{jcmtlineplots}.  The dotted vertical line marks the \hisoco\/
velocity from Table~\ref{jcmtfitstable}.}
\end{figure}

\begin{figure}
\plotone{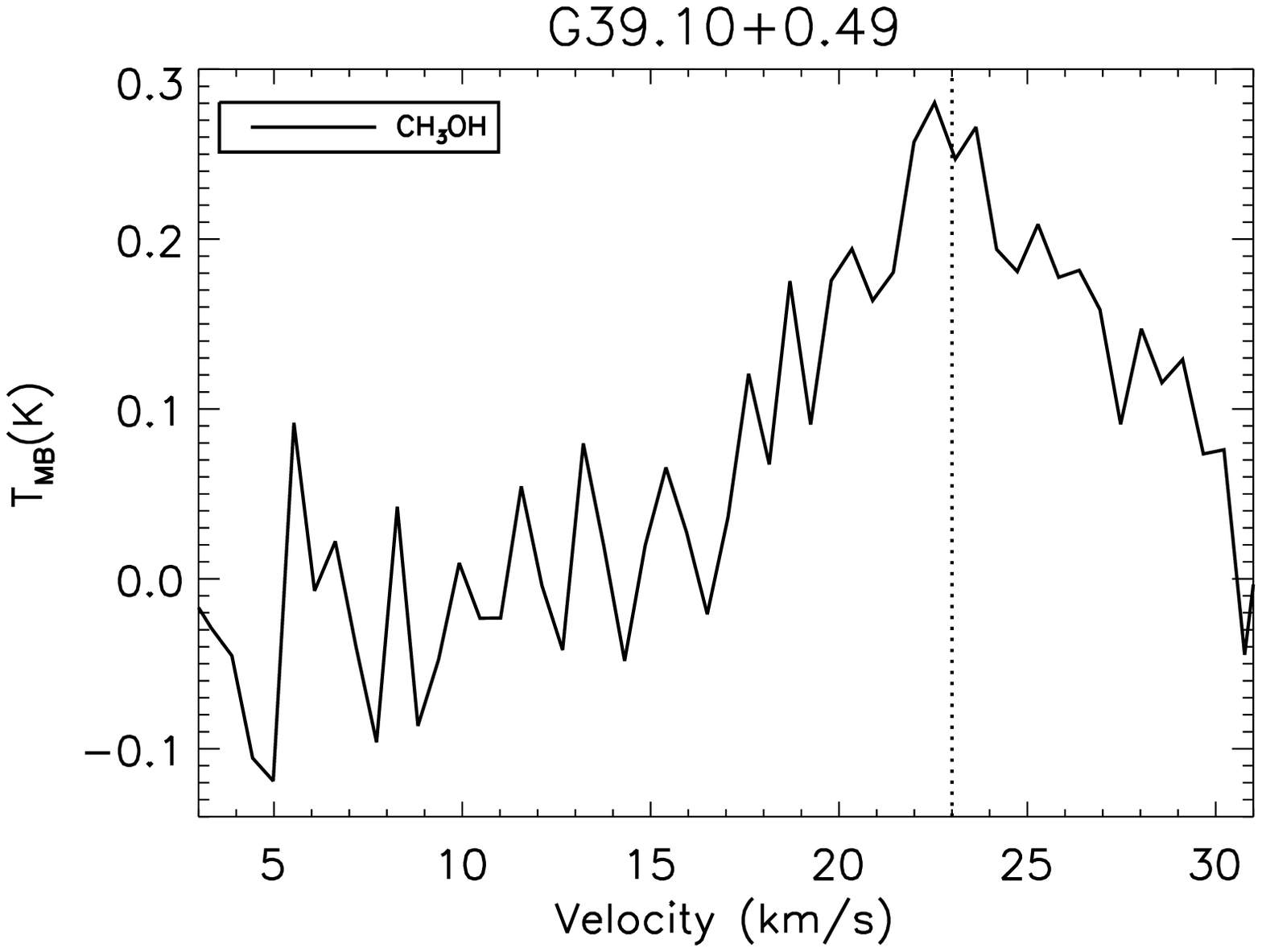}
\addtocounter{figure}{-1}
\caption{Figure Set 3: JCMT thermal CH$_{3}$OH(5$_{2,3}$-4$_{1,3}$,
\elow=44.3 K) spectrum.  
(See \S\ref{mollines} and Table~\ref{jcmtfitstable}.)
The velocity range shown for each EGO is the same as that 
in Figure~\ref{jcmtlineplots}.  The dotted vertical line marks the \hisoco\/
velocity from Table~\ref{jcmtfitstable}.}
\end{figure}

\begin{figure}
\plotone{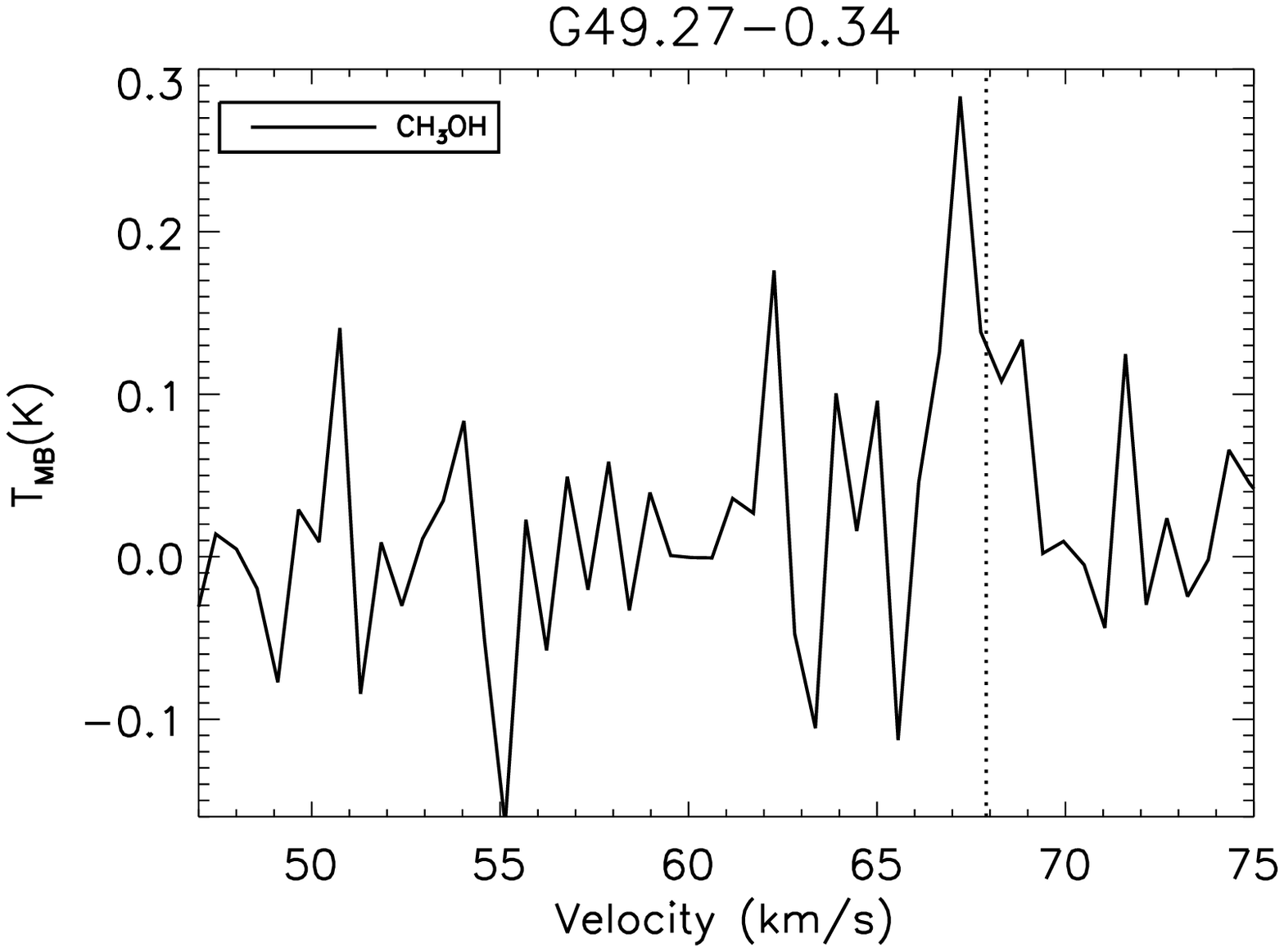}
\addtocounter{figure}{-1}
\caption{Figure Set 3: JCMT thermal CH$_{3}$OH(5$_{2,3}$-4$_{1,3}$,
\elow=44.3 K) spectrum.  
(See \S\ref{mollines} and Table~\ref{jcmtfitstable}.)
The velocity range shown for each EGO is the same as that 
in Figure~\ref{jcmtlineplots}. The dotted vertical line marks the \hisoco\/
velocity from Table~\ref{jcmtfitstable}. }
\end{figure}

\begin{figure}
\plotone{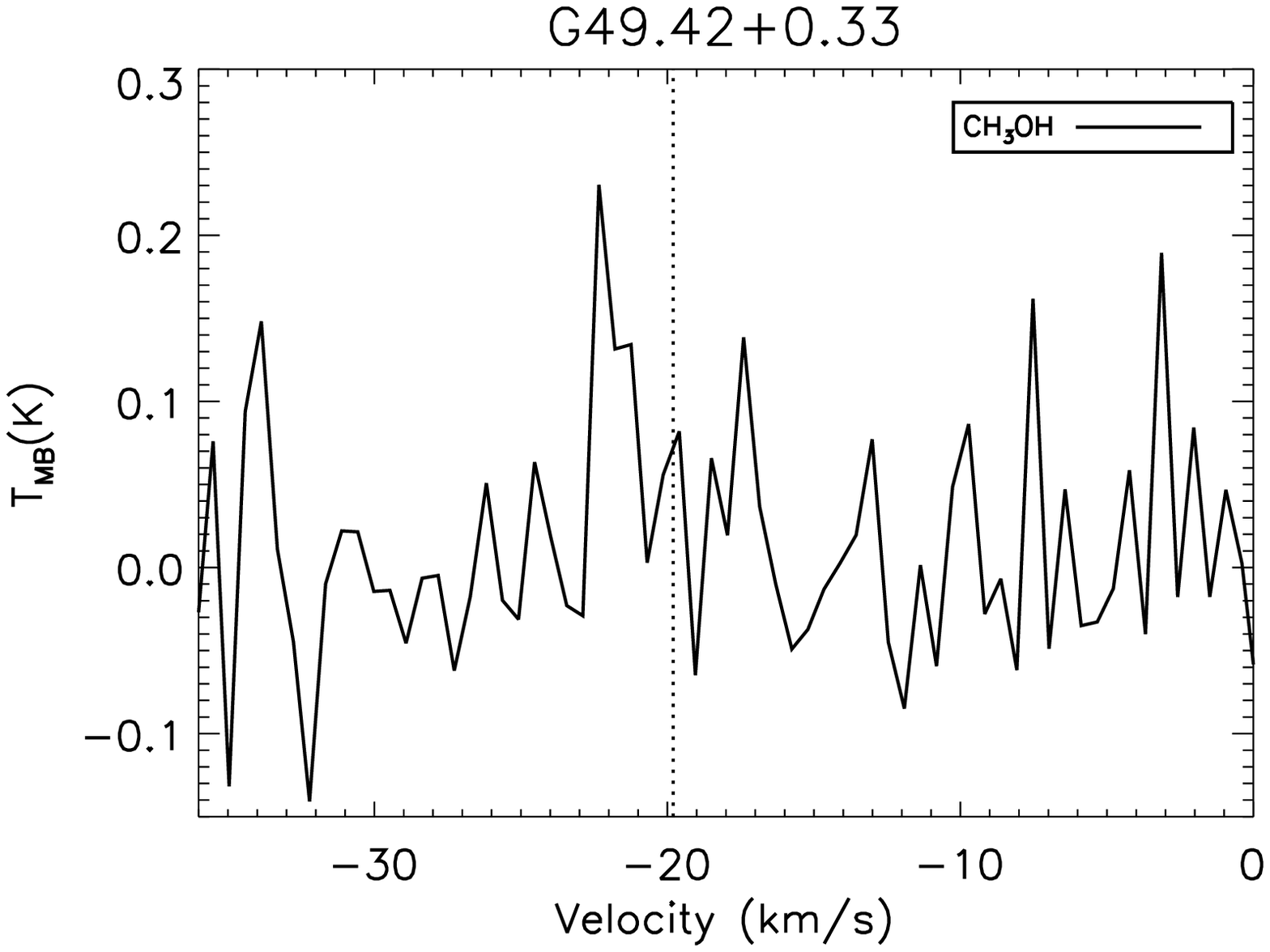}
\addtocounter{figure}{-1}
\caption{Figure Set 3: JCMT thermal CH$_{3}$OH(5$_{2,3}$-4$_{1,3}$,
\elow=44.3 K) spectrum.  
(See \S\ref{mollines} and Table~\ref{jcmtfitstable}.)
The velocity range shown for each EGO is the same as that 
in Figure~\ref{jcmtlineplots}.  The dotted vertical line marks the \hisoco\/
velocity from Table~\ref{jcmtfitstable}.}
\end{figure}

\clearpage

\begin{figure}
\plotone{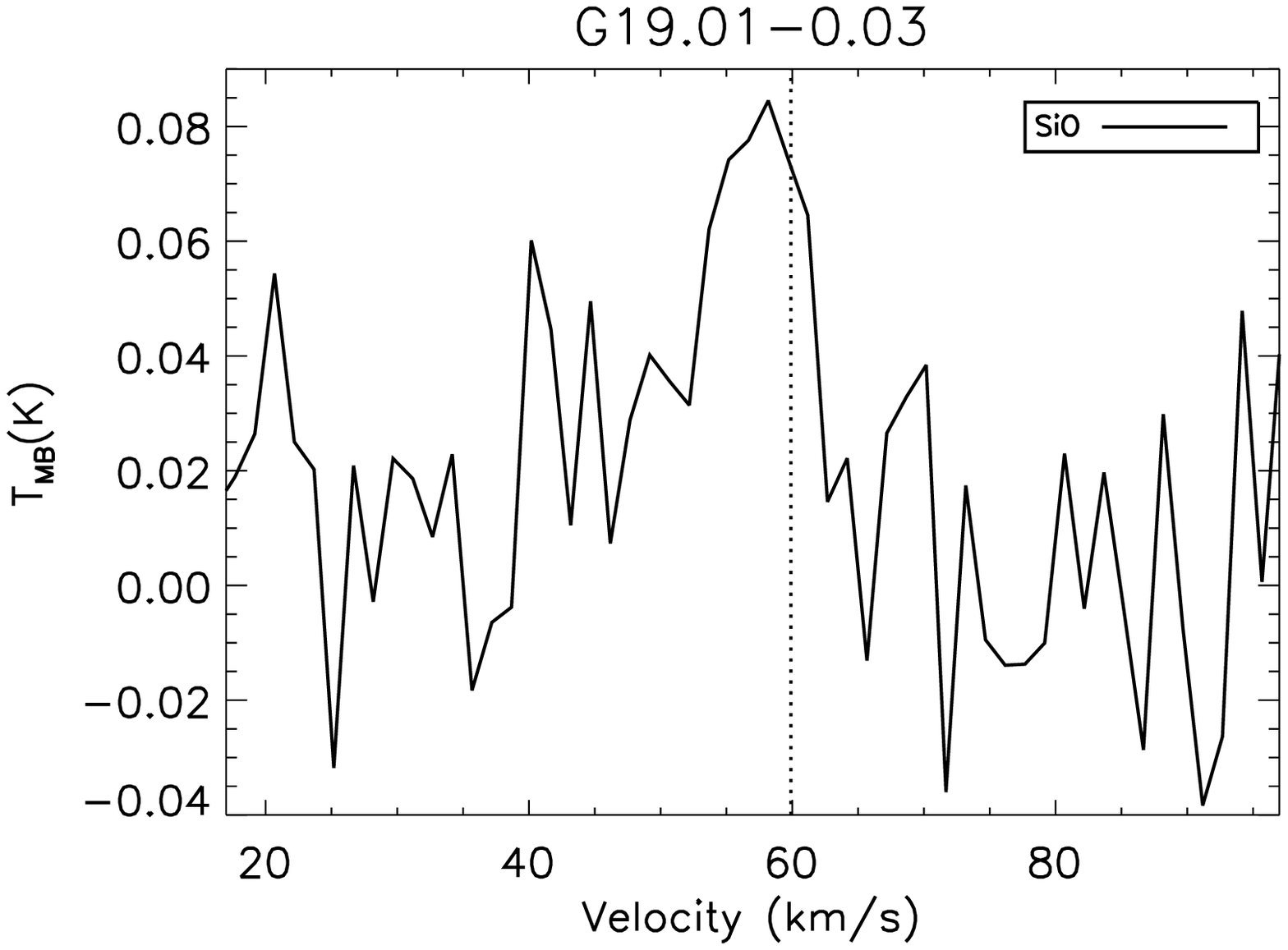}
\caption{Figure Set 4: JCMT thermal SiO (5-4, \elow=20.8 K) spectrum.  
(See \S\ref{mollines} and Table~\ref{jcmtfitstable}.) 
A velocity range of \q80\kms\/ is shown, centered on
the velocity of the SiO line.  The dotted vertical line marks the \hisoco\/
velocity from Table~\ref{jcmtfitstable}.   
}
\end{figure}

\begin{figure}
\plotone{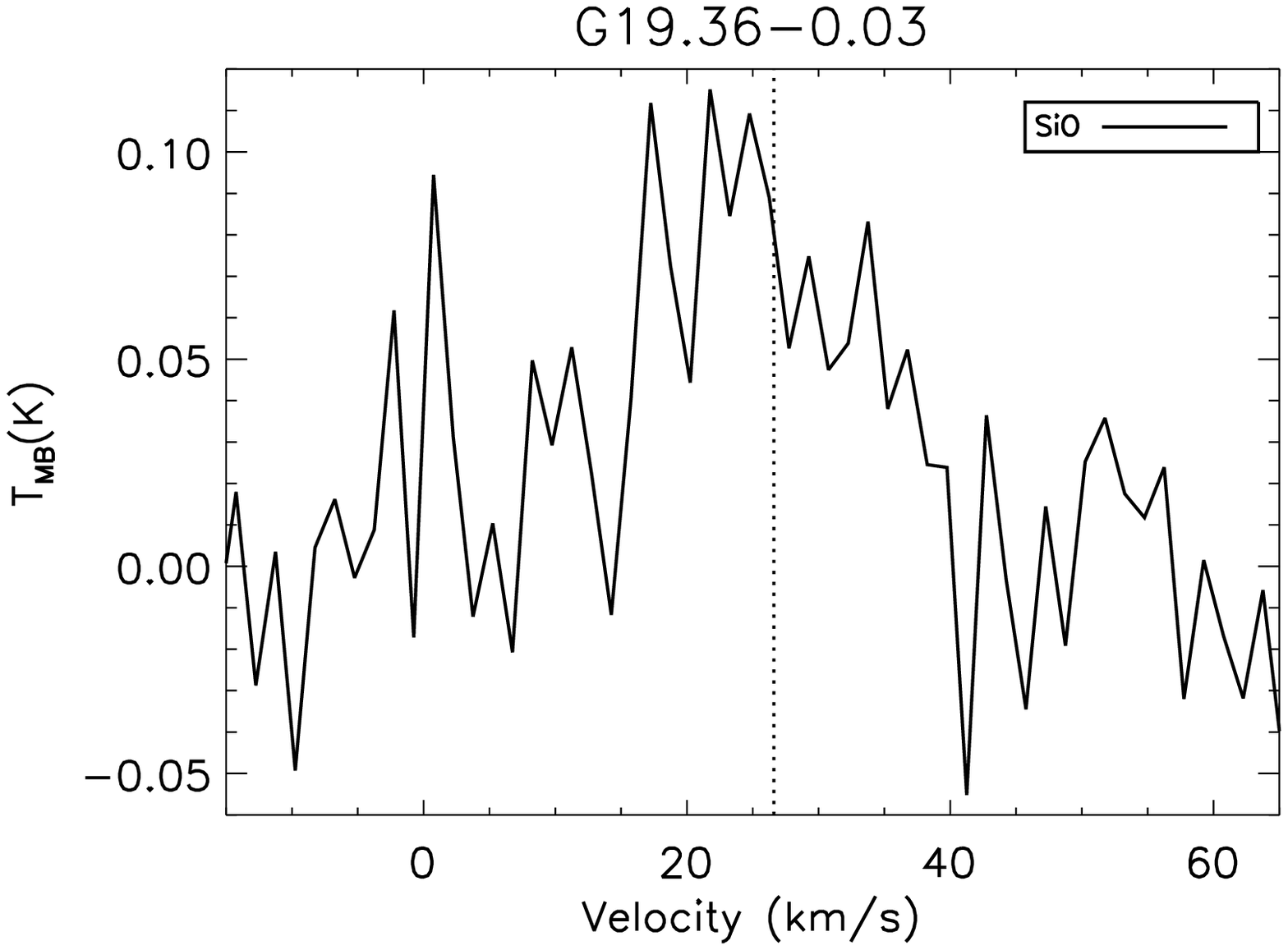}
\addtocounter{figure}{-1}
\caption{Figure Set 4: JCMT thermal SiO (5-4, \elow=20.8 K) spectrum.  
(See \S\ref{mollines} and Table~\ref{jcmtfitstable}.) 
A velocity range of \q80\kms\/ is shown, centered on
the velocity of the SiO line.  The dotted vertical line marks the \hisoco\/
velocity from Table~\ref{jcmtfitstable}.   
}
\end{figure}

\begin{figure}
\plotone{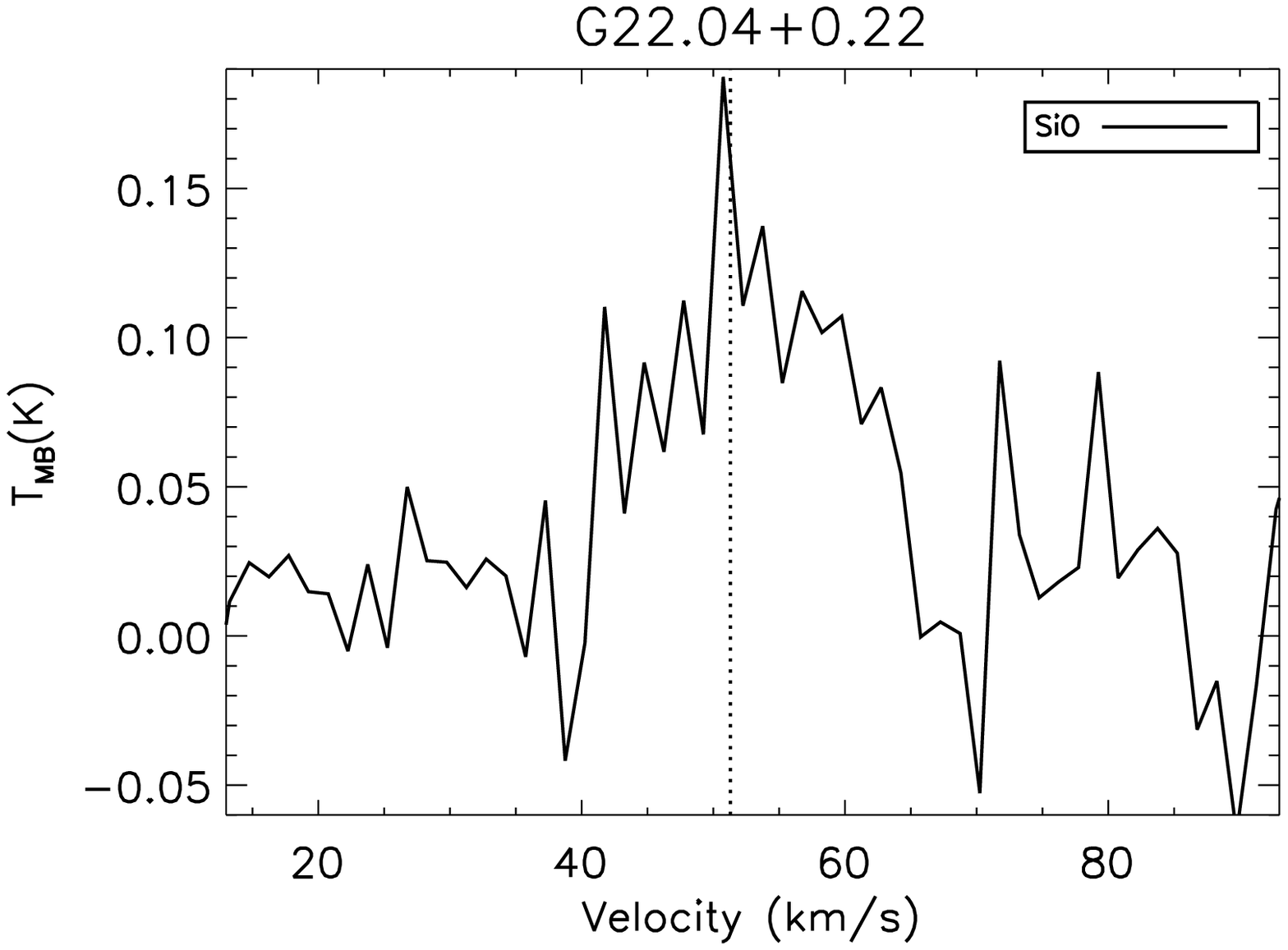}
\addtocounter{figure}{-1}
\caption{Figure Set 4: JCMT thermal SiO (5-4, \elow=20.8 K) spectrum.  
(See \S\ref{mollines} and Table~\ref{jcmtfitstable}.) 
A velocity range of \q80\kms\/ is shown, centered on
the velocity of the SiO line.  The dotted vertical line marks the \hisoco\/
velocity from Table~\ref{jcmtfitstable}.   
}
\end{figure}

\begin{figure}
\plotone{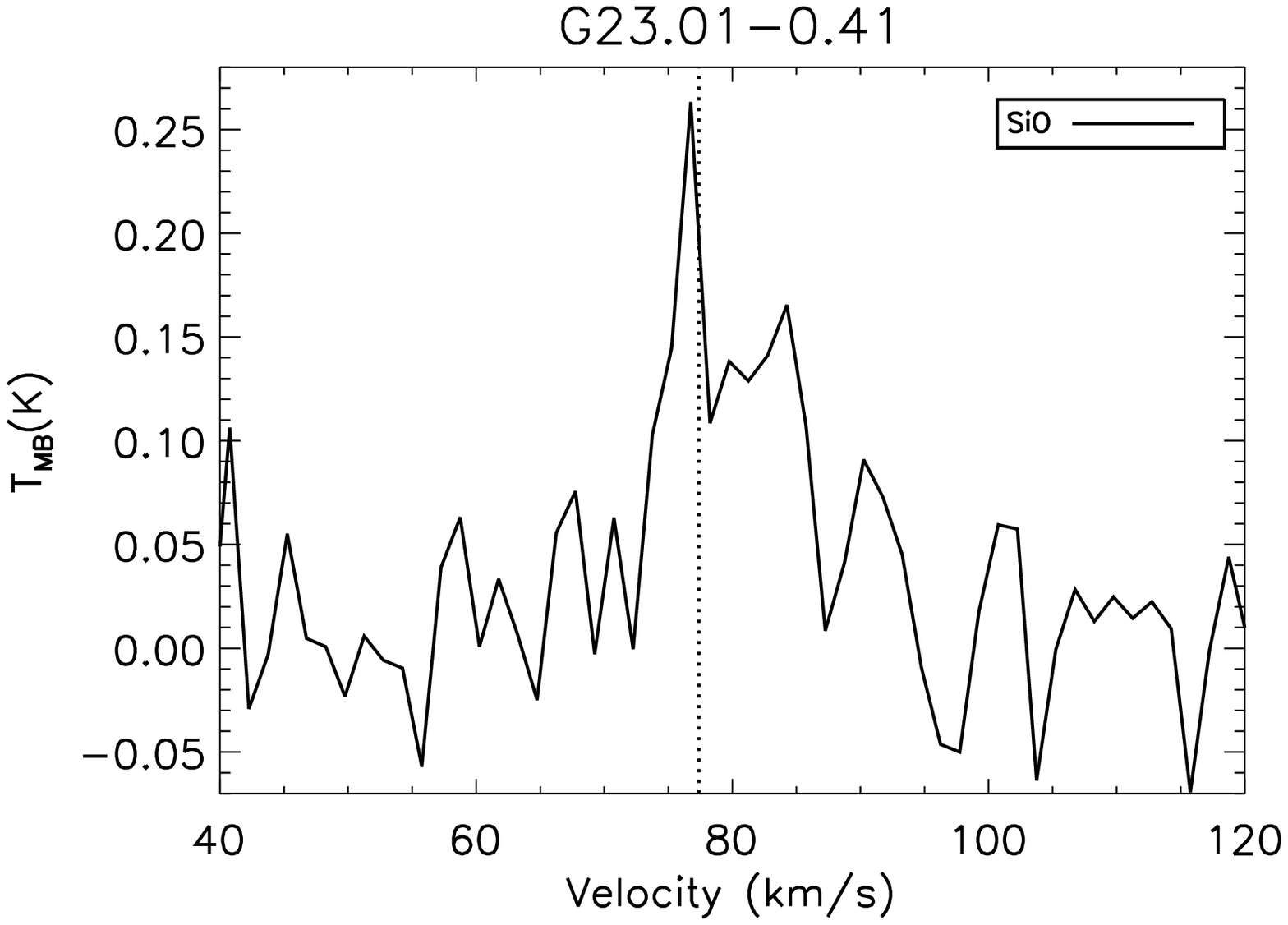}
\addtocounter{figure}{-1}
\caption{Figure Set 4: JCMT thermal SiO (5-4, \elow=20.8 K) spectrum.  
(See \S\ref{mollines} and Table~\ref{jcmtfitstable}.) 
A velocity range of \q80\kms\/ is shown, centered on
the velocity of the SiO line.  The dotted vertical line marks the \hisoco\/
velocity from Table~\ref{jcmtfitstable}.   
}
\end{figure}

\begin{figure}
\plotone{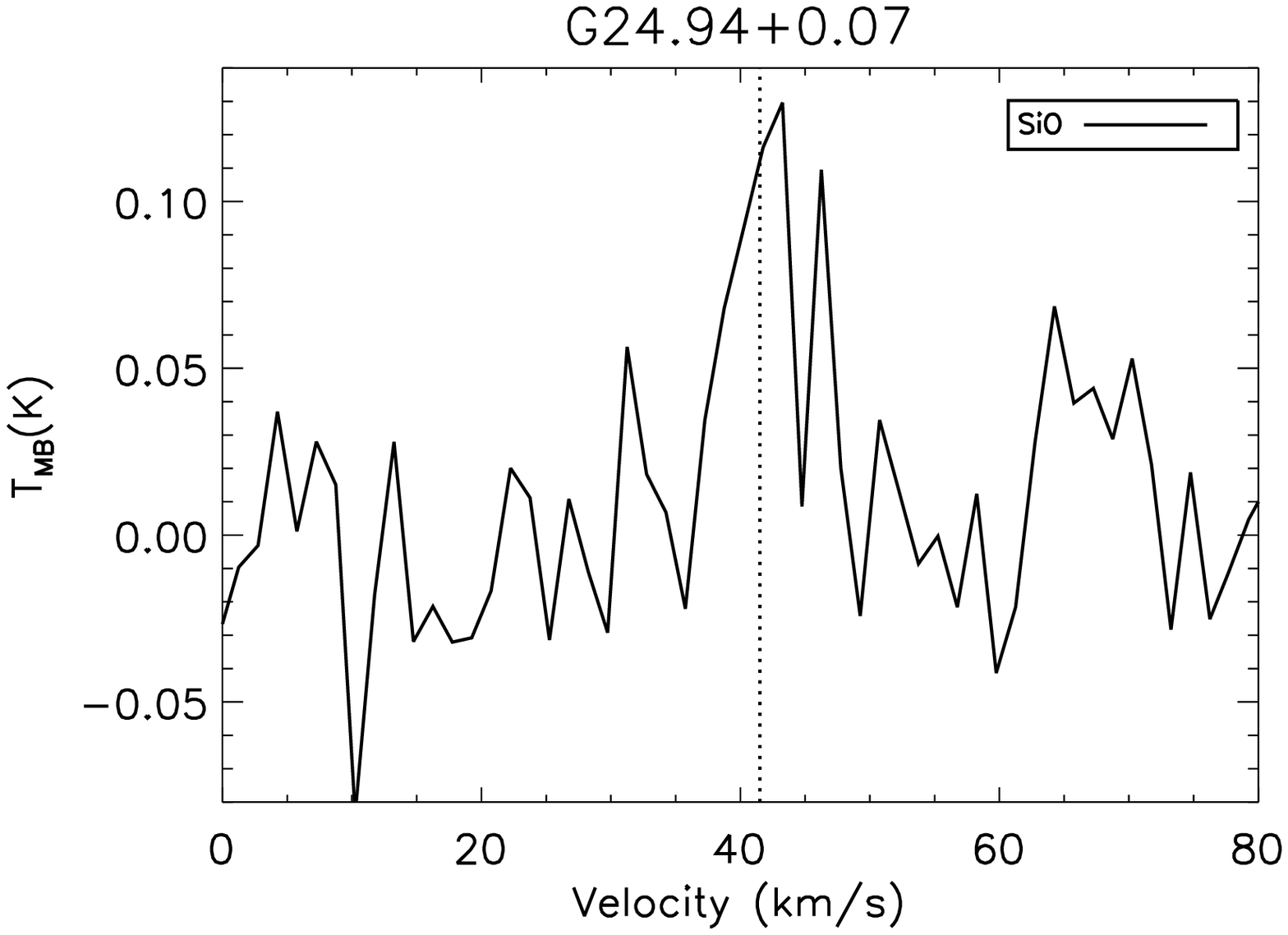}
\addtocounter{figure}{-1}
\caption{Figure Set 4: JCMT thermal SiO (5-4, \elow=20.8 K) spectrum.  
(See \S\ref{mollines} and Table~\ref{jcmtfitstable}.) 
A velocity range of \q80\kms\/ is shown, centered on
the velocity of the SiO line.  The dotted vertical line marks the \hisoco\/
velocity from Table~\ref{jcmtfitstable}.   
}
\end{figure}

\begin{figure}
\plotone{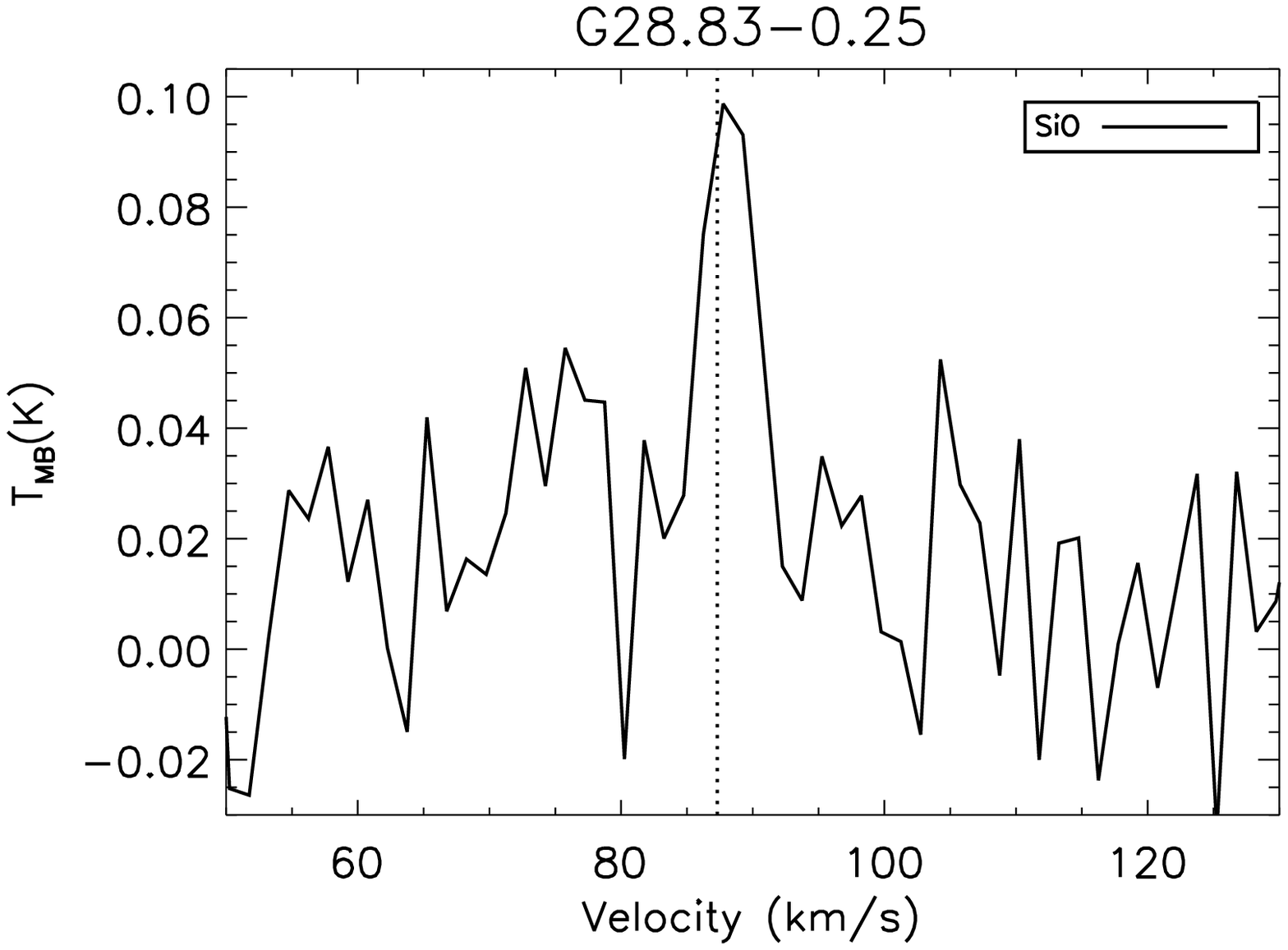}
\addtocounter{figure}{-1}
\caption{Figure Set 4: JCMT thermal SiO (5-4, \elow=20.8 K) spectrum.  
(See \S\ref{mollines} and Table~\ref{jcmtfitstable}.) 
A velocity range of \q80\kms\/ is shown, centered on
the velocity of the SiO line.  The dotted vertical line marks the \hisoco\/
velocity from Table~\ref{jcmtfitstable}.   
}
\end{figure}

\begin{figure}
\plotone{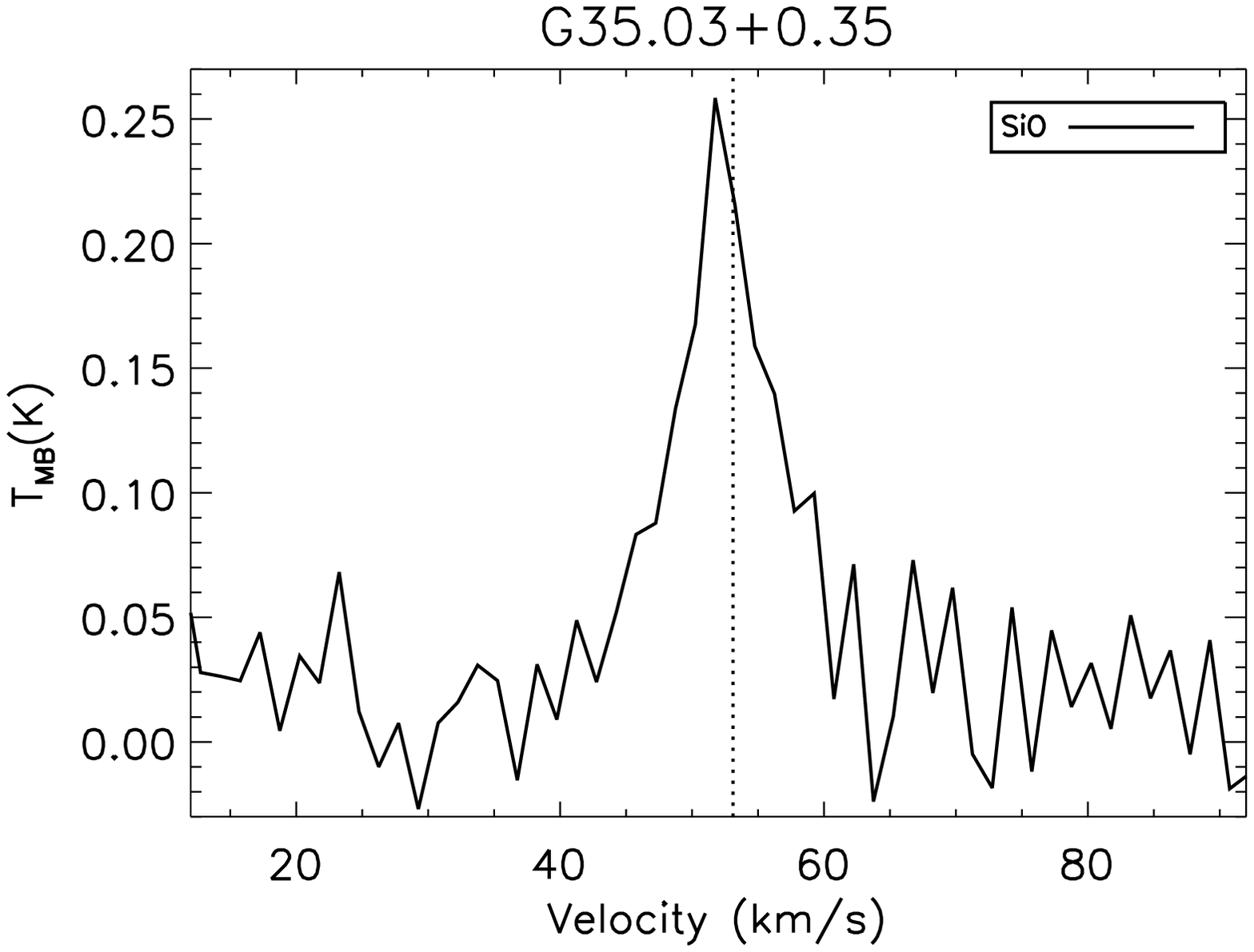}
\addtocounter{figure}{-1}
\caption{Figure Set 4: JCMT thermal SiO (5-4, \elow=20.8 K) spectrum.  
(See \S\ref{mollines} and Table~\ref{jcmtfitstable}.) 
A velocity range of \q80\kms\/ is shown, centered on
the velocity of the SiO line.  The dotted vertical line marks the \hisoco\/
velocity from Table~\ref{jcmtfitstable}.   
}
\end{figure}

\begin{figure}
\plotone{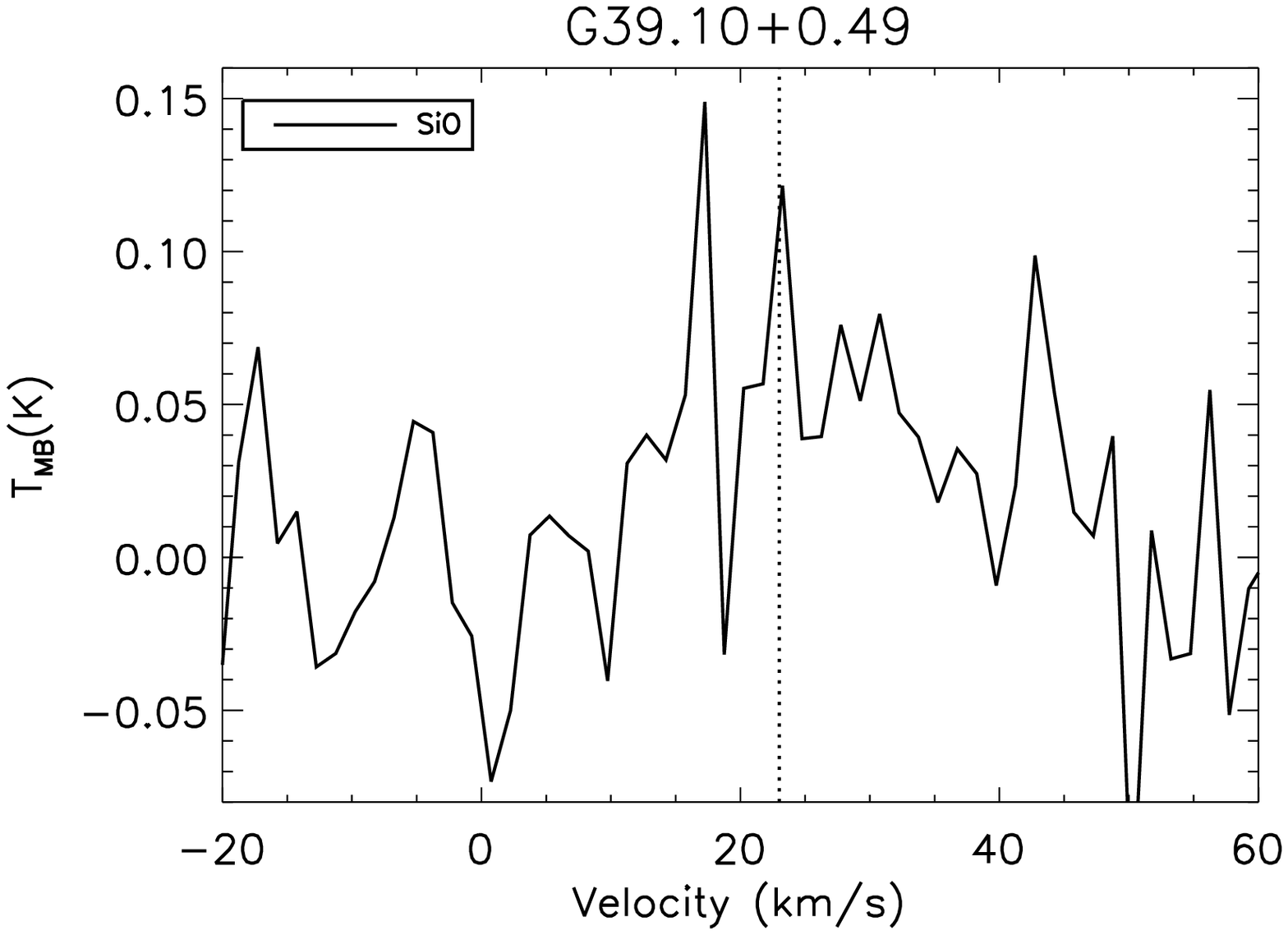}
\addtocounter{figure}{-1}
\caption{Figure Set 4: JCMT thermal SiO (5-4, \elow=20.8 K) spectrum.  
(See \S\ref{mollines} and Table~\ref{jcmtfitstable}.) 
A velocity range of \q80\kms\/ is shown, centered on
the velocity of the \hisoco\/ line, since SiO is not detected at the 3$\sigma$
level in this source.  The dotted vertical line marks the \hisoco\/
velocity from Table~\ref{jcmtfitstable}.   
}
\end{figure}

\begin{figure}
\plotone{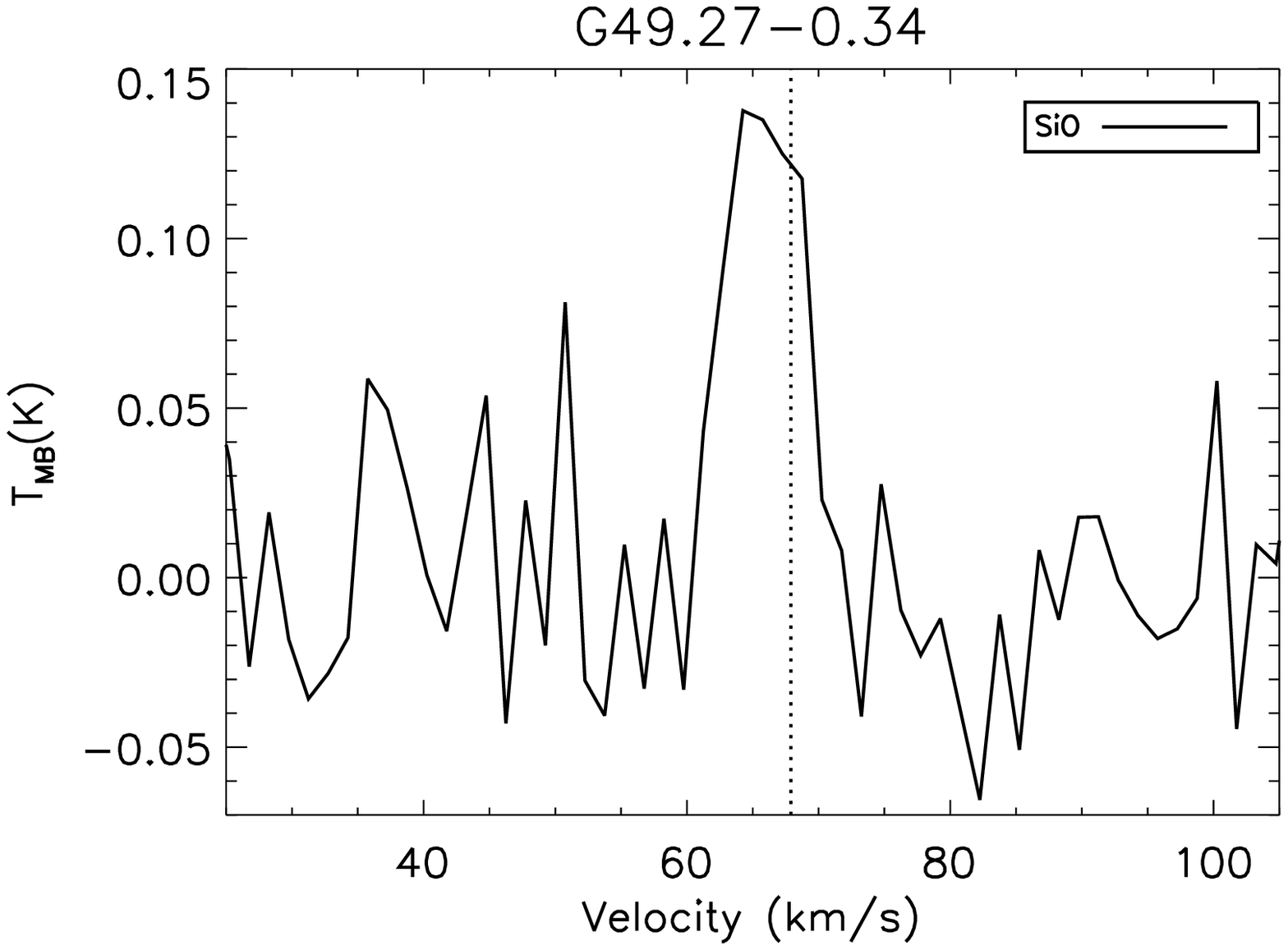}
\addtocounter{figure}{-1}
\caption{Figure Set 4: JCMT thermal SiO (5-4, \elow=20.8 K) spectrum.  
(See \S\ref{mollines} and Table~\ref{jcmtfitstable}.) 
A velocity range of \q80\kms\/ is shown, centered on
the velocity of the SiO line.  The dotted vertical line marks the \hisoco\/
velocity from Table~\ref{jcmtfitstable}.   
}
\end{figure}

\clearpage
\begin{deluxetable}{lccccccc}
\addtocounter{table}{-4}
\tablewidth{0pt}
\tablecaption{Full Table: Fitted Properties: 6.7 GHz Masers} 
\tablehead{ 
\colhead{Field Name} & 
\multicolumn{4}{c}{J2000.0 Coordinates} & 
\colhead{Intensity} &
\colhead{dI} &
\colhead{Velocity} \\
\colhead{(Target EGO)} &
\colhead{$\alpha$ ($^{\rm h}~~^{\rm m}~~^{\rm s}$)} &   
\colhead{dx (\pp)} & 
\colhead{$\delta$ ($^{\circ}~~{'}~~{''}$)} &
\colhead{dy (\pp)} & 
\colhead{(Jy beam$^{-1}$)} &
\colhead{(Jy beam$^{-1}$)} &
\colhead{(\kms)} 
}
\tablecolumns{8}
\tabletypesize{\scriptsize}
\setlength{\tabcolsep}{0.05in}
\startdata
G10.29-0.13 & 18 08 49.359 & 0.073 & -20 05 58.989 & 0.068 & 0.52 &0.03 & 1.70\\
G10.29-0.13 & 18 08 49.361 & 0.032 & -20 05 58.914 & 0.030 & 1.26 &0.03 & 1.83\\
G10.29-0.13 & 18 08 49.361 & 0.023 & -20 05 58.947 & 0.021 & 1.94 &0.03 & 1.97\\
G10.29-0.13 & 18 08 49.361 & 0.021 & -20 05 58.968 & 0.019 & 2.16 &0.03 & 2.11\\
G10.29-0.13 & 18 08 49.362 & 0.019 & -20 05 58.970 & 0.017 & 2.36 &0.03 & 2.25\\
G10.29-0.13 & 18 08 49.363 & 0.019 & -20 05 58.983 & 0.018 & 2.47 &0.04 & 2.38\\
G10.29-0.13 & 18 08 49.364 & 0.023 & -20 05 59.002 & 0.021 & 1.87 &0.03 & 2.52\\
G10.29-0.13 & 18 08 49.360 & 0.040 & -20 05 58.976 & 0.037 & 0.91 &0.03 & 2.66\\
\enddata
\end{deluxetable}

\clearpage
\begin{deluxetable}{lccccccc}
\tablewidth{0pt}
\tablecaption{Full Table: Fitted Properties: 44 GHz Masers} 
\tablehead{ 
\colhead{Field Name} & 
\multicolumn{4}{c}{J2000.0 Coordinates} & 
\colhead{Intensity} &
\colhead{dI} &
\colhead{Velocity} \\
\colhead{(Target EGO)} &
\colhead{$\alpha$ ($^{\rm h}~~^{\rm m}~~^{\rm s}$)} &   
\colhead{dx (\pp)} & 
\colhead{$\delta$ ($^{\circ}~~{'}~~{''}$)} &
\colhead{dy (\pp)} & 
\colhead{(Jy beam$^{-1}$)} &
\colhead{(Jy beam$^{-1}$)} &
\colhead{(\kms)} 
}
\tablecolumns{8}
\tabletypesize{\scriptsize}
\setlength{\tabcolsep}{0.05in}
\startdata
G10.29-0.13 & 18 08 49.113 & 0.040 & -20 05 58.185 &  0.041 &  0.24 &0.03 &10.43\\
G10.29-0.13 & 18 08 49.112 & 0.017 & -20 05 58.178 &  0.017 &  0.60 &0.03 &10.60\\
G10.29-0.13 & 18 08 49.113 & 0.012 & -20 05 58.198 &  0.013 &  0.79 &0.03 &10.76\\
G10.29-0.13 & 18 08 49.113 & 0.014 & -20 05 58.218 &  0.014 &  0.69 &0.03 &10.93\\
G10.29-0.13 & 18 08 49.115 & 0.023 & -20 05 58.233 &  0.024 &  0.42 &0.03 &11.10\\
G10.29-0.13 & 18 08 49.118 & 0.056 & -20 05 58.254 &  0.057 &  0.18 &0.03 &11.26\\
G10.29-0.13 & 18 08 46.802 & 0.052 & -20 05 49.354 &  0.053 &  0.57 &0.09 &13.26\\
G10.29-0.13 & 18 08 46.807 & 0.049 & -20 05 49.323 &  0.050 &  0.60 &0.09 &13.42\\
\enddata
\end{deluxetable}

\end{document}